\documentstyle[12pt,epsf]{article}
\newcommand{\om}{$\omega$}
\newcommand{\eps}{$\epsilon$}

\newcommand{\rl}{$R_L(q,\omega)$}
\newcommand{\rt}{$R_T(q,\omega)$}

\newcommand{\et}{{\em et al.}}
\newcommand{\bq}{\begin{eqnarray}}
\newcommand{\eq}{\end{eqnarray}}
\newcommand{\beq}{\begin{equation}}
\newcommand{\eeq}{\end{equation}}
\begin{document}
\thispagestyle{empty}    % remove 1st % to remove pagenumbers 

\begin{center}
\LARGE
Longitudinal and Transverse Quasi-Elastic Response Functions of
Light Nuclei\\[1cm]
\large
J. Carlson$^a$, J. Jourdan$^b$, R. Schiavilla$^{cd}$, I. Sick$^b$ \\[5mm]
\normalsize
$^a$Theoretical Division, Los Alamos National Laboratory, Los Alamos, New Mexico, USA \\
$^b$Departement f\"ur Physik und Astronomie, Universit\"at Basel, Basel, Switzerland \\
$^c$Jefferson Lab, Newport News, Virginia, USA \\
$^d$Physics Department, Old Dominion University, Norfolk, Virginia, USA \\[3cm]
\begin{minipage}{12cm}
{\bf Abstract.} The $^3$He and $^4$He longitudinal and transverse
response functions are determined from an analysis of the world data
on quasi-elastic inclusive electron scattering.  The corresponding
Euclidean response functions are derived and compared to those calculated
with Green's function Monte Carlo methods, using realistic interactions
and currents.  Large contributions associated with two-body currents
are found, particularly in the $^4$He transverse response, in agreement
with data.  The contributions of two-body charge and current operators
in the $^3$He, $^4$He, and $^6$Li response functions are also studied
via sum-rule techniques.  A semi-quantitative explanation for the
observed systematics in the excess of transverse
quasi-elastic strength, as function
of mass number and momentum transfer, is provided.  Finally, a number of
model studies with simplified interactions, currents, and wave functions
is carried out to elucidate the role played, in the full calculation, by
tensor interactions and correlations. 
\end{minipage} 

\end{center}
\vspace*{3cm}
\newpage
\setcounter{page}{1}
%%%%%%%%%%%%%%%%%%%%%%%%%%%%%%%%%%%%%%%%%%%%%%%%%%%%%%%%%%%%%%%%%%%%%%%%%%%%%%%%
%\newcommand{\s}{$S(k,E)$}

\section{Introduction \label{intro}}
Over the past thirty years or so, much effort has gone into
trying to understand quantitatively the roles that
short-range and tensor correlations, and two-body components of
the nuclear electromagnetic current play in the quasi-elastic
response of nuclei at intermediate momentum transfers.  Yet,
despite the considerable attention that has been devoted to
this topic, many open questions remain.  Complications arise,
in particular, as a consequence of the need of (and technical
difficulties associated with) providing an accurate
description of the initial bound- and the final scattering-state wave
functions, based on realistic Hamiltonians.
 
In part, the slow progress is also due to the confusing experimental
picture, particularly for medium- and heavy-weight nuclei, that for
some time obfuscated the interpretation of the data.
Early data \cite{Whitney74} had shown that, in comparison to an
impulse-approximation (IA) calculation using a simple (Fermi gas) model,
the inclusive cross section showed an excess of transverse strength, mainly
in the region of the \lq\lq dip\rq\rq between the quasi-elastic
and the $\Delta$ peaks.  This excess was attributed to two-body
currents and $\pi$-production, but not quantitatively understood.
 
The longitudinal and transverse response functions, obtained
during the eighties from a Rosenbluth separation of experimental
cross sections, seemed to indicate that, in addition,
there was a gross (up to 40\%) lack of {\em longitudinal}
strength in the main quasi-elastic peak, and a correspondingly too low 
Coulomb sum rule \cite{Altemus80,Meziani84}.  While this state of affairs
is yet to be resolved satisfactorily, particularly
for very heavy nuclei like lead,
where Coulomb corrections are difficult
%in treatments of Coulomb distortion effects --- negligible
%for light nuclei --- based on effective-momentum~\cite{Traini88,Traini95}
%and distorted-wave~\cite{Onley92,Kim96} approximations, there
\cite{Onley92,Kim96,Traini88,Traini95}
there are nevertheless clear indications for $A \leq 56$ from the work of 
Jourdan \cite{Jourdan96c}, who carefully analyzed the {\em world} data on
quasi-elastic scattering including all the known corrections, that there
is no missing strength in the longitudinal response of medium-A nuclei
%(this conclusion in the views of two of the authors
%of the present work, J.C. and R.S., is still controversial, however,
(for very heavy nuclei, there are still controversial issues on the Coulomb
corrections,  
see Refs.~\cite{Traini01,Morgenstern01,Kim01}).
This apparent lack of longitudinal strength has absorbed much of the
theoretical effort of the past two decades.

On the other hand, the excess of transverse strength --- presumably due to two-body
currents --- observed in the quasi-elastic region appears to be a genuine
problem.  In this respect, the experimental situation has been put in sharp
focus by the work on super-scaling by Donnelly and Sick \cite{Donnelly99a}
which allowed to systematically compare the longitudinal and
transverse response functions.  This work showed in the most clear way that the 
transverse strength for nuclei with mass number $A$=12, $\dots$, 56 
exceeds the longitudinal one already in the main quasi-elastic
peak by 20-40\%, in addition to the excess of strength occurring
in the \lq\lq dip\rq\rq between the quasi-elastic and $\Delta$-peaks.
This excess of strength in the region of the quasi-elastic peak is the
main subject of this paper.  The region of
the dip, which has attracted the attention in the past, will be
largely ignored as the understanding of this region is clouded by issues
relating to pion production and the $\Delta$-tail. 

Theoretical calculations of two-body contributions in the region of
the quasi-elastic peak have been performed by many groups
\cite{Donnelly78}--\nocite{VanOrden81,Kohno81,Blunden89,Leidemann90,Dekker94,Amaro92,Alberico84}
\nocite{Carlson94,Vandersluys95,Anguiano96,Fabrocini97a,Gadiyak98}
\cite{Bauer00} using different approaches.
Some of these calculations find appreciable contributions, 20--40\% of the 
transverse response, due to the dominant two-body terms (pion contact and
in-flight, and $\Delta$-excitation diagrams), other calculations find small,
$<$10\%, effects.  It is not always clear why calculations with similar
starting assumptions give very different results. 

In general, calculations based on an independent-particle initial state 
(shell model, Fermi gas model, possibly with RPA correlations added) give very
small two-body contributions in the quasi-elastic peak
\cite{Donnelly78}--\nocite{VanOrden81,Kohno81,Blunden89,Dekker94,Amaro92}\cite{Alberico84},
\cite{Anguiano96,Bauer00}: the pion and $\Delta$ terms 
tend to cancel to produce a small overall effect.  The origin of bigger 
(20--50\%) contributions to the transverse response as obtained in 
Refs.~\cite{Vandersluys95,Gadiyak98} is not entirely understood: the 
different treatment of the $\Delta$ in matter in Ref.~\cite{Vandersluys95}
may be partly responsible \cite{Ryckebusch01}. 

The model study of Leidemann and Orlandini \cite{Leidemann90},
in which the nuclear response was expressed in terms of the
response of deuteron-like pairs of nuclear density, first
pointed out that it is important to account in the initial
state for the {\em tensor} correlations between $n$$p$ 
pairs.  Only when these (rather short-range) tensor correlations
were included would the two-body terms give appreciable contributions
to the quasi-elastic response.  This insight was 
quantitatively confirmed by Fabrocini \cite{Fabrocini97a}, who calculated the 
transverse response of infinite nuclear matter using correlated 
basis function theory including one-particle-one-hole
intermediate states.  This calculation
is based on a realistic nucleon-nucleon (N-N) interaction and two-body terms 
derived  consistently from the N-N interaction, and accounts for the 
interactions in both the initial and final states.  It also found that
substantial two-body contributions in the quasi-elastic peak are obtained
only if the tensor correlations, predominantly induced by pion exchange,
are retained. 

The calculation of Carlson and Schiavilla \cite{Carlson94} was
performed for $^4 {\rm He}$ using Green's function Monte Carlo (GFMC)
techniques, a realistic (Argonne $v_8^\prime$) N-N interaction and
again consistently constructed two-body terms.  The inelastic response
could be accurately calculated in terms of the
Euclidean response (an integral over the response function, see below).
These\lq\lq exact\rq\rq calculations found that the charge-exchange
character of the N-N interaction leads to shifts of
both the longitudinal and transverse strength to higher excitation
energies, thus producing a quenching of the response in the
region of the quasi-elastic peak.  This mechanism, however, is
more than offset in the transverse channel by two-body currents, in
particular those associated with pion exchange (required by gauge
invariance), and hence the response is enhanced over the entire
quasi-elastic spectrum.  This enhancement was found to be substantial,
and in agreement with that observed experimentally.  The study of
Ref.~\cite{Carlson94}, while providing a qualitative understanding
of the $^4$He quasi-elastic response, did not identify quantitatively, however,
those aspects of the calculation responsible for the successful prediction.

In the present paper we study the longitudinal and transverse response
functions of light nuclei, $^3$He and $^4$He, using GFMC theory.  Accurate
data for these responses in the region of the quasi-elastic peak are
determined via an analysis of the world data.  A simultaneous study
of $^3$He and $^4$He is particularly interesting as the predicted
two-body contributions in the transverse channel increase very rapidly between 
$A$=3 and $A$=4, a feature which can give us a further handle for the 
understanding of two-body effects.  The study of $^4 {\rm He}$, including
the higher momentum transfers now available, is especially promising, since
the available data \cite{Jourdan96c} seem to indicate 
that the relative excess of transverse strength in the quasi-elastic 
peak is largest for this nucleus.  In order to include heavier nuclei and
hence examine the evolution with mass number of this excess transverse
strength, we also study via sum-rules the two-body contributions for p-shell
nuclei, for which variational Monte Carlo (VMC) wave functions are available.

The layout of this paper is as follows.  In Sec. \ref{exp} we discuss the
determination of the longitudinal and transverse response functions starting 
from the world data on inclusive electron scattering, while in Sec. \ref{scaling}
we perform a scaling analysis in order to investigate the global properties of the 
experimental response.  In Sec. \ref{euclidean} we describe the theory of the 
Euclidean responses which link the nuclear ground-state properties to integral
properties of the electromagnetic response as measured in (e,e$^\prime$), and
in Sec \ref{sec:MEC} present the model adopted for the nuclear electromagnetic
current.  Before comparing theory with experiment (Sec. \ref{results}),
we carry out in Sec. \ref{model} a model study of the relation between
the inclusive cross section and the Euclidean response in order to better
understand the characteristics of the latter.  An extension of the study
to heavier nuclei via sum rules is given in Sec. \ref{sec:sum}, in which the
dependence of the excess transverse strength upon mass number is examined.
In Sec. \ref{ingredients} we further analyze the calculated results by 
introducing various simplifications, so as to identify the most important
aspects of the calculations.  Finally, in Sec. \ref{conclusions} we summarize
our conclusions.

\section{Experimental response functions \label{exp}}
In order to determine the longitudinal ($L$) and transverse ($T$) responses, we have 
analyzed the (e,e$^\prime$) world data on $^3$He and $^4$He.
A determination of the response functions in inclusive quasi-elastic 
scattering from the world cross section data has many advantages over
the traditional approach of using data from a single experiment only.
Particularly for medium-$A$ nuclei the limitations of the traditional 
approach was partly responsible for the misleading conclusions mentioned
in the introduction and discussed in \cite{Jourdan96c}. 

For the extraction of the response functions, the difference of  
cross sections at high-energy/forward-angle and at 
low-energy/backward-angle is used.
Kinematics dependent systematic errors do 
not cancel in this difference even for measurements performed at a single 
facility, except perhaps for the errors in the 
overall normalization  of the cross sections.
The dominant systematic errors, {\em i.e.} uncertainties in the spectrometer 
acceptance, detector efficiencies, background contributions, 
re-scattering, and radiative corrections are strongly dependent on the 
specific kinematics.

To improve the determination of the response functions
the difference of the $L$- and $T$-contributions to the cross sections has to be 
maximized by including data over the largest possible angular range.
This can only be achieved by including all available world
cross section data. For $^3$He and $^4$He the use of the world data 
not only expands the range of available data in scattering angle
but also increases the  range of momentum transfer $q$ where a
separation can be done, thus leading to new information on the
response functions.

At low $q$ the extensive sets of data for $^3$He \cite{Marchand85,Dow88}
and for $^4$He \cite{Zghiche94,Reden90} with good angular coverage have been
used in the present analysis.  For $^3$He at high $q$ the data by 
Marchand {\em et al.} \cite{Marchand85} and
by Dow {\em et al.} \cite{Dow88}, which both cover the angular region
from 90$^{o}$ to 144$^{o}$, are combined
with complementary cross sections by Day {\em et al.} \cite{Day79} which provide
high-energy/forward-angle data with energies up to 7.2~GeV at 
scattering angles of 8$^{o}$.
Similarly for $^4$He the data by Zghiche {\em et al.} \cite{Zghiche94}, which cover the 
angular region from 75$^{o}$ to 145$^{o}$, and by Dytman {\em et al.} 
\cite{Reden90} which contribute data at 60$^{o}$, 
are complemented with the forward angle cross sections
by Rock {\em et al.} \cite{Rock82a}, Day {\em et al.} \cite{Day93}, 
Sealock {\em et al.} \cite{Sealock89}, and Meziani {\em et al.} \cite{Meziani92}
covering the angular range from 8$^{o}$ to 37$^{o}$.

In contrast to the analysis performed for medium-$A$ nuclei \cite{Jourdan96c},
Coulomb distortions play a negligible role for $^3$He and $^4$He and no
corrections need to be applied.  The following expression, valid 
in the plane wave Born approximation (PWBA), is used for the
Rosenbluth or $L/T$-separation:
\begin{equation}
\Sigma(q,\omega,\epsilon) = 
\frac{d^2\sigma}{d\Omega d\omega} \, \frac{1}{\sigma_{Mott}}
\, \epsilon \, \left(\frac{q}{Q}\right)^4 = 
\epsilon \,  R_L(q,\omega) + \frac{1}{2}\left(\frac{q}{Q}\right)^2
\, R_T(q,\omega) \label{e:resp_tot} \>\>,
\end{equation}
where the longitudinal virtual photon polarization \eps~ is defined as
\begin{equation}
\epsilon = \left(1 + \frac{2q^2}{Q^2} \tan^2{\frac{\vartheta}{2}}\right)^{-1} \>\>,
\end{equation}
and varies between 0 to 1 as the electron scattering angle $\vartheta$ 
ranges from 180 to 0 degrees.  Here, $d^2\sigma/d\Omega d\omega$ are the 
experimental cross sections, \om, $q$ and $Q$ are the energy transfer,
3- and 4-momentum transfers, respectively, and $\sigma_{Mott}$ is the
Mott cross section.  The structure of Eq.~(\ref{e:resp_tot}) shows that measurements
of the cross section at fixed \om~ and $q$ but different \eps~ 
allow for a separation of the two response functions \rl\ and \rt. 

In practice, the experimental spectra of the various experiments were  
measured for a given incident energy and scattering angle as a function 
of the energy loss of the scattered electron, varied by changing 
the magnetic field of the spectrometer.  To determine the cross section
at given values of $q$ and \om, the data have to be interpolated.
This traditionally was done by dividing out $\sigma_{Mott}$ from the
measured cross sections and interpolating the responses along \om/$E$. 

\begin{figure}[tbh]
\centerline{\mbox{\epsfysize=55mm\epsffile{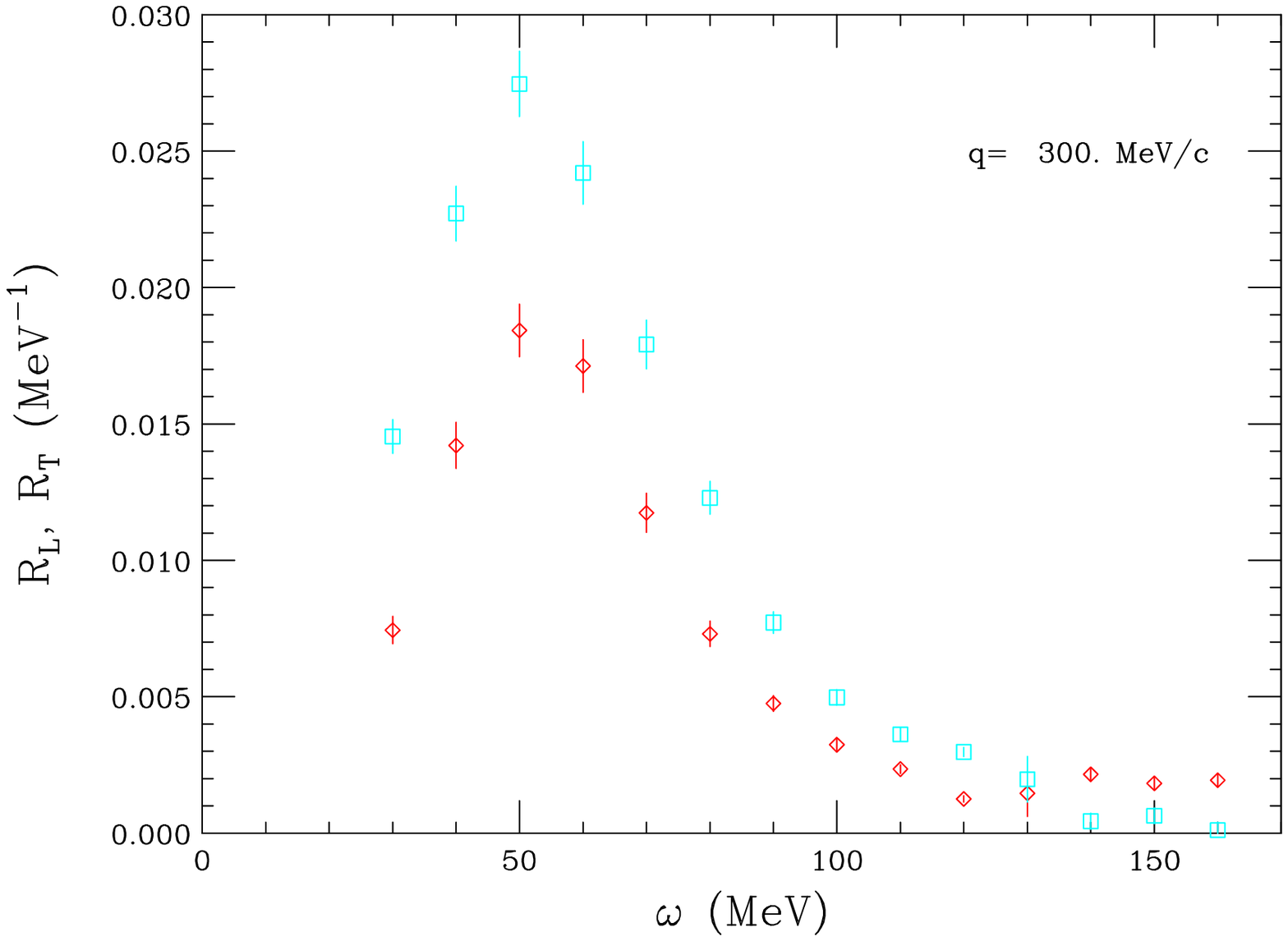}
\hspace{7mm} \epsfysize=55mm\epsffile{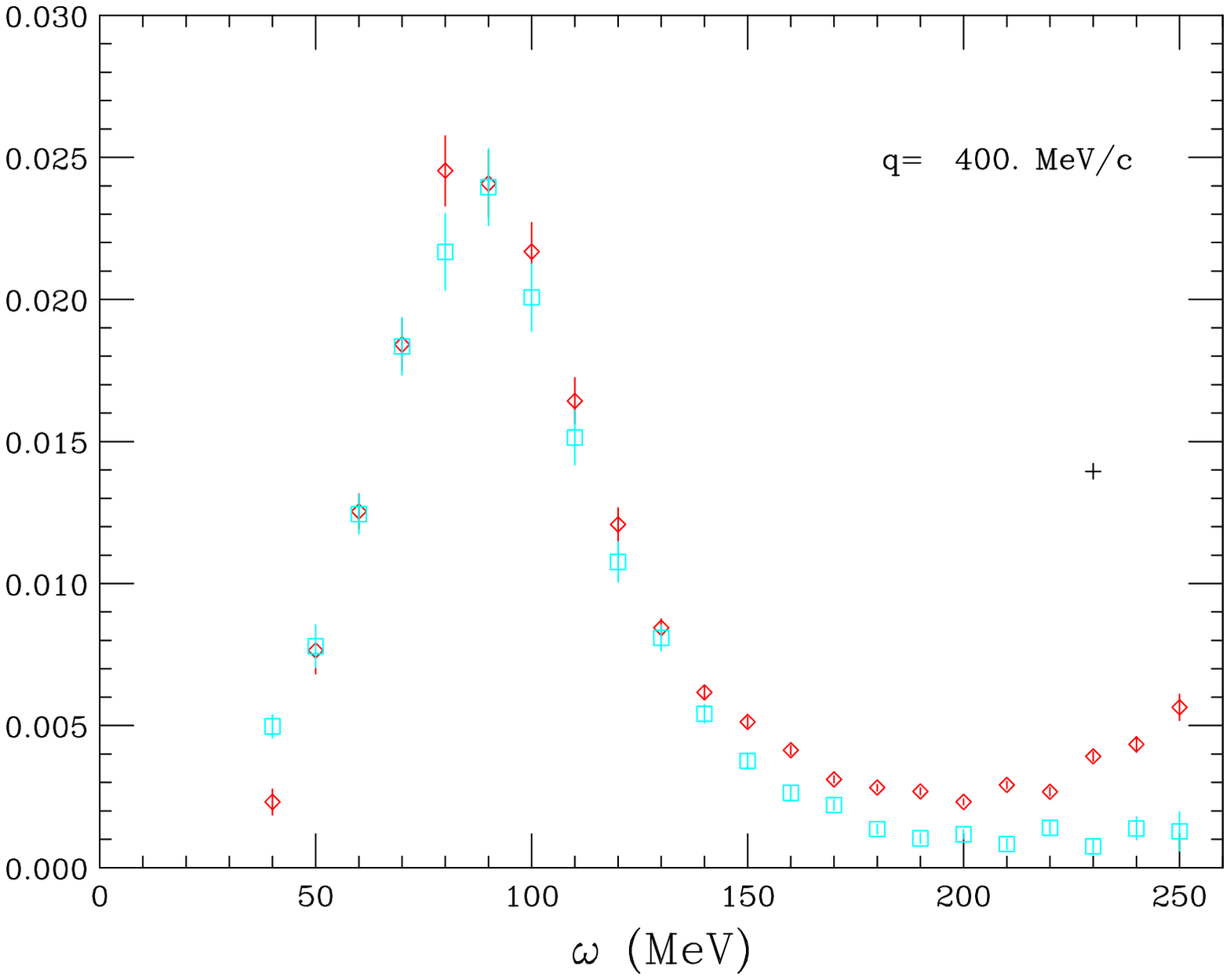}}}
\centerline{\mbox{\epsfysize=55mm\epsffile{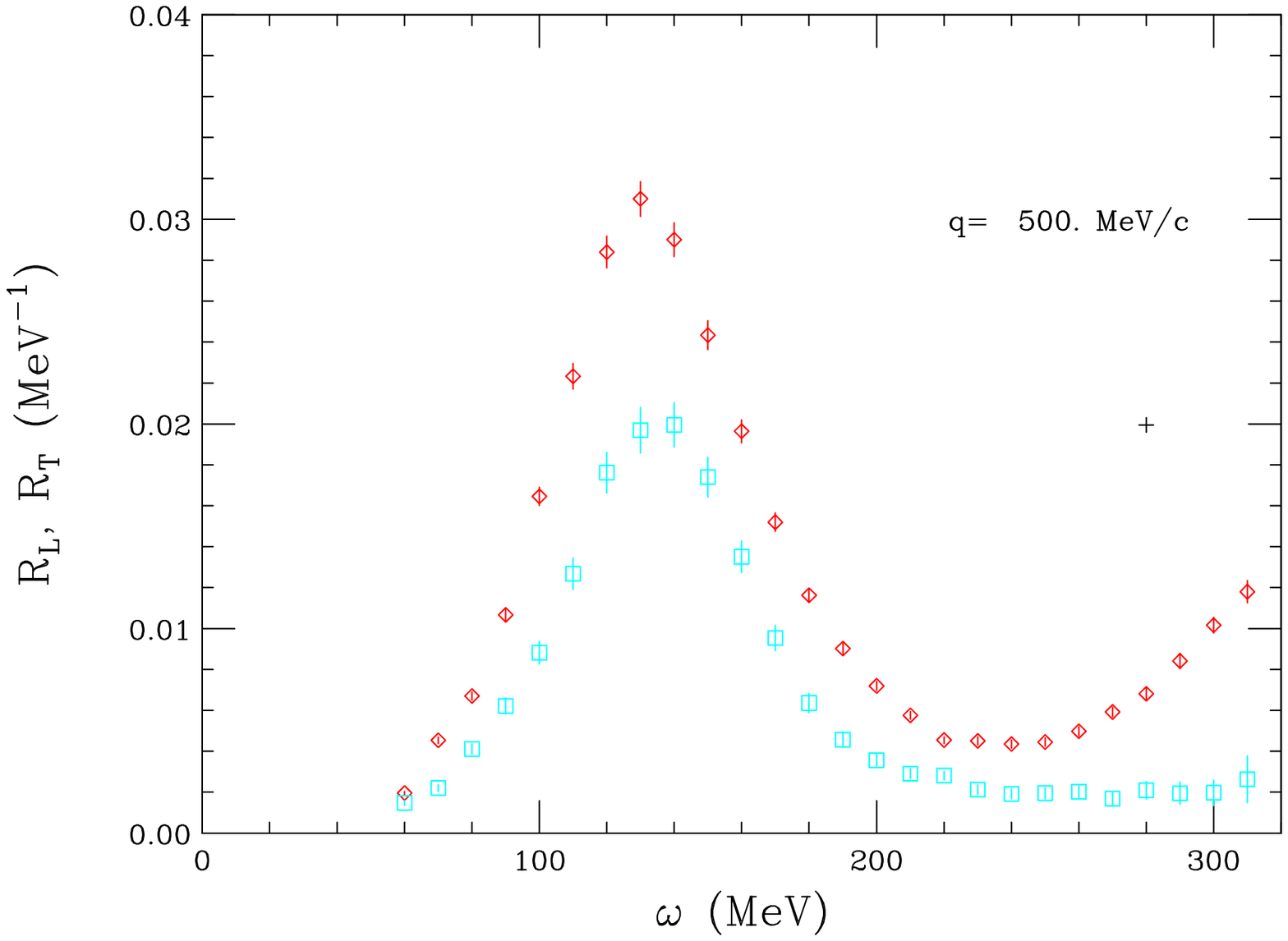}
\hspace{7mm} \epsfysize=55mm\epsffile{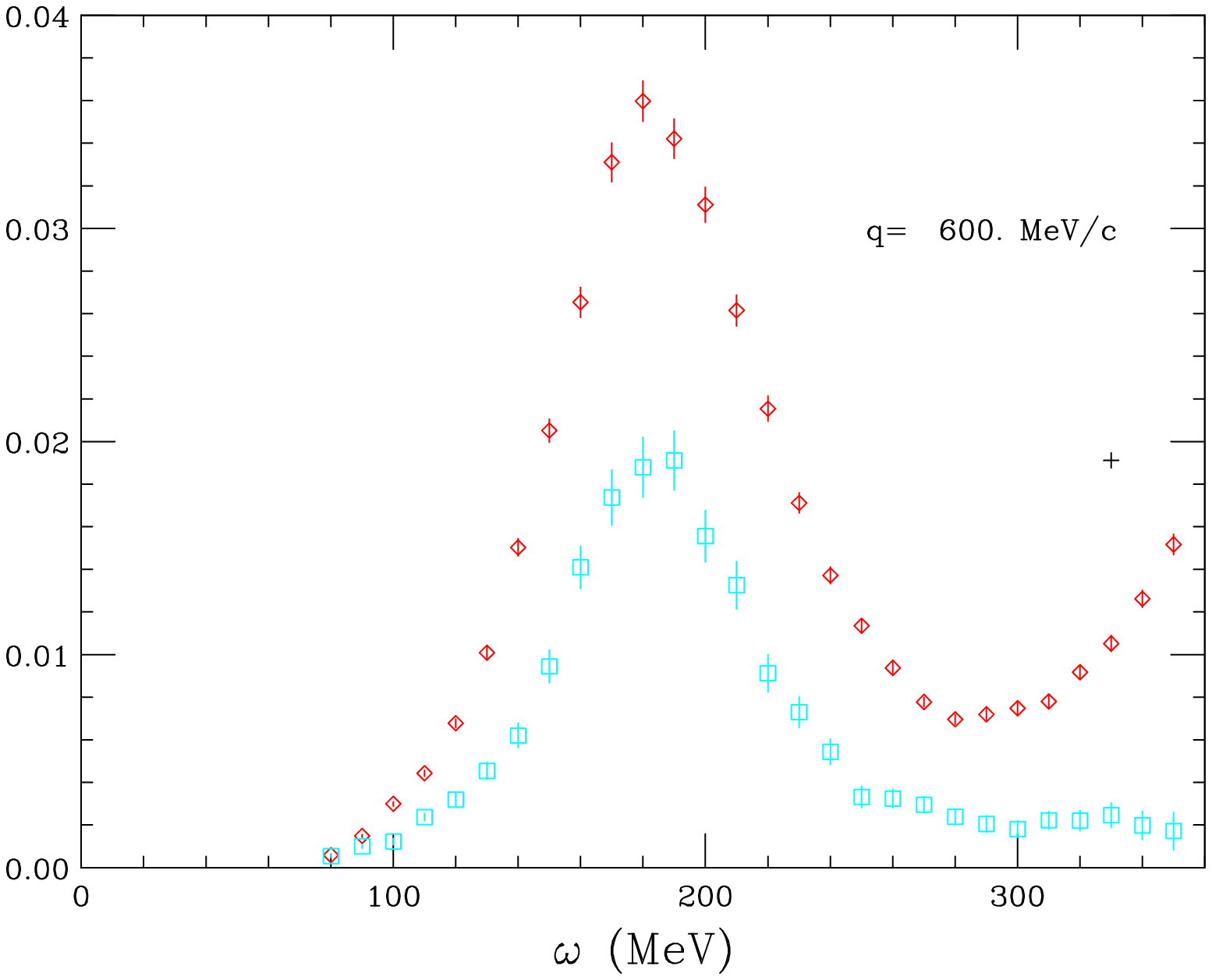}}}
\centerline{\mbox{\epsfysize=55mm\epsffile{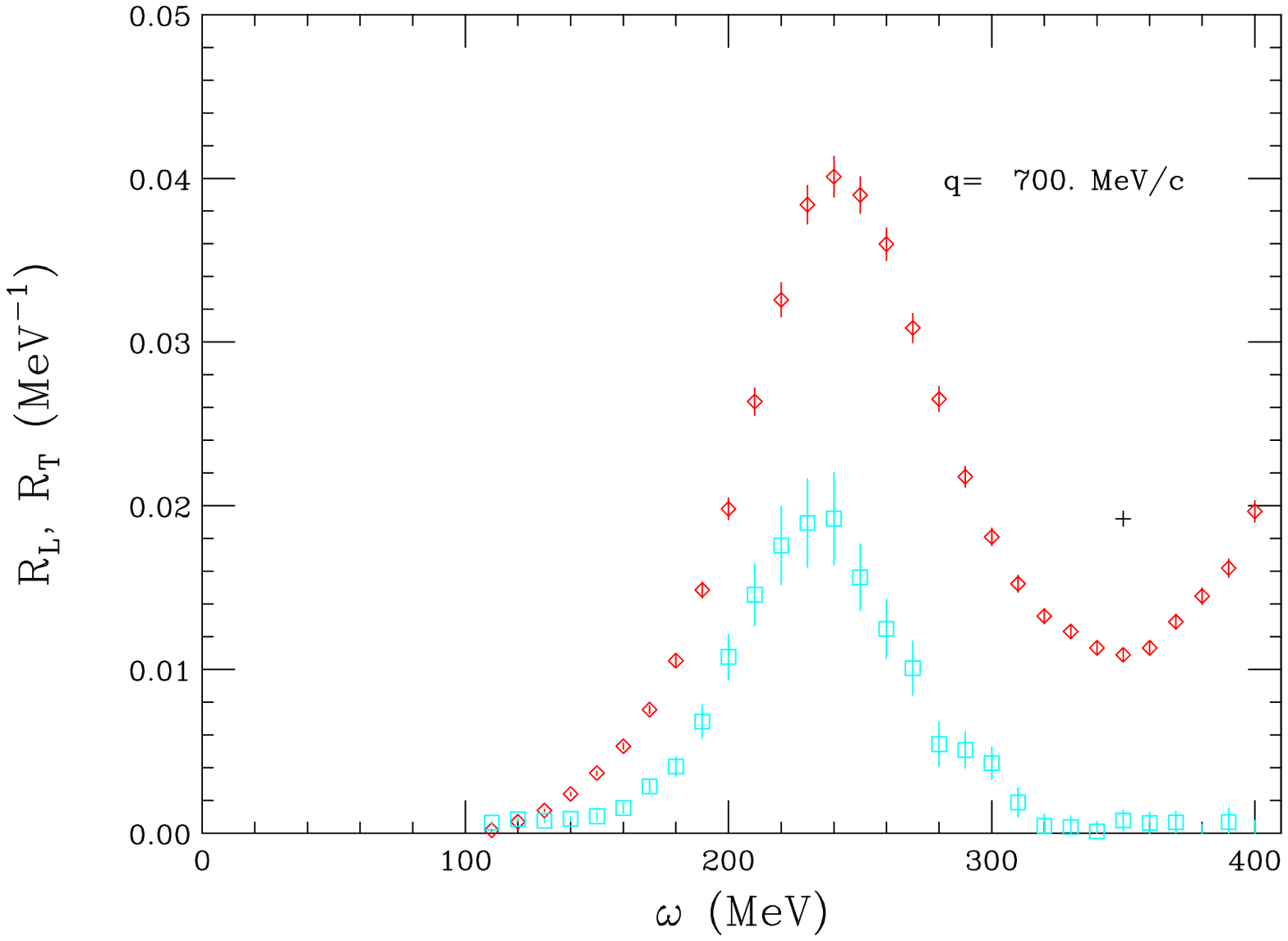}
 \hspace*{7.3cm}}}
\vspace{-5cm}  \hspace{9.cm}
 \parbox{6cm}
{\caption[]{Longitudinal ($\Box$) and transverse ($\Diamond$) response functions
 of $^3$He at momentum transfers of 300, 400, 500, 600 and 700 MeV/c.
Indicated with a + are the upper integration limits
used for the Euclidean response (section \ref{results}).
}\label{rhe3}}
%\vspace*{2cm}
\end{figure}

\begin{figure}[tbh]
\centerline{\mbox{\epsfysize=55mm\epsffile{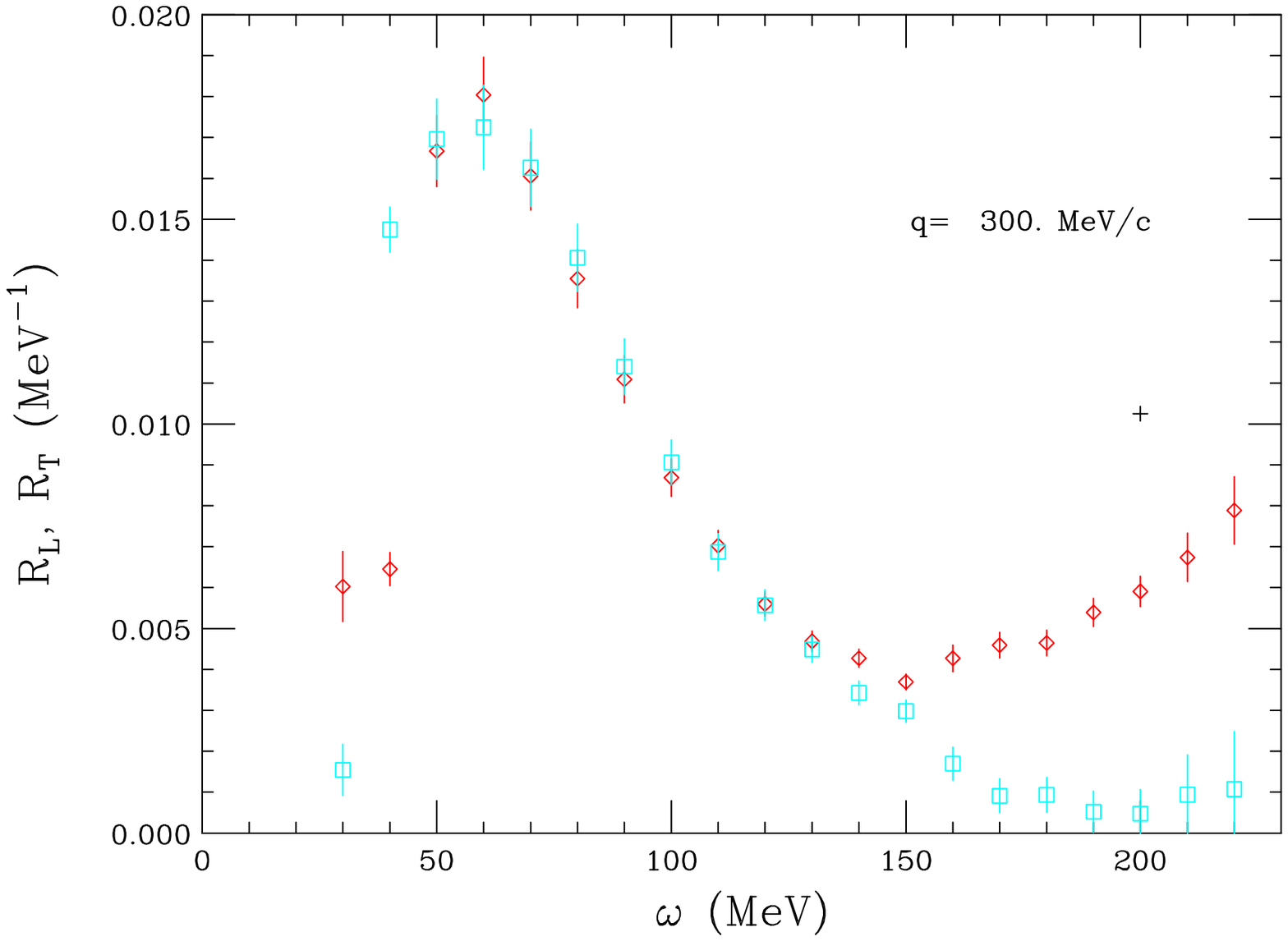}
\hspace{7mm} \epsfysize=55mm\epsffile{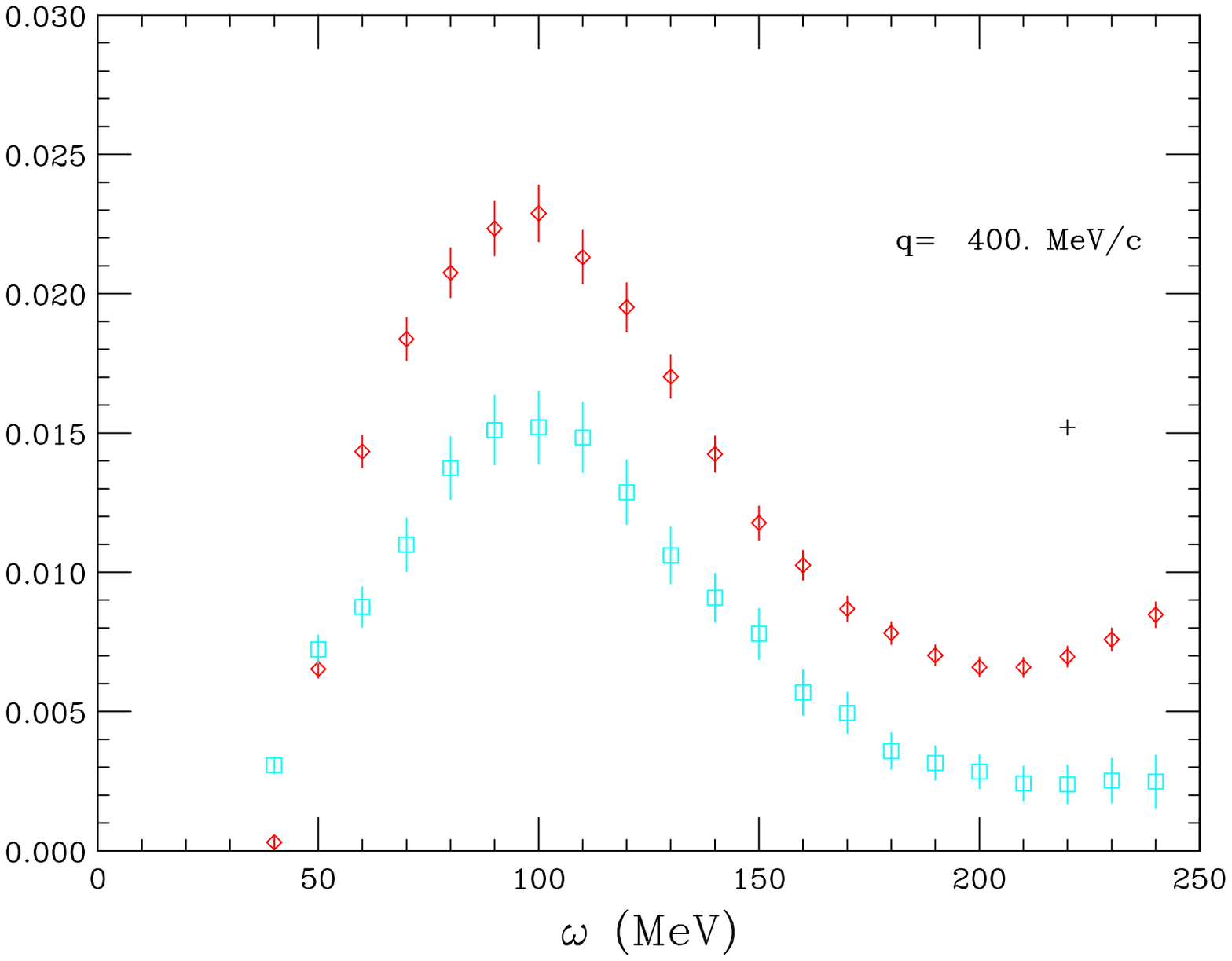}}}
\centerline{\mbox{\epsfysize=55mm\epsffile{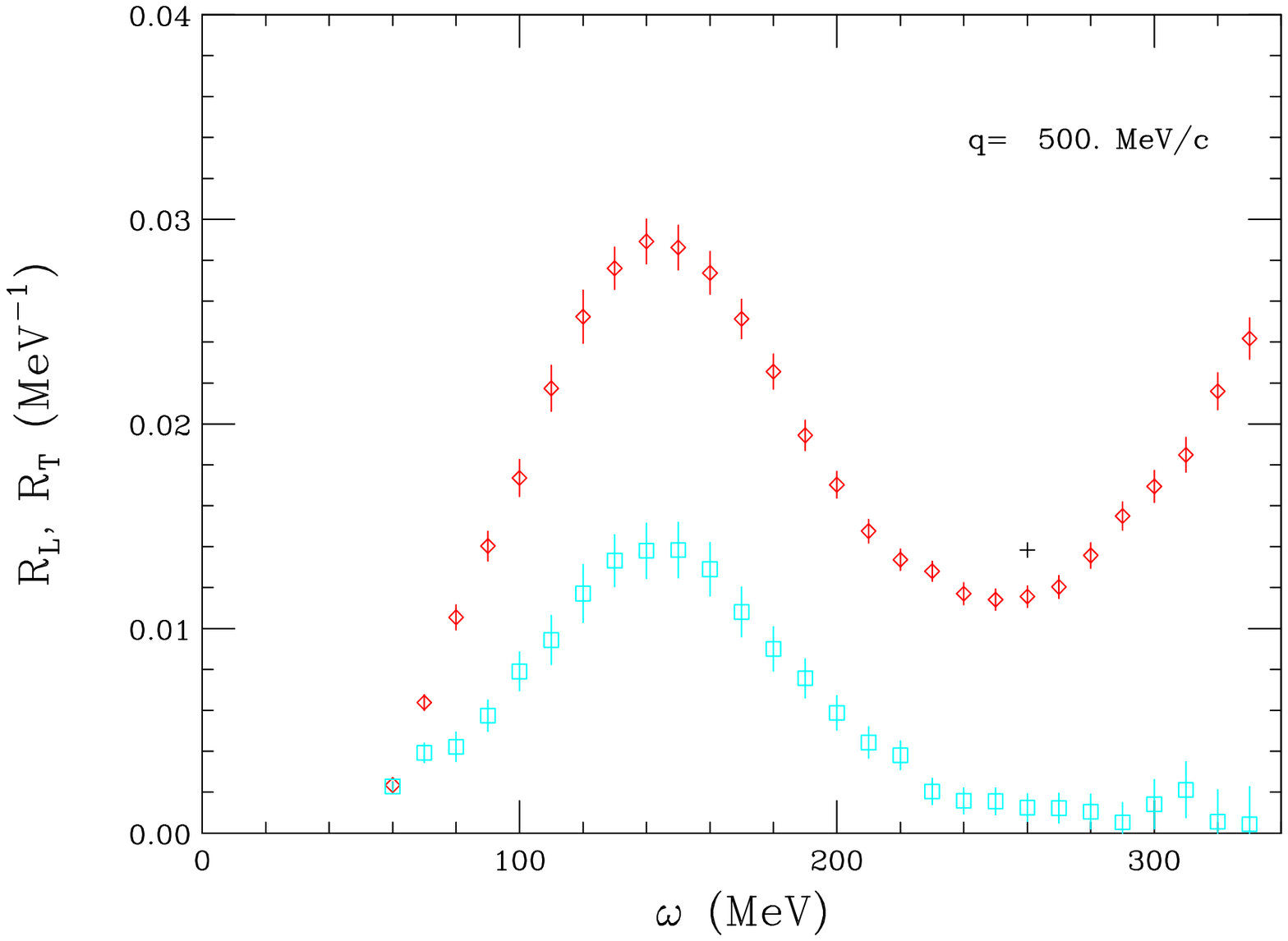}
\hspace{7mm} \epsfysize=55mm\epsffile{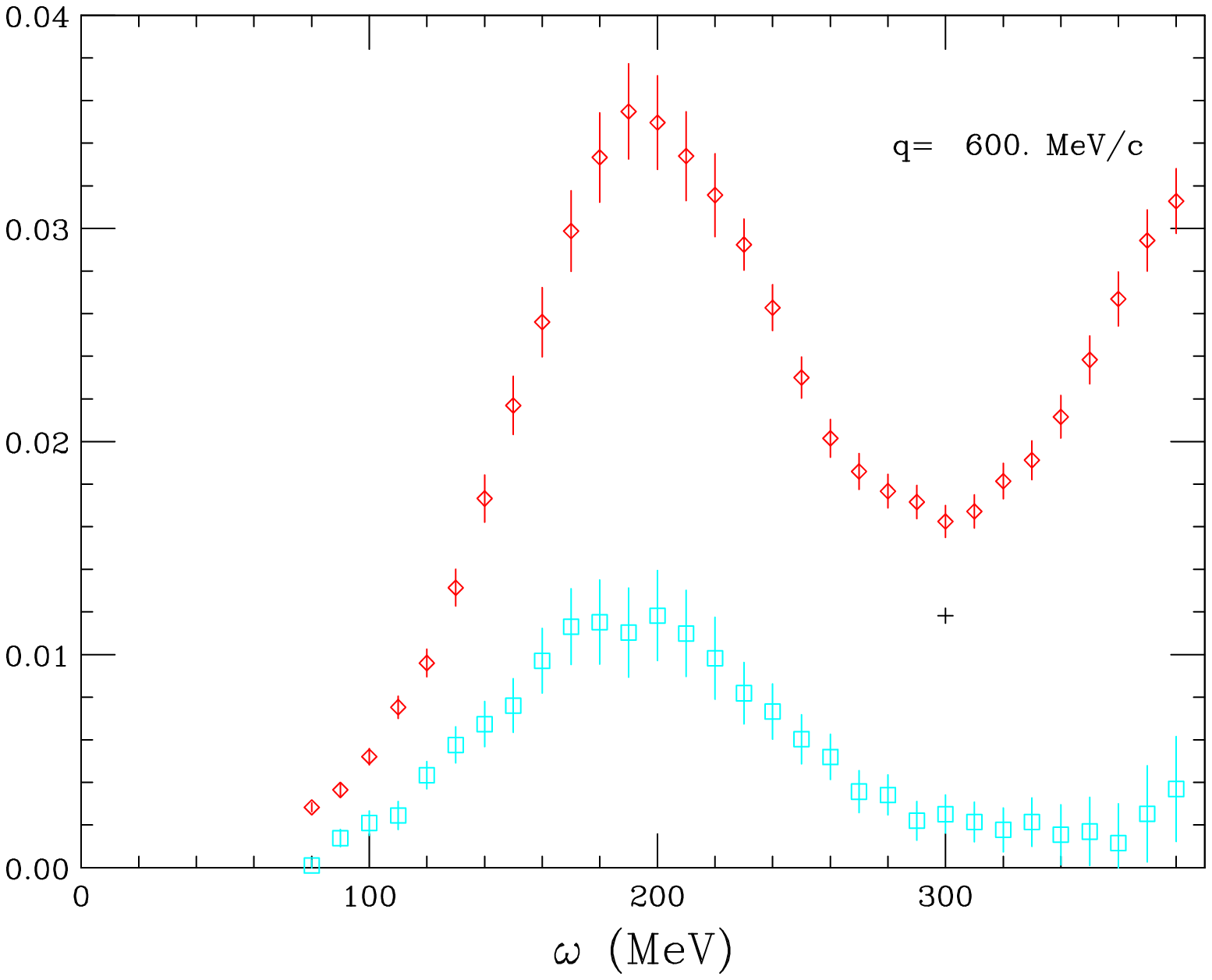}}}
\centerline{\mbox{\epsfysize=55mm\epsffile{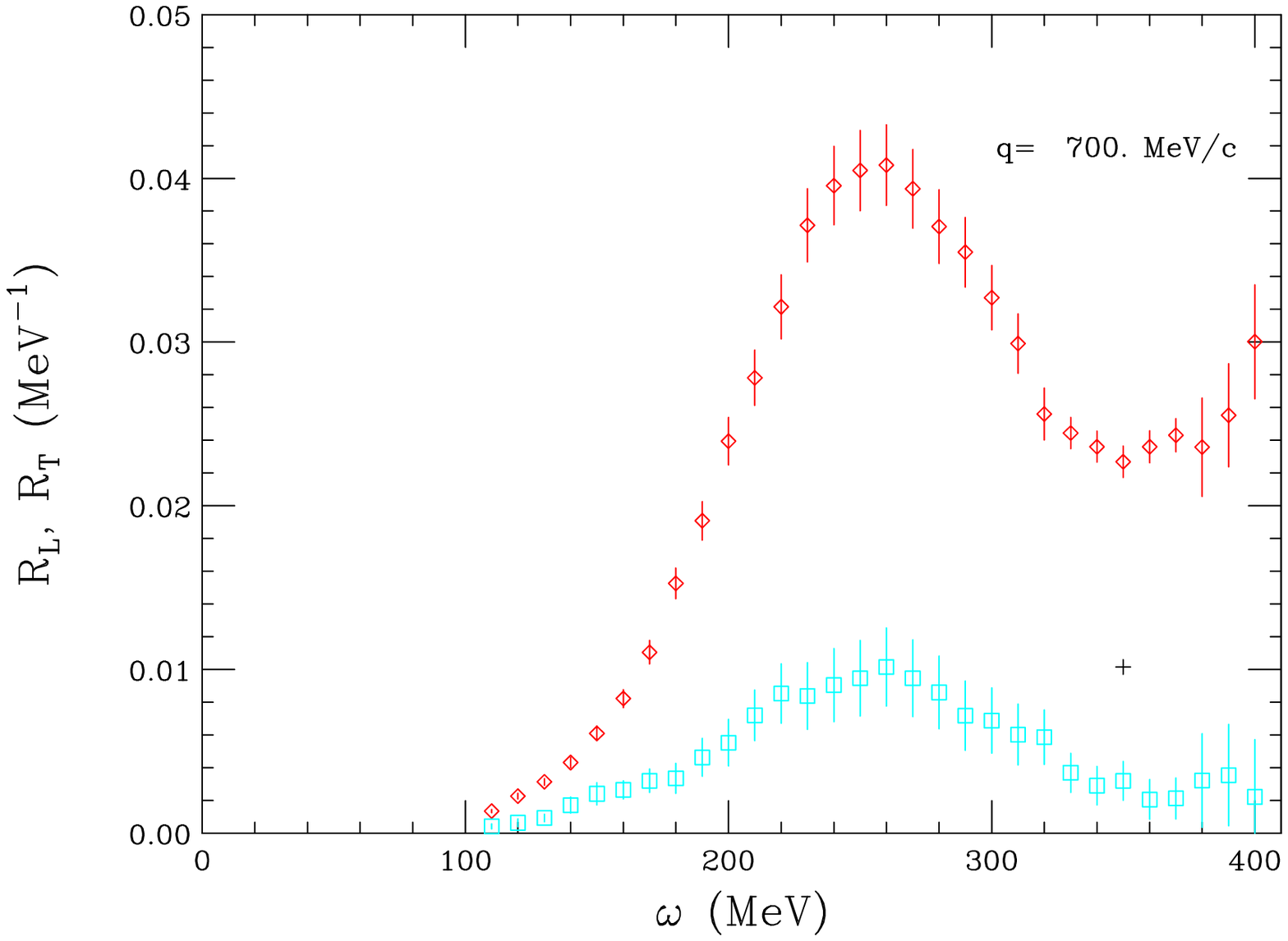}
 \hspace*{7.3cm}}}
\vspace{-5cm}  \hspace{9.cm}
 \parbox{6cm}
{\caption[]{Longitudinal ($\Box$) and transverse ($\Diamond$) response functions
 of $^4$He at momentum transfers of 300, 400, 500, 600 and 700 MeV/c.
Indicated with a + are the upper integration limits
used for the Euclidean response (section \ref{results}).
%}
}\label{rhe4}}
%\vspace*{2cm}
\end{figure}                                                                    

In an analysis of the world data, where the various experiments 
were not planned for an ideal coverage of the $q$-\om--plane,
the usual scheme is unreliable due to occasionally large 
spacings between various spectra.  Thus, in the present analysis 
an improved scheme is employed by first dividing out 
an appropriate sum of elementary electron-nucleon cross sections,
{\em i.e.} $\sigma_{ep}$ for the proton and $\sigma_{en}$ 
for the neutron, and removing kinematical dependencies.  Essentially what is 
calculated from the data is the scaling function $F(y,q)$ defined by 
\begin{equation}
F(y,q) = \frac{d^2\sigma}{d\omega d\Omega} \, \frac{1}
{Z\sigma_{ep}(q) + N\sigma_{en}(q)}\frac{d\omega}{dy}.
\label{e:fy}
\end{equation}
The scaling variable $y$ is fixed by energy and momentum conservation via
\begin{equation}
 y = -q + \sqrt{\omega^2 + 2\omega \, m} \>\>,
\label{e:y}
\end{equation}
neglecting small contributions from the binding energy, the 
perpendicular component of the nucleon momentum, and the recoil energy of the
residual nucleus.  In the next phase of the analysis, the extracted
$F(y,q)$ are then used to determine $F(y,q_o)$ at the desired value
$q_o$ by interpolating $F(y,q)$ along lines of constant $y$.

For $y<0$, $F(y,q)$ is known to be nearly independent of $q$ over a large 
range.  This makes the present interpolation as reliable as the conventional
interpolation scheme even if the data are separated by large values of $q$.
For $y>0$, the dependence of $F(y,q)$ on $q$ is relatively more severe, since 
inelastic processes contribute to the cross section.  Thus, the $q$-value 
of the $L/T$-separation has been chosen to minimize the correction
due to the interpolation at large \om. 

The interpolation procedure and the separation has been tested with 
the data of the Saclay experiments \cite{Zghiche94,Marchand85} alone. 
Provided the same interpolation scheme is used, the published values 
of \rl~ and \rt~ are reproduced exactly.  The improved interpolation
scheme, using $y$-scaling, gives results which also are identical
within the statistical errors.

With the interpolated cross section data, the response functions are extracted
for $q$=300--700 MeV/c in steps of 100 MeV/c for both nuclei.
The combined world data cover almost the full \eps-range, with typical 
values ranging from 0.05 to 0.95 for most $q$-sets.  At high $q$ this 
has to be compared with the results of the low energy data alone
which only cover the region from 0.05 to 0.55.  In addition, with a
global analysis it is possible to determine for the first time 
the response functions at $q$=700 MeV/c.

If the interpolated responses of data are plotted as a function of \eps,
a linear dependence is expected.  In contrast to the analysis of
medium-$A$ nuclei, in which important deviations were observed for high $q$ and \om,
no significant deviations were observed in the present analysis
once the quoted systematic errors of the individual data sets are included. 

The longitudinal and transverse response functions resulting from this
analysis of the {\em world} data are shown in Figs.~\ref{rhe3} and \ref{rhe4}.

\section{Scaling analysis \label{scaling}}
In order to show the excess strength of \rt, in this section we study the 
scaling properties of the present response functions. 
Barbaro \et~\cite{Barbaro98} have discussed the close connection between
the Coulomb sum rule and the notion of $y$-scaling.
More recently the notion of $\psi^\prime$-scaling was introduced by 
Alberico \cite{Alberico88} while studying the properties of the
relativistic Fermi gas model.  The application of this notion
to finite nuclear systems, requiring the inclusion of binding effects, has 
been discussed by Cenni \et\ \cite{Cenni97}.
Guided by these results, Sick and Donnelly have applied $\psi^\prime$-scaling 
to a large body of inclusive scattering data \cite{Donnelly99a}.
Scaling in $\psi^\prime$ has the merit to allow the study of the scaling 
properties for a combined set of {\em different} nuclei.  The only 
relevant scale parameter in the quasi-free scattering regime is 
the Fermi momentum of the nucleus which is taken into account
in the definition of the dimensionless scaling variable $\psi^\prime$ 
(approximately given by $y/k_F$), and the scaling function $f(\psi^\prime)$. 

%\vspace*{-10mm}
\begin{figure}[htb]
\centerline{\mbox{\epsfysize=70mm\epsffile{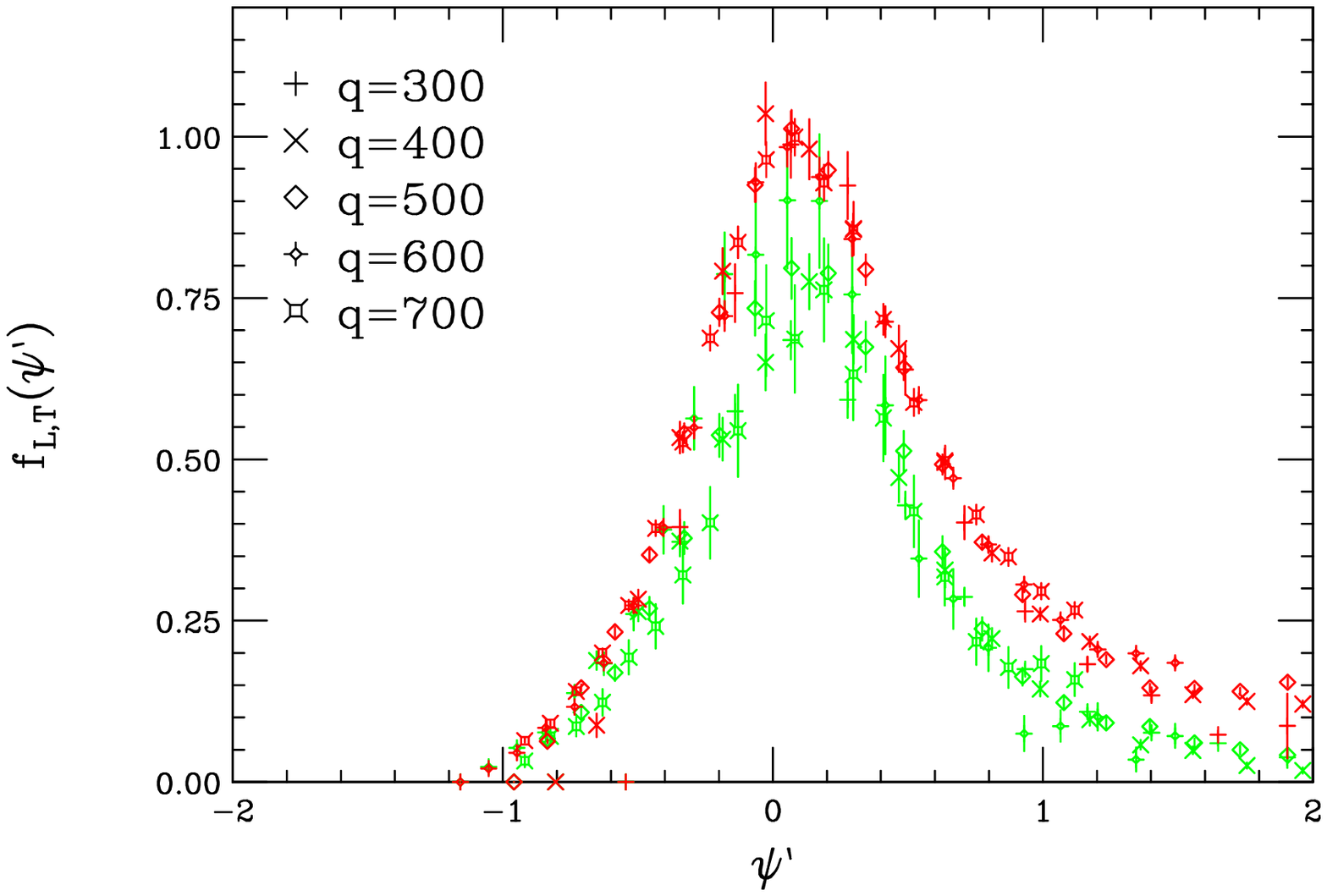}}}
\centerline{\mbox{\epsfysize=70mm\epsffile{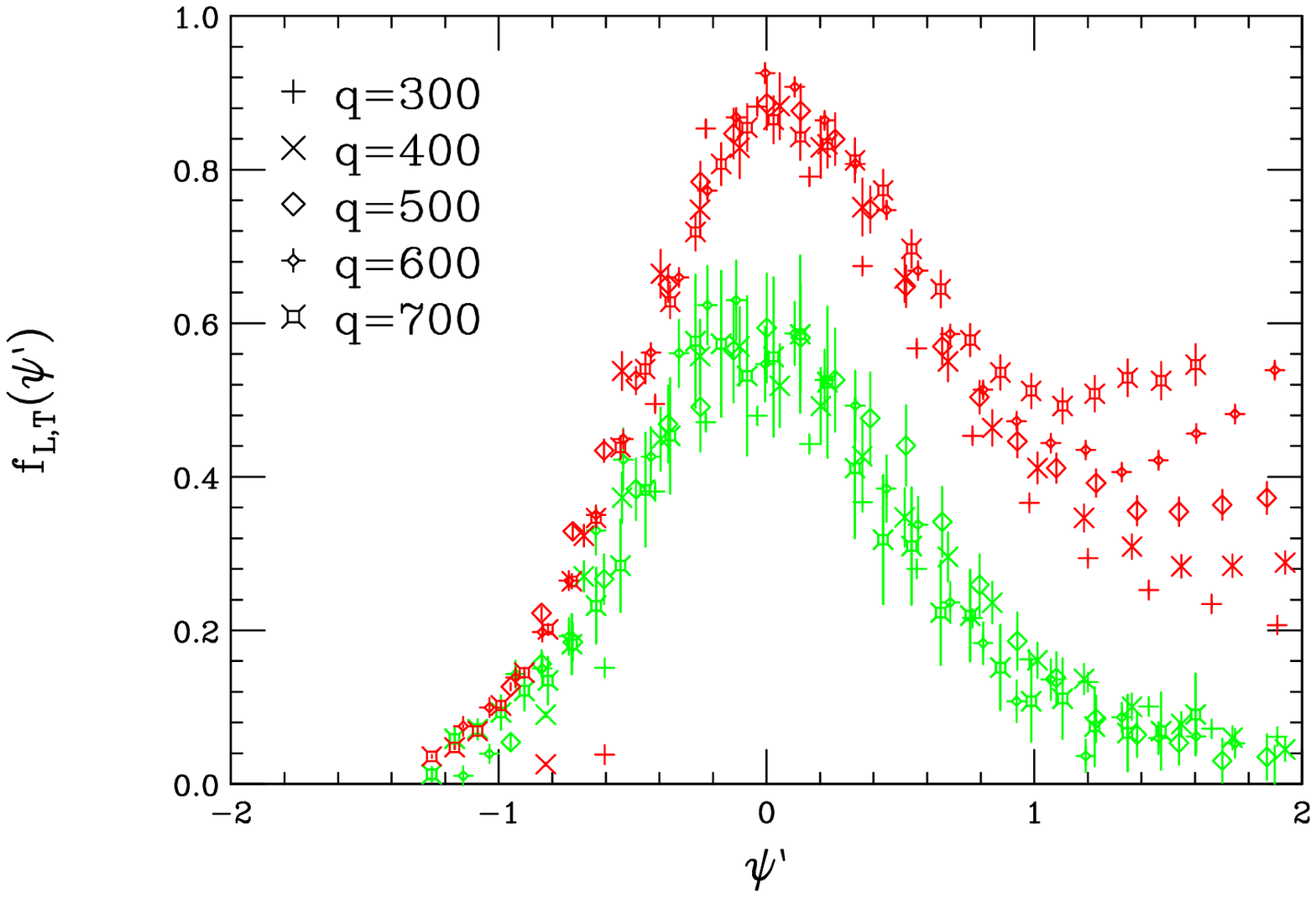}}}
% \epsfysize=55mm\epsffile{4he_flft.ps}}}
%\centerline{\mbox{\includegraphics[scale=0.45]{3he_flft.eps}
% \includegraphics[scale=0.45]{4he_flft.eps}}}
\begin{center} \parbox{13cm}
{\caption[]{The scaling functions $f_L$ and $f_T$ are shown for all $q$
values on the top for $^3$He and on the bottom for $^4$He. The upper bands of 
points correspond to $f_T$, the lower bands to $f_L$. 
}\label{flft}} \end{center}
\vspace*{-0.5cm}
\end{figure}

As discussed in \cite{Donnelly99a}, $\psi^\prime$-scaling can also be studied
for separated response functions.  The dimensionless 
scaling functions $f_{L,T}$ are defined in \cite{Donnelly99a} as
\begin{equation}
 f_{L,T} \equiv k_F \frac{R_{L,T}}{G_{L,T}} \>\>,
\end{equation}
with the factors $G_{L,T}$ given in \cite{Donnelly99a}.  For the relativistic 
Fermi gas model and in IA, the universal relation 
\begin{equation}
f_L=f_T=f
\end{equation}
is predicted.  Neglecting powers higher than two in $\eta_F = k_F/m$,
a relation between $f_L$ and the Coulomb sum rule is obtained as
\begin{equation}
\int d\psi f^{RFG} (\psi) = 1 + \frac{1}{20} \eta_F^2 + \cdots \>\>.
\end{equation}

In Fig.~\ref{flft} we compare the scaling functions $f_L(\psi^\prime)$ and 
$f_T(\psi^\prime)$ obtained for all response functions extracted from
the global analysis of the $^3$He and $^4$He data.
Within the error bars of the separated data, the longitudinal
response functions scale to a universal curve over the entire quasi-elastic peak.
Scaling of \rl\ is expected and provides a consistency check for the
Coulomb sum rule.  The results for \rt\ confirm that
the basic problem in quasi-elastic electron-nucleus 
scattering is the {\em excess strength in the transverse response}.
This excess is much larger for $^4$He than for $^3$He.  Scaling is also
observed for \rt\ at negative values of $\psi^\prime$, thus suggesting
that processes other than quasi-free knock out can also lead to scaling.

The excess of transverse strength is particularly large for $^4$He.
It exceeds the longitudinal strength at all momentum transfers, 
and does not seem to be limited to the\lq\lq dip\rq\rq region, but affects
the whole quasi-elastic peak region, extending
below the $\pi$-production threshold.  The transverse strength in the dip, which
increases with increasing $q$, is related to the growing overlap between
the high-energy side of the quasi-elastic peak and the 
tail of the $\Delta$-peak.

\begin{figure}[htb]
\centerline{\mbox{\epsfysize=80mm\epsffile{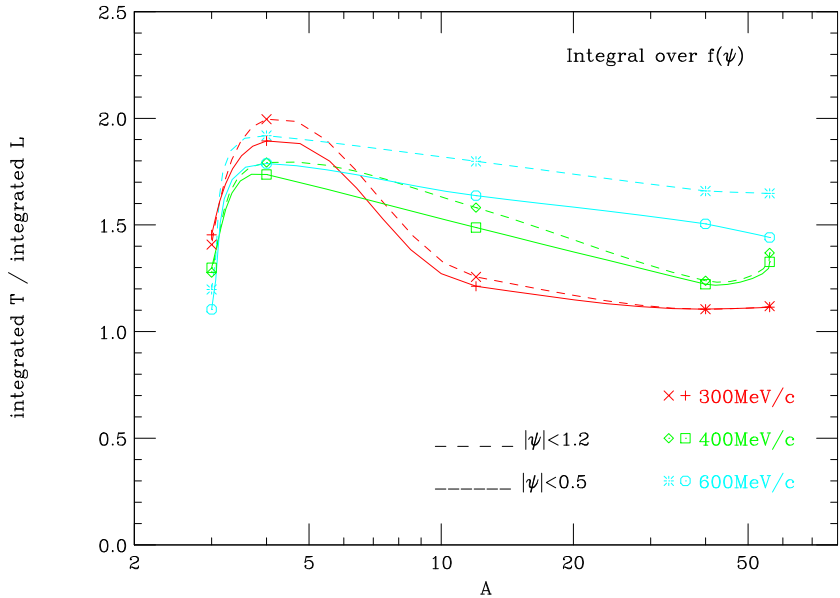}}}
\caption[]{Ratio of transverse to longitudinal integrated strength for 
$^3$He, $^4$He, $^{12}$C, $^{40}$Ca, and $^{56}$Fe:
300 MeV/c: x and +, 400 MeV/c: $\Diamond$ and $\Box$,
600 MeV/c: $\ast$ and $\circ$.  Points at the same $q$ are joined by lines. 
The integrations are over the indicated ranges of $\psi^\prime$.
}\label{rat}
\end{figure}

In order to study the $A$-dependence of this excess, we can look at the 
longitudinal and transverse responses integrated over $\psi^\prime$ --- those
for $^{12}$C, $^{40}$Ca, and $^{56}$Fe have been determined in
Ref.~\cite{Donnelly99a}.  We have integrated these responses over the region
of $\psi^\prime$ that essentially covers the quasi-elastic peak ($|\psi^\prime|<1.2$).
When limiting the integration range to $|\psi^\prime| < 0.5$ much of the
contribution from the tail of the $\Delta$ is eliminated, at least for the
light nuclei. The ratio of transverse to longitudinal integrated strength
is shown in Fig. \ref{rat}.

Figure \ref{rat} makes it clear that: i) the excess of transverse
strength rises very rapidly between $^3$He and $^4$He, and is indeed largest 
for $^4$He; ii) it is already large at the lowest $q$, the increase at the larger $q$ 
for the heavier nuclei is mainly due to the fact that the tail of the 
$\Delta$ peak contributes appreciably despite the restricted range of integration in 
$\psi^\prime$.
   
\section{Calculations of Euclidean response \label{euclidean}}
Since we are primarily concerned with the overall strength of
the longitudinal and transverse response, we consider the
Euclidean response functions, defined as \cite{Carlson94,Carlson92}
\begin{equation}
\tilde{E}_{T,L} (q,\tau) \ = \ \int_{\omega_{\rm th}}^\infty \ 
 \exp[-( \omega - E_0) \tau ] \  \ R_{T,L} (q,\omega)\>\>,
\end{equation}
where the $R_{T,L} (q, \omega)$ are the standard responses, $E_0$ is the
ground-state energy of the nucleus, and $\omega_{\rm th}$ is the
threshold for the response of the system excluding the elastic contribution.
The longitudinal and transverse Euclidean response functions represent
weighted sums of the corresponding $R_L(q,\omega)$ and $R_T(q,\omega)$:
at $\tau$=0 they correspond to the Coulomb and transverse sum rules,
respectively, while their derivatives with respect to $\tau$ evaluated
at $\tau$=0 correspond to the energy-weighted sum rules.  Larger values of
$\tau$ correspond to integrals over progressively lower-energy regions
of the response.

In a non-relativistic picture, the ${E}_{T,L}$ can be simply obtained from:
\begin{eqnarray}
\tilde{E}_L (q,\tau)&\! =\! & \langle 0 | \rho^\dagger ({\bf q})
\exp [ -(H - E_0) \tau] \rho({\bf q}) | 0 \rangle -  
\exp\left(-\frac{q^2 \tau}{2 A m}\right) |\langle 0({\bf q}) |\rho({\bf q})|0\rangle |^2,
\\ 
\tilde{E}_T (q,\tau) &\! =\! & \langle 0 |{\bf j}_T^\dagger({\bf q}) \exp[ -(H - E_0) \tau ]
{\bf j}_T({\bf q}) | 0 \rangle - 
\exp \left(-\frac{q^2 \tau}{2 A m}\right)
|\langle 0 |({\bf q}) |{\bf j}_T({\bf q})| 0 \rangle |^2 
\end{eqnarray}
where the elastic contributions have been explicitly subtracted, $|0({\bf q})\rangle$
represents the ground state recoiling with momentum ${\bf q}$, and sums over
spin projections are understood. 

In this paper we present results for the scaled Euclidean responses
\begin{equation}
E_{L,T}(q,\tau) = \frac{{\rm exp}\left[q^2 \tau/(2m)\right]}{[G_{E,p}(\tilde{Q}^2)]^2}\, 
 \tilde{E}_{L,T}(q,\tau)\>\>,
\end{equation}
where $\tilde{Q}^2$ is the squared four-momentum transfer evaluated at the
quasi-elastic peak.  This removes the trivial energy dependence obtained
from scattering off an isolated (non-relativistic) nucleon, and the $q$ dependence
associated with the nucleon form factors.  The longitudinal
response $E_L (q,\tau)$ is unity for an isolated proton, and the transverse
response $E_T (q,\tau)$ is simply the square of its magnetic moment.

The chief advantage of formulating the Euclidean response is that
it can be calculated exactly using Green's function or path integral
Monte Carlo techniques, including both final state interactions
and two-nucleon currents.  While the present calculations consider
only $A$ up to 4, they can be very simply extended to mass up to $A$=10
in direct analogy with ground-state calculations \cite{Wiringa00}.  In the future
it may be possible to use the Auxiliary-Field Diffusion Monte Carlo
technique developed by Schmidt and Fantoni \cite{Schmidt99} to calculate the
response for much heavier systems.

Other techniques have also been used to calculate the response
in few-nucleon systems, including Faddeev methods \cite{Golak95,Ishikawa98} and
Lorentz integral transform techniques \cite{Martinelli95,Efros97}.
Faddeev methods sum explicitly over the final states in the system, and hence
are directly applicable to inclusive and exclusive responses
for all possible final states.  This essentially corresponds
to a complete real-time calculation of the propagation of the system.
Lorentz integral techniques introduce a small imaginary time
component in the propagation of the response, directly summing over a
limited region of $\omega$.  For systems in which  a precise calculation
is possible, the full response can be calculated \cite{Efros97}.
The Euclidean response is the fully imaginary-time response,
and hence is a more integrated quantity.  While detailed dynamical
information is more limited, it is possible to perform calculations
in much heavier systems \cite{Bon96}.

The ground-state wave functions used in this study
are obtained with variational Monte Carlo.
They are of the general form \cite{Wiringa00}:
\begin{equation}
| \Psi_T \rangle = \prod_{i<j<k} \left[ 1 - {\tilde U}_0 (ijk) \right] \, 
{\cal S} \prod_{i<j} \Bigg[ \Big[ 1 - \sum_{k\ne i,j}{\tilde U}_{2 \pi} (ij;k)\Big]
\, {F}_{ij} \Bigg]\, |\Phi \rangle \>\>, \label{eq:varwvfn}
\end{equation}
where for 3- and 4-nucleon systems $| \Phi \rangle$ is
simply an anti-symmetrized product of spins and isospins.
The central three-nucleon correlation $\tilde{U}_0 (ijk)$
is a scaled version of the repulsive central component
of the Urbana-IX (UIX) three-nucleon interaction.  The magnitude
of the correlation and its range are scaled via variational
parameters.  The pair
correlations ${F}_{ij}$ depend upon the pair separation
$r_{ij}$ and the spins and isospins of the pair:
\begin{equation}
 {F}_{ij} = f^c (r_{ij}) \Big[ 1 + u^\sigma (r) \sigma_i \cdot \sigma_j
+ u^t (r) S_{ij} + u^{\sigma\tau} (r) \sigma_i \cdot \sigma_j
\tau_i \cdot \tau_j + u^{t \tau} S_{ij} \tau_i \cdot \tau_j \Big] \>\>. \label{eq:fij}
\end{equation}
The correlation $\tilde U_{2 \pi} (ij;k) $ is similarly scaled from
the anti-commutator part of the two-pion exchange three-nucleon
interaction.  The anti-commutator depends upon the spins and
isospins of only the two nucleons $i$ and $j$, but the 
spatial positions of all three.  Similarly the magnitude of
the spin-isospin dependent correlations $u$ for pair $ij$ are quenched
by the presence of other nucleons.  Both the two-nucleon correlation
${F}_{ij}$ and the $\tilde{U}_{2 \pi}$ correlation arising
from the three-nucleon interaction contain tensor-like terms
correlating the spins and orientations of the nucleons.  The 
contributions of these correlations to the response are discussed
below.

While these wave functions are not exact, they offer a rather
precise characterization of the Euclidean response, as evidenced
by comparisons with calculations using the correlated-hyperspherical-harmonics
wave functions~\cite{Viviani95} in $A$=3.  These comparisons
are presented in Sec.~\ref{ingredients}.
The Hamiltonian used in these studies is the Argonne model $v_8^\prime$ 
\cite{Wiringa00} N-N interaction plus the UIX three-nucleon
interaction.  This interaction reproduces many known properties
of the alpha particle, including its binding energy and charge
form factor.
 
Calculation of the Euclidean response is a straightforward extension of the ground-state
techniques employed in Green's function Monte Carlo.   
We wish to calculate matrix elements of the following type:
\begin{equation}
\tilde{M} (\tau) = \frac{\langle 0 | O_2 \, \exp[-(H - E_0) \tau] \, 
O_1 |0 \rangle}{\langle 0 | \exp [-(H-E_0) \tau] | 0 \rangle} \>\>.
\end{equation}
For a ground-state calculation of the energy ($O_1$=1, $O_2$=$H$)
the matrix element is evaluated by
a Monte Carlo sampling of the coordinate-space paths.  
The denominator is exactly one for an exact ground-state wave function,
otherwise there is a correction for finite $\tau$.
For each path a complete set of
$2^A A!/(N!\ Z!)$ amplitudes is kept corresponding to all possible
spin-isospin components of the ground state wave function.
Since the operators do not in general conserve isospin we cannot
use the most compact isospin basis used in ground-state calculations.

For a more general matrix element $\tilde M$ we simply keep another
complete set of amplitudes for each operator $O_1$,
each set of amplitudes corresponding to the full operator
acting on the ground state.  The paths are sampled
precisely as in the ground-state calculation \cite{Pudliner97},
and hence unaffected by the operators $O_1$, $O_2$. 
This allows us to calculate the response to a variety of operators
(charge, current, different momenta, etc.) simultaneously.

We have found it computationally advantageous to 
calculate the response simultaneously for several
different directions of momentum transfer.  A randomly picked 
set of three orthogonal axes are chosen, with $\hat{\bf q}$
directions along both the positive and negative directions
of each axis.  This method yields much lower statistical errors
in calculating the response, and along with the more efficient methods
for sampling path integrals recently applied to ground-state
calculations \cite{Pudliner97}, allows
for much more precise results than obtained previously.
It is also possible to calculate the response at several different
momentum transfers simultaneously.

It is certainly possible to extract more detailed information
from the Euclidean response.  Most efforts in this
direction proceed under Maximum Entropy techniques employing
Bayesian statistics \cite{Jarrell96}.  These techniques make
use of the correlated error estimates in $\tilde{R} (\tau)$
for different $\tau$.  Given the enhanced precision of the
present calculations we are exploring these possibilities.
These considerations are beyond the scope of the
present investigations, though, where we are primarily concerned
with the total strength in the longitudinal and transverse channels.

\section{Electromagnetic current operator \label{sec:MEC} }
The model for the nuclear electromagnetic current adopted in the
present study is briefly reviewed in this section for completeness,
for a more complete description see Ref.~\cite{Carlson98}.
The charge and current operators consist of one- and two-body terms:
\begin{eqnarray}
\rho({\bf q})&=& \sum_i \rho^{(1)}_i({\bf q})
             +\sum_{i<j} \rho^{(2)}_{ij}({\bf q}) \>\>, \label{eq1}\\
{\bf j}({\bf q})&=& \sum_i {\bf j}^{(1)}_i({\bf q})
             +\sum_{i<j} {\bf j}^{(2)}_{ij}({\bf q}) \label{eq2} \>\>,
\end{eqnarray}
where ${\bf q}$ is the momentum transfer.  The one-body
operators $\rho^{(1)}_i$ and ${\bf j}^{(1)}_i$ have the
standard expressions obtained from a relativistic reduction of the
covariant single-nucleon current, and are listed below for convenience.
The charge operator is written as
\begin{equation}
\rho^{(1)}_i({\bf q})= \rho^{(1)}_{i,{\rm NR}}({\bf q})+
                       \rho^{(1)}_{i,{\rm RC}}({\bf q}) \>\>, \label{eq6}
\end{equation}
with
\begin{equation}
\rho^{(1)}_{i,{\rm NR}}({\bf q})= \epsilon_i \>
 {\rm e}^{{\rm i}{\bf q}\cdot {\bf r}_i} \label{eq7} \>\>,
\end{equation}
\begin{equation}
\rho^{(1)}_{i,{\rm RC}}({\bf q})= \left
( \frac {1}{\sqrt{1+Q^2/4m^2} }-1\right )
\epsilon_i \> {\rm e}^{{\rm i}{\bf q}\cdot {\bf r}_i}
- {\frac {{\rm i}}{4m^2}} \left ( 2\, \mu_i-\epsilon_i \right )
{\bf q} \cdot (\sigma_i \times {\bf p}_i) \>
{\rm e}^{ {\rm i} {\bf q} \cdot {\bf r}_i } \>\>,
\label{eq8}
\end{equation}
where $Q^2=q^2-\omega^2$ is the four-momentum transfer, and
$\omega$ is the energy transfer.  The current operator is expressed as
\begin{equation}
     {\bf j}^{(1)}_i({\bf q})={\frac {1} {2m}} \epsilon_i \>
   \bigl[ {\bf p}_i\>,\>{\rm e}^{{\rm i} {\bf q} \cdot {\bf r}_i} \bigr ]_+
   -{\frac {{\rm i}} {2m}} \mu_i \>
     {\bf q} \times \sigma_i \> {\rm e}^{{\rm i} {\bf q} \cdot
   {\bf r}_i}  \label{eq9}\>\>\>,
\end{equation}
where $[ \cdots \, ,\, \cdots ]_+$ denotes the anticommutator.  The following
definitions have been introduced:
\begin{eqnarray}
\epsilon_i &\equiv& G_{E,p}(Q^2) \frac {1}{2}\left ( 1 + \tau_{z,i} \right )
                   +G_{E,n}(Q^2) \frac {1}{2}\left ( 1 - \tau_{z,i} \right )
 \>\>,\label{eq8a}\\
\mu_i &\equiv& G_{M,p}(Q^2) \frac {1}{2}\left ( 1 + \tau_{z,i} \right )
              +G_{M,n}(Q^2) \frac {1}{2}\left ( 1 - \tau_{z,i} \right )
\label{eq9a} \>\>,
\end{eqnarray}
and ${\bf p}$, $\sigma$, and $\tau$ are
the nucleon's momentum, Pauli spin
and isospin operators, respectively.  The two terms
proportional to $1/m^2$ in $\rho^{(1)}_{i,{\rm RC}}$
are the well known Darwin-Foldy and spin-orbit relativistic
corrections~\cite{Deforest66,Friar73}, respectively.

The calculations of the response functions discussed
in the previous section have been carried out using
the dipole parameterization of the nucleon form factors
\begin{eqnarray}
G_{E,p}(Q^2)&=&G_D(Q^2) \>\>, \\
G_{E,n}(Q^2)&=&-\mu_n \frac{Q^2}{4 m^2} \frac{G_D(Q^2)}{1+Q^2/m^2} \>\>, \\
G_{M,p}(Q^2)&=&\mu_p G_D(Q^2) \>\>, \\
G_{M,n}(Q^2)&=&\mu_n G_D(Q^2) \>\>, 
\end{eqnarray}
where 
\begin{equation}
G_D(Q^2) = \frac{1}{(1+Q^2/\Lambda^2)^2} \>\>,
\end{equation}
with $\Lambda=0.834$ GeV/c, and where $\mu_p$ and $\mu_n$ are
the proton ($\mu_p=2.793$ n.m.) and neutron ($\mu_n=-1.913$ n.m.)
magnetic moments, respectively.
It is worth emphasizing that the available
semi-empirical parameterizations of the proton electric
and magnetic, and neutron magnetic form factors
do not differ significantly --- less than a couple of \% --- in the
low momentum transfer range of interest here,
$Q^2 \le 0.4$ (GeV/c)$^2$, and that uncertainties in the neutron electric
form factor have a negligible impact on the present results.
Finally, we should note that in the actual calculations of the
Euclidean responses the value of the four-momentum transfer occurring
in the nucleon form factors (as well as in the electromagnetic $N$$\Delta$
transition form factor, see below) is kept fixed at the quasi-elastic peak, 
as already mentioned in Sec.~\ref{euclidean}.   

The most important features of the two-body parts of the
electromagnetic current operator are summarized below.  The reader
is referred to Refs.~\cite{Viviani96,Carlson98} for a derivation and listing
of their explicit expressions.

\subsection{Two-body current operators}

The two-body current operator consists of \lq\lq model-independent\rq\rq and
\lq\lq model-dependent\rq\rq \  components, in the classification scheme
of Riska~\cite{Riska89}.  The model-independent terms
are obtained~\cite{Schiavilla89a} from the nucleon-nucleon interaction
(the charge-independent part of the Argonne $v_{18}$ in the
present study), and by construction
satisfy current conservation with it.
The leading operator is the isovector
\lq\lq $\pi$-like \rq\rq current derived
from the isospin-dependent spin-spin ($\sigma \tau$)
and tensor ($t \tau$) interactions.
The latter also generate an isovector \lq\lq $\rho$-like\rq\rq current, while
additional model-independent isoscalar and isovector currents arise from the
isospin-independent and isospin-dependent central and momentum-dependent
interactions.  These currents are short-ranged and numerically
far less important than the $\pi$-like current.  For the purpose of
later discussions, we list below the explicit expression for the latter:
\begin{eqnarray}
{\bf j}^{(2)}_{ij}({\bf q};\pi)
&=&G_E^V(Q^2) (\tau_i \times \tau_j)_z
\Bigg[ 
{\rm e}^{{\rm i}{\bf q}\cdot {\bf r}_i}
 f_{PS}(r) \,\sigma_i\, (\sigma_j \cdot \hat{\bf r})
+{\rm e}^{{\rm i}{\bf q}\cdot {\bf r}_j}
 f_{PS}(r)\, \sigma_j\, (\sigma_i \cdot \hat{\bf r}) \nonumber \\
&-& (\sigma_i \cdot \nabla_i) (\sigma_j \cdot \nabla_j)
(\nabla_i - \nabla_j) g_{PS}({\bf q};{\bf R},{\bf r})
\Bigg ] \>\>,
\end{eqnarray}
where $G_E^V(Q^2)=G_{E,p}(Q^2)+G_{E,n}(Q^2)$ is the isovector
combination of the nucleon electric form factors, and ${\bf R}$
and ${\bf r}$ are the center-of-mass and relative positions of
nucleons $i$ and $j$, ${\bf R}=({\bf r}_i+{\bf r}_j)/2$ and
${\bf r}={\bf r}_i-{\bf r}_j$, respectively.  The functions
$f_{PS}$ and $g_{PS}$ are defined as 
\begin{equation}
f_{PS}(r) = \frac{\rm d}{\rm dr} \int\frac{d{\bf k}}{(2\pi)^3}
{\rm e}^{{\rm i}{\bf k}\cdot {\bf r}}\, v_{PS}(k) \>\>,
\end{equation}
\begin{equation}
g_{PS}({\bf q};{\bf R},{\bf r}) = 
\int\frac{d{\bf k}_i}{(2\pi)^3} \frac{d{\bf k}_j}{(2\pi)^3}
{\rm e}^{{\rm i}{\bf k}_i\cdot {\bf r}_i}
{\rm e}^{{\rm i}{\bf k}_j\cdot {\bf r}_j}
\,(2\pi)^3\delta({\bf q}-{\bf k}_i-{\bf k}_j)
\, \frac{v_{PS}(k_i)-v_{PS}(k_j)}{k_i^2 - k_j^2} \>\>,
\end{equation}
where $v_{PS}(k)$ is obtained from the $\sigma\tau$ and $t\tau$
components of the interaction,
\begin{equation}
v_{PS}(k)= v^{\sigma \tau}(k)- 2 \> v^{t\tau}(k) \>\>,
\end{equation}
with
\begin{eqnarray}
v^{\sigma \tau}(k)&=&\frac{4\pi}{k^2} \int_0^\infty r^2 dr \,
\left [ j_0(kr)-1 \right ] v^{\sigma \tau}(r) \>\>, \\
v^{t \tau}(k)&=&\frac{4\pi}{k^2}
\int_0^\infty r^2 dr\, j_2(kr)v^{t\tau}(r) \>\>.
\end{eqnarray}
The factor $j_0(kr)-1$ in the
expression for $v^{\sigma \tau}(k)$ ensures that its volume integral
vanishes~\cite{Schiavilla89a}.  

In a one-boson-exchange (OBE) model, in which
the isospin-dependent spin-spin and tensor interactions
are due to $\pi$-meson (and $\rho$-meson) exchanges, the function
$v_{PS}(k)$ simply reduces to
\begin{equation}
v_{PS}(k) \rightarrow v_\pi(k) \equiv -{\frac {f_\pi^2} {m_\pi^2} }
{\frac {f^2_\pi(k)} { k^2+m_\pi^2} } \ , 
\end{equation}
where $m_\pi$, $f_\pi$, and $f_\pi(k)$ denote,
respectively, the pion mass, $\pi$$N$$N$ coupling constant
and form factor.  In this limit, the functions $f_{PS}$ and
$g_{PS}$ read:
\begin{equation}
f_{PS}(r)\rightarrow f_\pi(r)= \frac{f_\pi^2}{4\pi}
\frac{{\rm e}^{-m_\pi r}} {(m_\pi r)^2} (1+m_\pi r)
\end{equation}
\begin{equation}
g_{PS}({\bf q};{\bf R},{\bf r}) \rightarrow 
g_\pi({\bf q};{\bf R},{\bf r})= 
\frac{ {\rm e}^{{\rm i}{\bf q}\cdot {\bf R}} }{8\pi}
\int_{-1/2}^{+1/2} dx\, {\rm e}^{-{\rm i} x {\bf q}\cdot {\bf r}}
\,\frac{{\rm e}^{-L_\pi(x) r}}{L_\pi(x)} \>\>,
\end{equation}
with 
\begin{equation}
L_\pi(x)=\sqrt{ m_\pi^2+q^2(1-4 x^2)/4 } \>\>,
\end{equation}
where for simplicity the $\pi$$N$$N$ form factor
has been set equal to one.  The resulting current
is then identical to that commonly used in the literature.

The model-dependent currents are purely transverse
and therefore cannot be directly linked
to the underlying two-nucleon interaction.
The present calculation includes the isoscalar $\rho \pi \gamma$ and
isovector $\omega \pi \gamma$ transition currents as well as the isovector
current associated with excitation of intermediate $\Delta$-isobar resonances.
The $\rho \pi \gamma$ and $\omega \pi \gamma$ couplings are known from the
measured widths of the radiative decays
$\rho \rightarrow \pi \gamma$~\cite{Berg80} and
$\omega \rightarrow \pi \gamma$~\cite{Chemtob71,Chemtob73}, respectively, while their
momentum-transfer dependence is modeled using vector-meson-dominance.
Monopole form factors are introduced at the meson-baryon vertices with
cutoff values of $\Lambda_\pi$=3.8 fm$^{-1}$ and
$\Lambda_\rho$=$\Lambda_\omega$=6.3 fm$^{-1}$
at the $\pi N$$N$, $\rho N$$N$ and $\omega N$$N$ vertices, respectively.

Among the model-dependent currents, however,
those associated with the $\Delta$-isobar
are the most important ones.  In the present calculation,
these currents are treated in the static $\Delta$ approximation rather
than in the more accurate transition-correlation-operator
scheme, developed in Ref.~\cite{Schiavilla92} and applied to the
calculation of the trinucleon form factors~\cite{Marcucci98},
$n$$d$ and $p$$d$ radiative capture cross sections at
low energies~\cite{Viviani96,Viviani00}, and $S$-factor of
the proton weak capture on $^3${\rm He}~\cite{Marcucci01}.
Again for later convenience, it is useful to list explicitly
the two-body $\Delta$-excitation current used in the present work:
\begin{eqnarray}
{\bf j}^{(2)}_{ij}({\bf q};\Delta) =&&
{\bf  j}_i({\bf q};\Delta \rightarrow N)
\frac{ v_{NN \rightarrow \Delta N,ij} }
{m-m_\Delta} \nonumber \\
&+& \frac{v_{\Delta N \rightarrow NN,ij} }{m-m_\Delta}
{\bf j}_i({\bf q};N \rightarrow \Delta) +
i \rightleftharpoons j \>\>,
\end{eqnarray}
where the $N \rightleftharpoons \Delta$
electromagnetic current is modeled as
\begin{equation}
{\bf j}_i({\bf q}; N \rightarrow \Delta)=
-{\frac {{\rm i}} {2m} } G_{\gamma N \Delta}(Q^2)
{\rm e}^{{\rm i} {\bf q} \cdot {\bf r}_i} {\bf q} \times {\bf S}_i\, T_{z,i}
\>\>,   \label{jnd}
\end{equation}
and the expression for
${\bf j}_i({\bf q}; \Delta \rightarrow N)$ is obtained from that
for ${\bf j}_i({\bf q}; N \rightarrow \Delta)$ by replacing the
transition spin and isospin operators ${\bf S}$
and ${\bf T}$ with their hermitian conjugates.
The electromagnetic form factor $G_{\gamma N \Delta}(Q^2)$
is parameterized as
\begin{equation}
G_{\gamma N \Delta}(Q^2)= \frac{ \mu^* }
 { (1+Q^2/\Lambda_{N\Delta,1}^2 )^2
\sqrt{1+Q^2/\Lambda_{N\Delta,2}^2} } \>\>,
\end{equation}
where the $N \rightarrow \Delta$
transition magnetic moment $\mu^*$
is taken here to be equal to 3 n.m., as
obtained in an analysis of $\gamma N$ data
in the $\Delta$-resonance region~\cite{Carlson86}.
This analysis also gives $\Lambda_{N\Delta,1}$=0.84 GeV/c and
$\Lambda_{N\Delta,2}$=1.2 GeV/c.  It is important
to point out, however, that the quark-model
value for $\mu^*$,
$\mu^*= (3\sqrt{2}/5) \mu_N^V=3.993$ n.m. ($\mu_N^V$
is the nucleon isovector magnetic moment), is often used in the
literature.  This value is significantly larger than
that adopted above.  Finally, the transition
interaction $v_{NN \rightarrow \Delta N,ij}$ is given by  
\begin{equation}
v_{NN\rightarrow \Delta N,ij}=\lbrack v^{\sigma \tau {\rm II}}(r)
{\bf S}_i \cdot \sigma_j  +v^{t \tau {\rm II}}(r) S^{\rm II}_{ij}
\rbrack {\bf T}_i \cdot \tau_j \>\>,
\end{equation}
and $v_{\Delta N \rightarrow N N,ij}$ is the
hermitian conjugate of the expression above.  The $S^{\rm II}_{ij}$
is the tensor operator where the Pauli spin $\sigma_i$ has
been replaced by the transition spin ${\bf S}_i$, and the
functions $v^{\sigma\tau{\rm II}}(r)$ and $v^{t\tau{\rm II}}(r)$ 
are defined as
\begin{eqnarray}
v^{\sigma\tau {\rm II}}(r)&=&
\frac{f_\pi f^*_\pi}{4\pi}\frac{m_{\pi}}{3}\frac{{\rm{e}}^{-x}}{x}\,C(x) \ ,
\label{eq:uvst} \\
v^{t\tau\alpha}(r)&=&
\frac{f_\pi f^*_\pi}{4\pi}\frac{m_{\pi}}{3}\left(1+\frac{3}{x}+\frac{3}{x^2}
\right)\frac{{\rm{e}}^{-x}}{x}\,C^{2}(x) \>\>,
\label{eq:uvtt}
\end{eqnarray}
where $x\equiv m_{\pi}r$, 
$f_\pi^*=(6\sqrt{2}/5) f_{\pi}$ is the quark-model value
for the $\pi N\Delta$ coupling constant (adopted in the present
work), and the cutoff function $C(x)\,=\,1-e^{-\lambda x^{2}}$,
with $\lambda$=4.09.

Standard manipulations of the product of
spin and isospin transition operators~\cite{Schiavilla92} lead to the
following expression for the $\Delta$-excitation current:
\begin{eqnarray}
{\bf j}^{(2)}_{ij}({\bf q};\Delta)
&=&{\rm i} \frac {G_{\gamma N \Delta}(Q^2)}{9m}
 {\rm e}^{{\rm i} {\bf q} \cdot {\bf r}_i}
 \Bigg [ 4\, \tau_{z,j} \Big [ f_\Delta(r) \sigma_j
+ g_\Delta(r) \hat {\bf r} (\sigma_j \cdot \hat {\bf r})
\Big ] \nonumber \\
&-&(\tau_i \times \tau_j)_z \Big [ f_\Delta(r)
(\sigma_i \times \sigma_j)
+ g_\Delta(r) (\sigma_i \times \hat {\bf r})
(\sigma_j \cdot \hat {\bf r}) \Big ] \Bigg ] \times {\bf q} \nonumber \\
&+& i \rightleftharpoons j  \>\>,
\label{j1d}
\end{eqnarray}
where
\begin{eqnarray}
f_\Delta(r)&\equiv&
\frac{v^{\sigma \tau II}(r)-v^{t \tau II}(r) }{m-m_\Delta} \>\>,\label{xf3} \\
g_\Delta(r)&\equiv& 3 \, \frac{ v^{t \tau II}(r) }{m-m_\Delta} \label{xf4}\>\>.
\end{eqnarray}
The expression above reduces to that commonly used in the literature,
if the quark-model values for the $\pi N\Delta$ and $\gamma N\Delta$
coupling constants are adopted.

\subsection{Two-body charge operators}

While the main parts of the two-body currents are linked to the form of the
two-nucleon interaction through the continuity equation, the most important
two-body charge operators are model-dependent, and should be considered as
relativistic corrections.
Indeed, a consistent calculation of two-body charge effects in nuclei would
require the inclusion of relativistic effects in both the interaction models
and nuclear wave functions.  Such a program is yet to be carried out
for systems with $A\geq 3$.  There are nevertheless
rather clear indications for the relevance of two-body
charge operators from the failure of the impulse approximation (IA) in
predicting the deuteron tensor polarization
observable~\cite{Abbott99}, and charge form factors of the three- and
four-nucleon systems~\cite{Marcucci98,Schiavilla01}.
The model commonly used~\cite{Schiavilla90} includes the $\pi$-, $\rho$-, and
$\omega$-meson exchange charge operators with both isoscalar and isovector
components, as well as the (isoscalar) $\rho \pi \gamma$ and (isovector)
$\omega \pi \gamma$ charge transition couplings, in addition to the
single-nucleon Darwin-Foldy and spin-orbit relativistic corrections.
The $\pi$- and $\rho$-meson exchange charge operators are constructed from the
isospin-dependent spin-spin and tensor components
of the two-nucleon interaction (again, the Argonne
$v_{18}$ model), using the same
prescription adopted for the corresponding current operators.
Explicit expressions for these operators can be found
in Ref.~\cite{Schiavilla90}.  Here, we only emphasize
that for $Q \leq 1$ GeV/c the
contribution due to the $\pi$-exchange charge operator is typically an order
of magnitude larger than that of any of the remaining two-body mechanisms
and one-body relativistic corrections.

\section{Model studies \label{model}}
The Euclidean response is an excellent tool to test our understanding of
inclusive quasi-elastic scattering, since it incorporates an exact treatment 
of the states in the continuum.  The Euclidean response does have the
disadvantage, however, that it corresponds to a weighted integral over
the energy loss $\omega$ and, as a consequence, the interpretation of
potential differences between calculated results and experimental data
is not so straightforward.

\begin{figure}[htb]
\centerline{\mbox{\epsfysize=80mm\epsffile{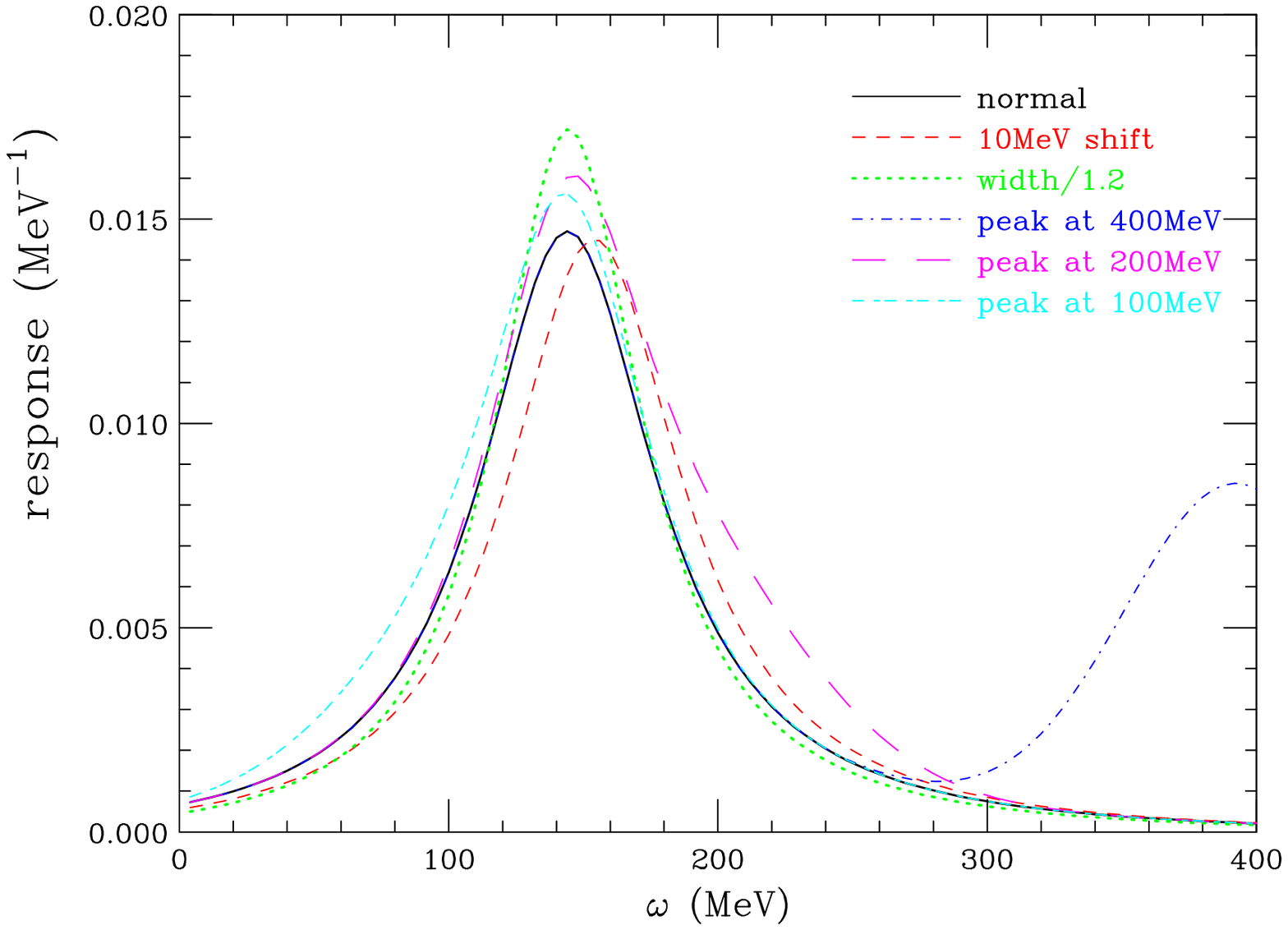}}}
\centerline{\mbox{\hspace*{2mm}\epsfysize=78mm\epsffile{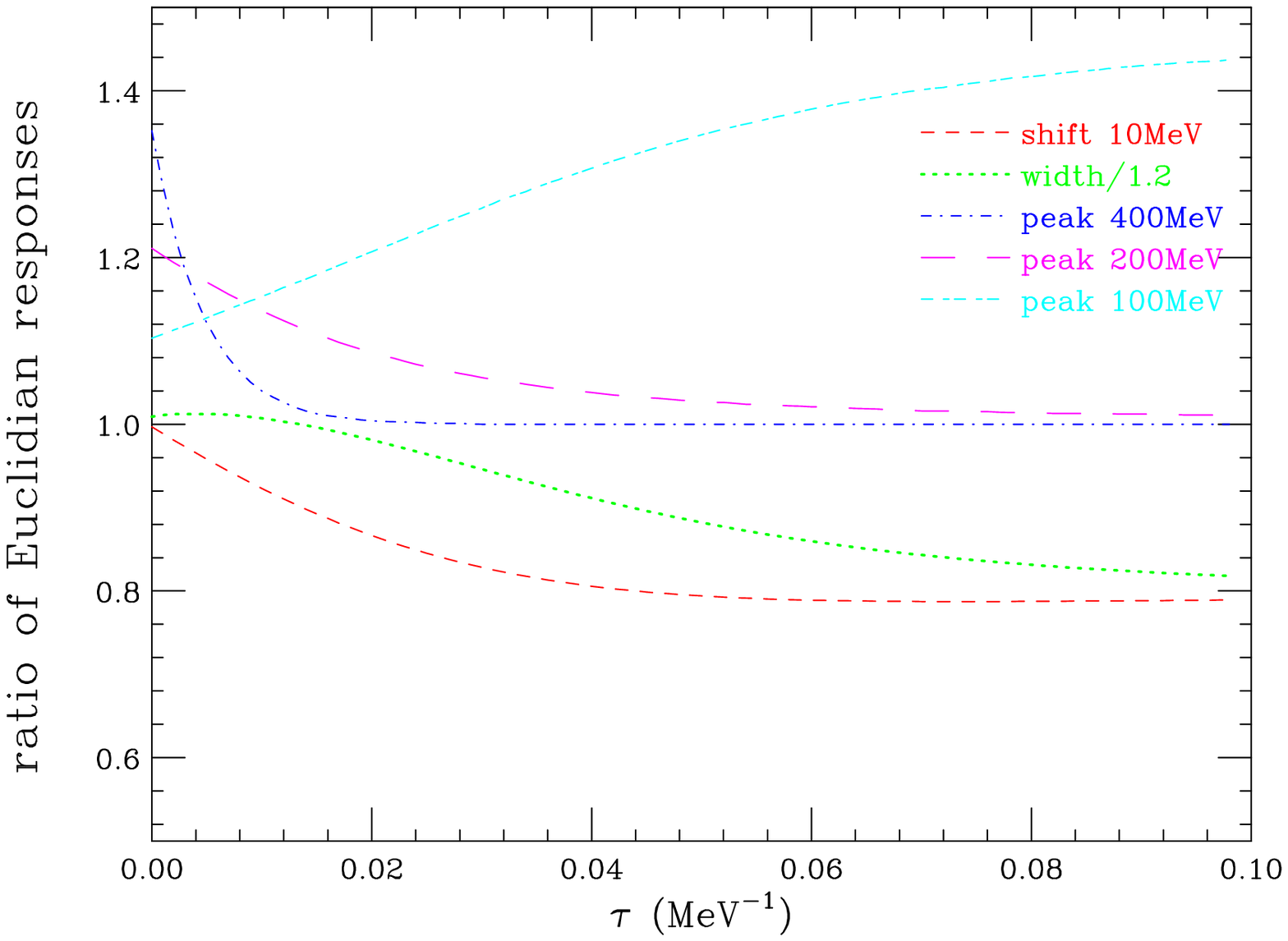}}}
\begin{center} \parbox{12cm}{ \caption[]{Top: model response 
(solid line) changed  by various modifications (see
text). Bottom:  corresponding ratio of modified and unmodified 
Euclidean responses.  
\label{respw} }}
\end{center}
\end{figure}

In order to develop a better feeling for the properties of the Euclidean 
response, in this section we discuss a simple-minded model calculation.
We use a parameterized cross section --- {\em de facto} a fit to the longitudinal
$R_L (\omega)$ at one momentum transfer --- and study the change in the Euclidean
response upon various changes of the cross section as a function of energy loss. 

In the top panel of Fig.~\ref{respw} we show the model quasi-elastic peak
as a function of energy loss (solid curve) and a selection of modifications.
The changes have in general been made by adding a gaussian with arbitrary
amplitude and selected position in energy loss. Figure~\ref{respw} shows
the quasi-elastic peak a) with a gaussian placed at very large energy loss
(400 MeV), b) a gaussian placed on the high-energy loss tail of the quasi-elastic
peak (200 MeV) and c) a gaussian placed on the low-$\omega$ side of the peak
(100 MeV).  It also displays a curve where d) the width of the quasi-elastic
response has been decreased by 20\% (with the overall amplitude adjusted to
conserve the area) and one for the case where e) the quasi-elastic peak is
shifted by 10 MeV.

The lower panel of Fig.~\ref{respw} shows the resulting changes in 
terms of the ratio of modified to original Euclidean responses.  The
value at $\tau$=0 reflects the (arbitrary) integral over $\omega$  of
the added modification.  Several features are noteworthy: 
\begin{itemize}
\item
The Euclidean response at finite $\tau$ very quickly suppresses the
contribution from large energy loss.  The dash-dot curve shows that
already at $\tau > 0.01$ the contribution from the large peak added
at $\omega$=400 MeV is suppressed.  For the experimental transverse
response function $R_T (\omega)$ this implies that the contribution
from pion production in the $\Delta$ peak (which is not included in
the theory we are going to compare to) is only affecting the results
for very small $\tau$.  We will therefore ignore this region.
\item
The region of the quasi-elastic cross section at low $\omega$ comes
in very prominently at the larger values of $\tau$, as indicated by
the curve labeled \lq\lq peak at 100 MeV\rq\rq.   
\item
A shift of the quasi-elastic peak to larger $\omega$ leads to an
Euclidean response that quickly falls with increasing $\tau$, reaching
saturation by the time $\tau$ gets to values approaching 0.05.
\end{itemize}

\section{Results \label{results}}
\begin{figure}[p]
\vspace*{-1cm}
\centerline{\mbox{\epsfysize=53mm\epsffile{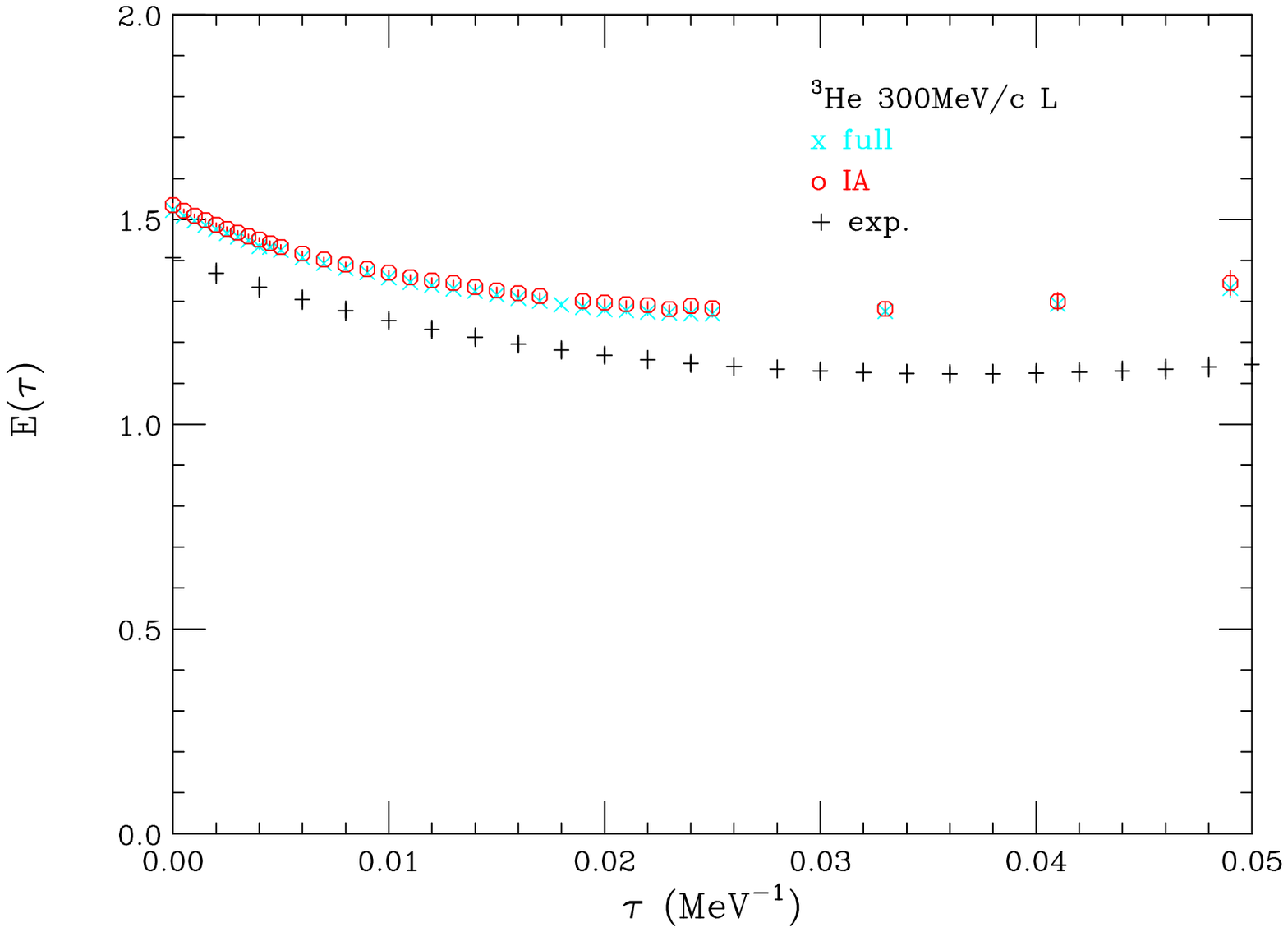}
\hspace{7mm} \epsfysize=53mm\epsffile{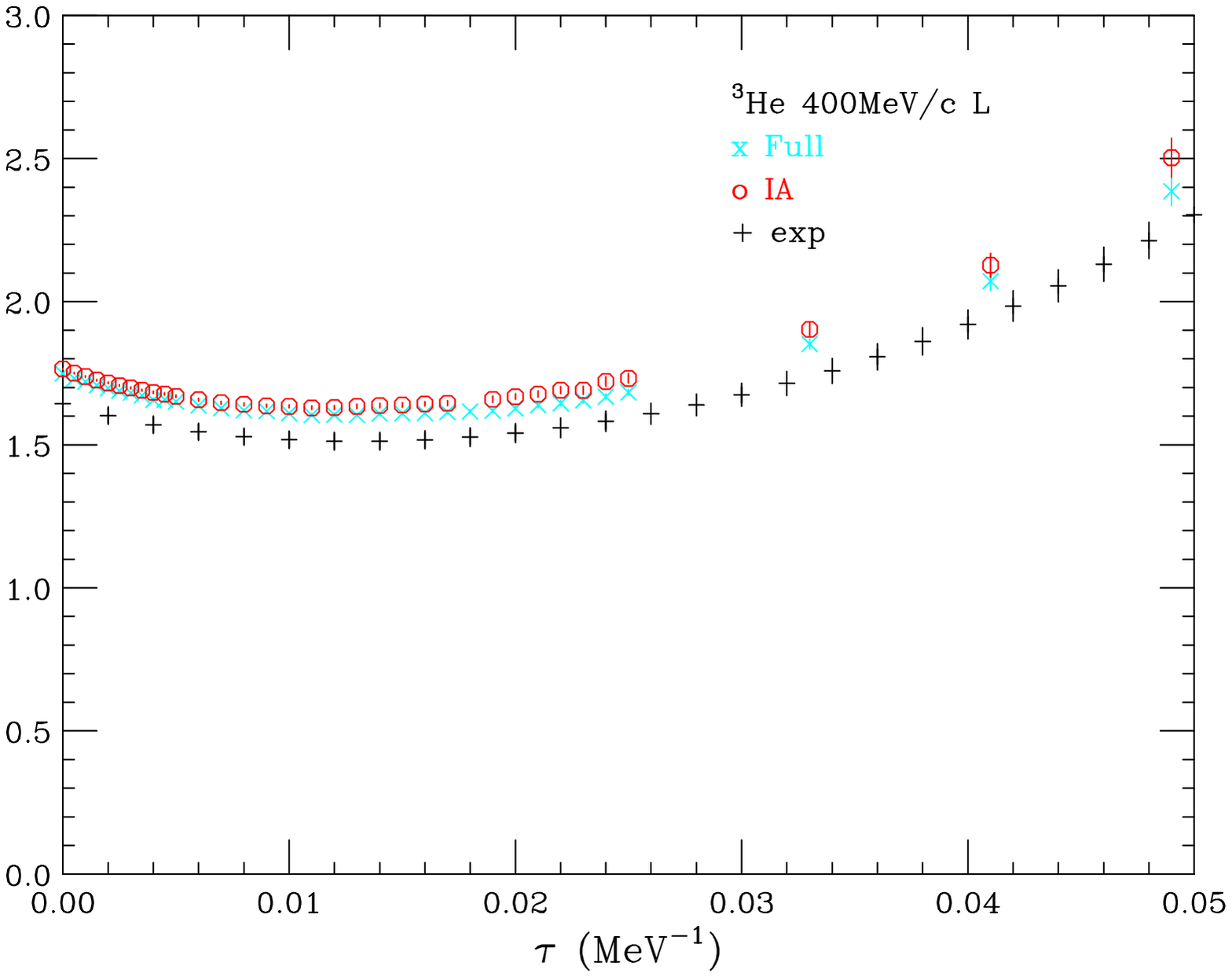}}}
\centerline{\mbox{\epsfysize=53mm\epsffile{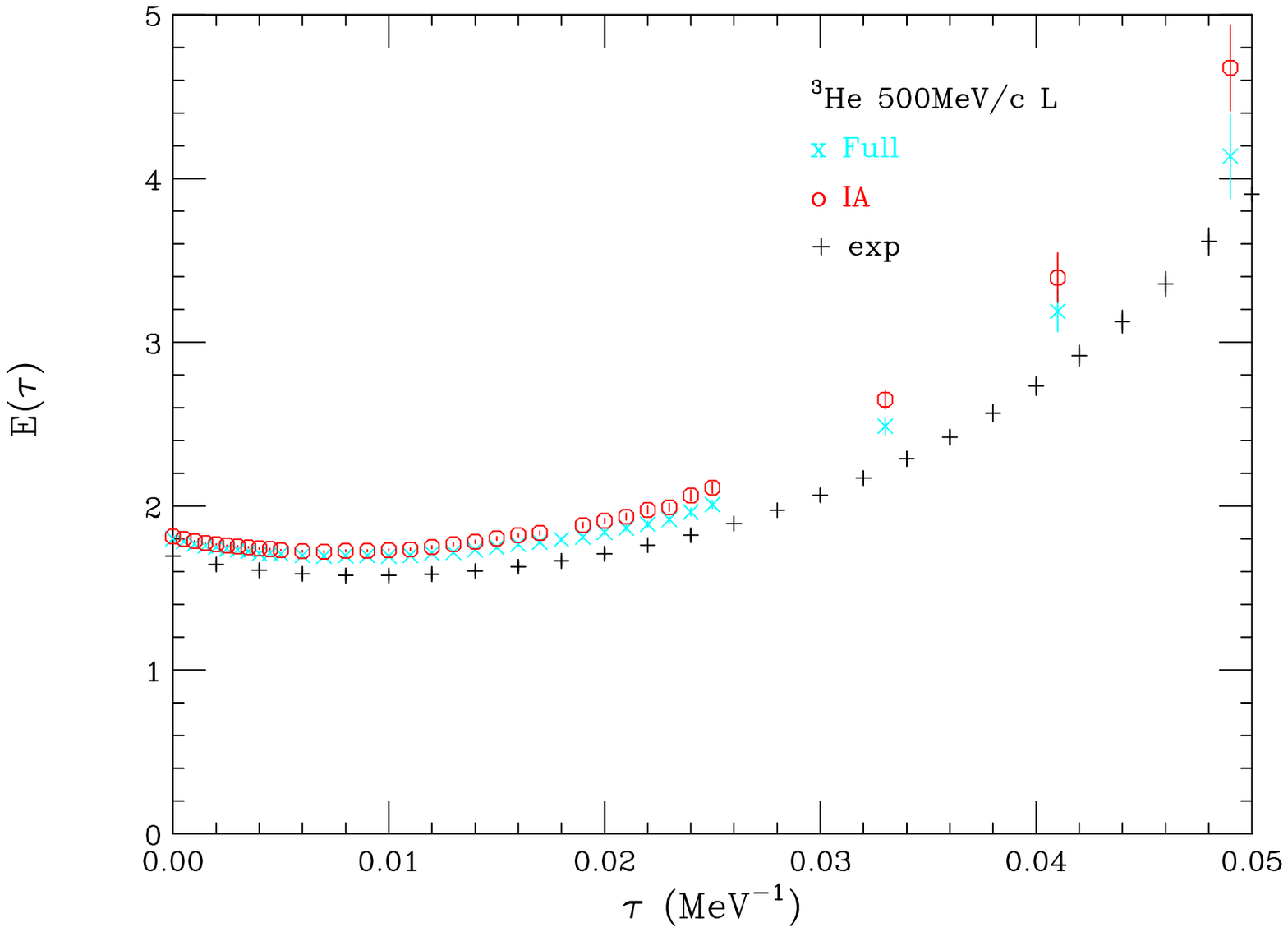}
\hspace{7mm} \epsfysize=53mm\epsffile{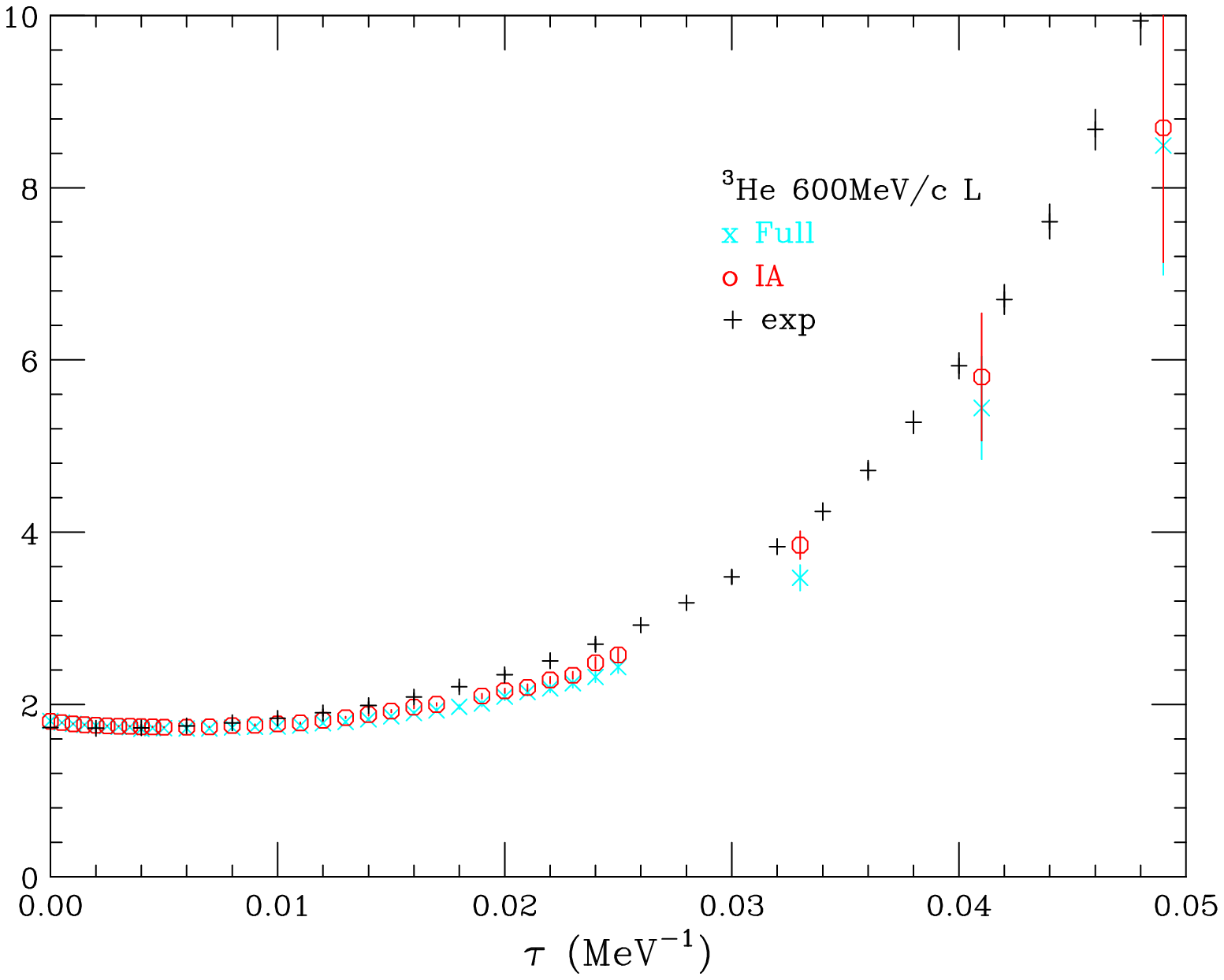}}}
\centerline{\mbox{\epsfysize=53mm\epsffile{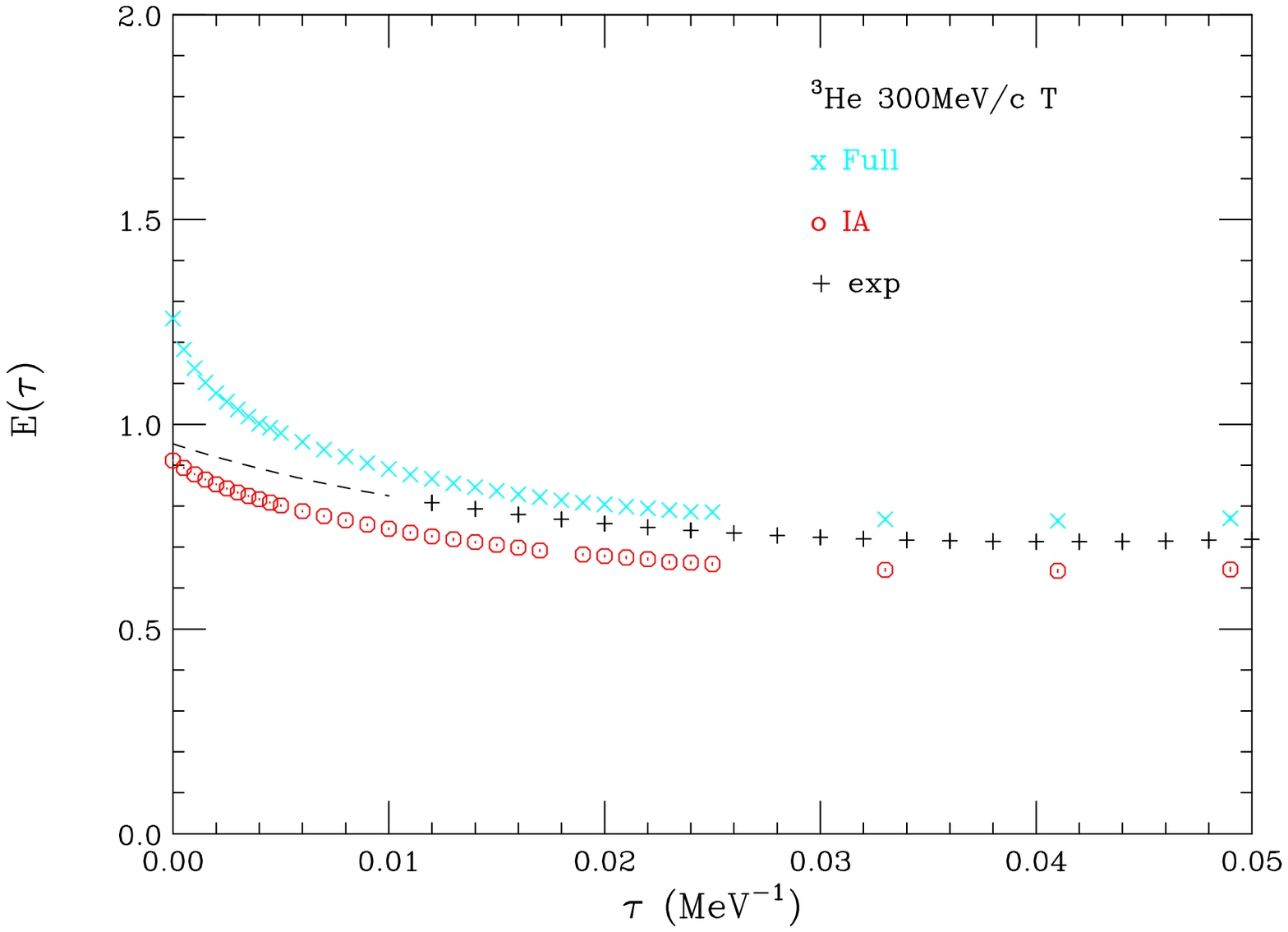}
\hspace{7mm} \epsfysize=53mm\epsffile{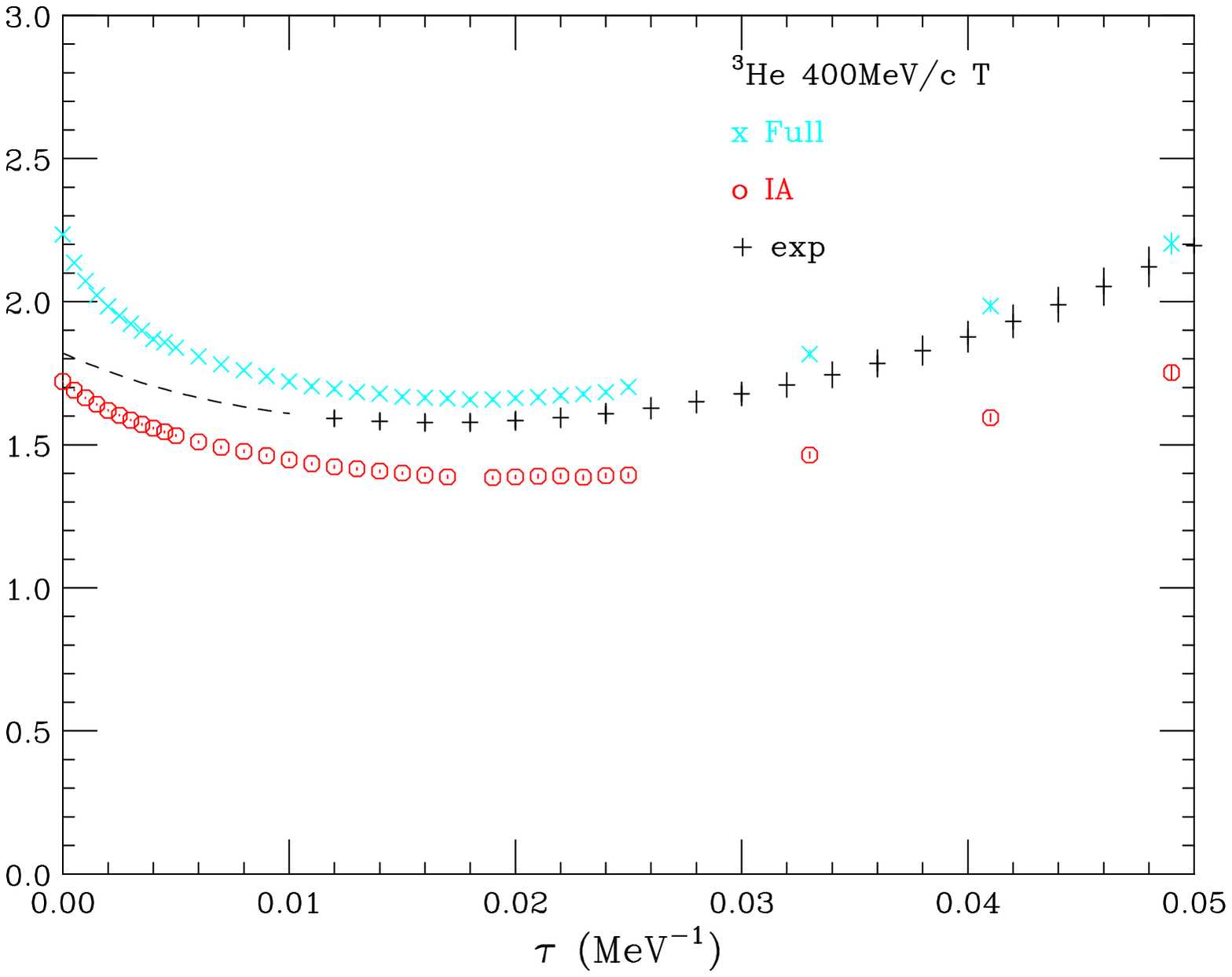}}}
\centerline{\mbox{\epsfysize=53mm\epsffile{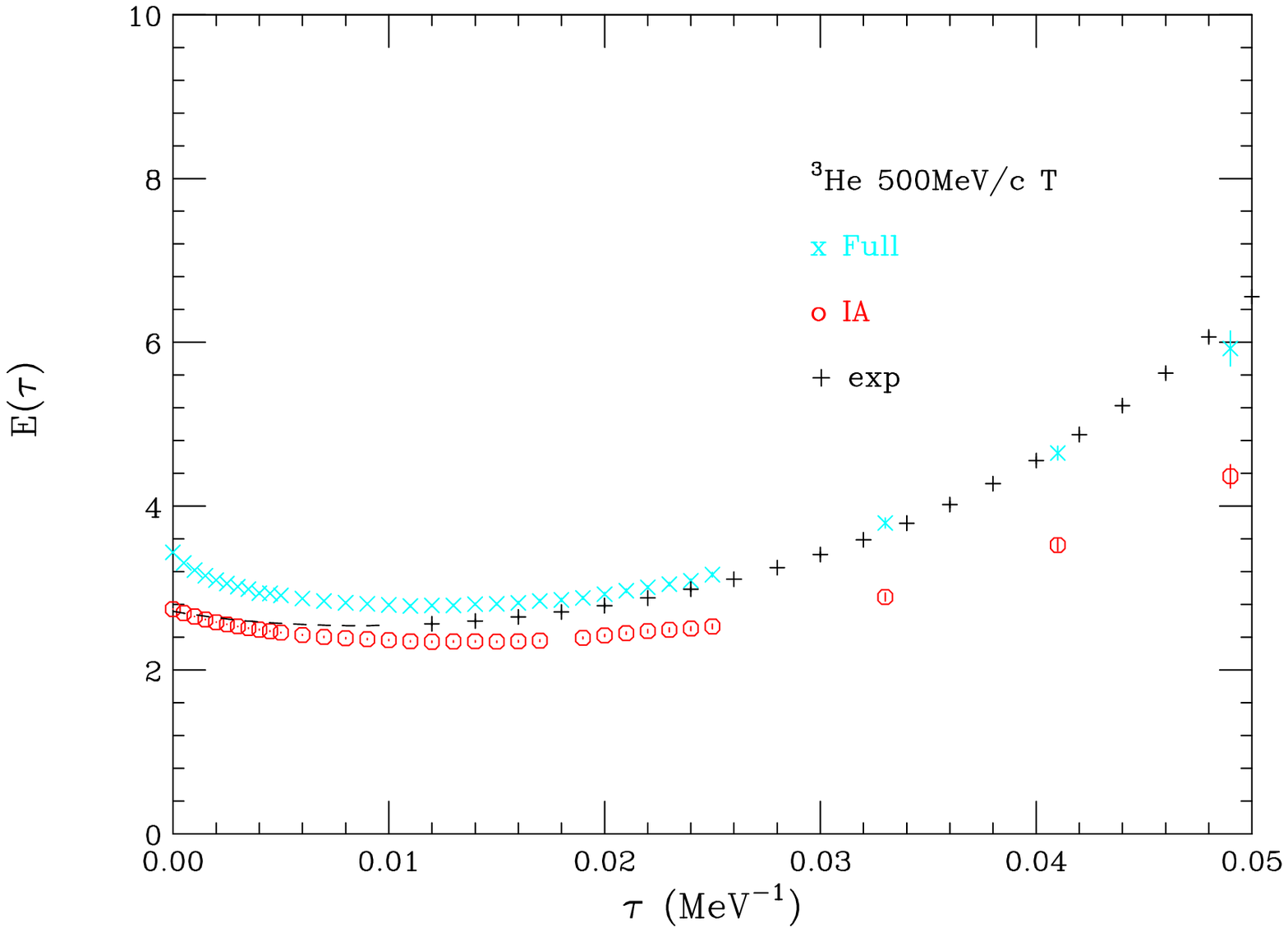}
\hspace{7mm} \epsfysize=53mm\epsffile{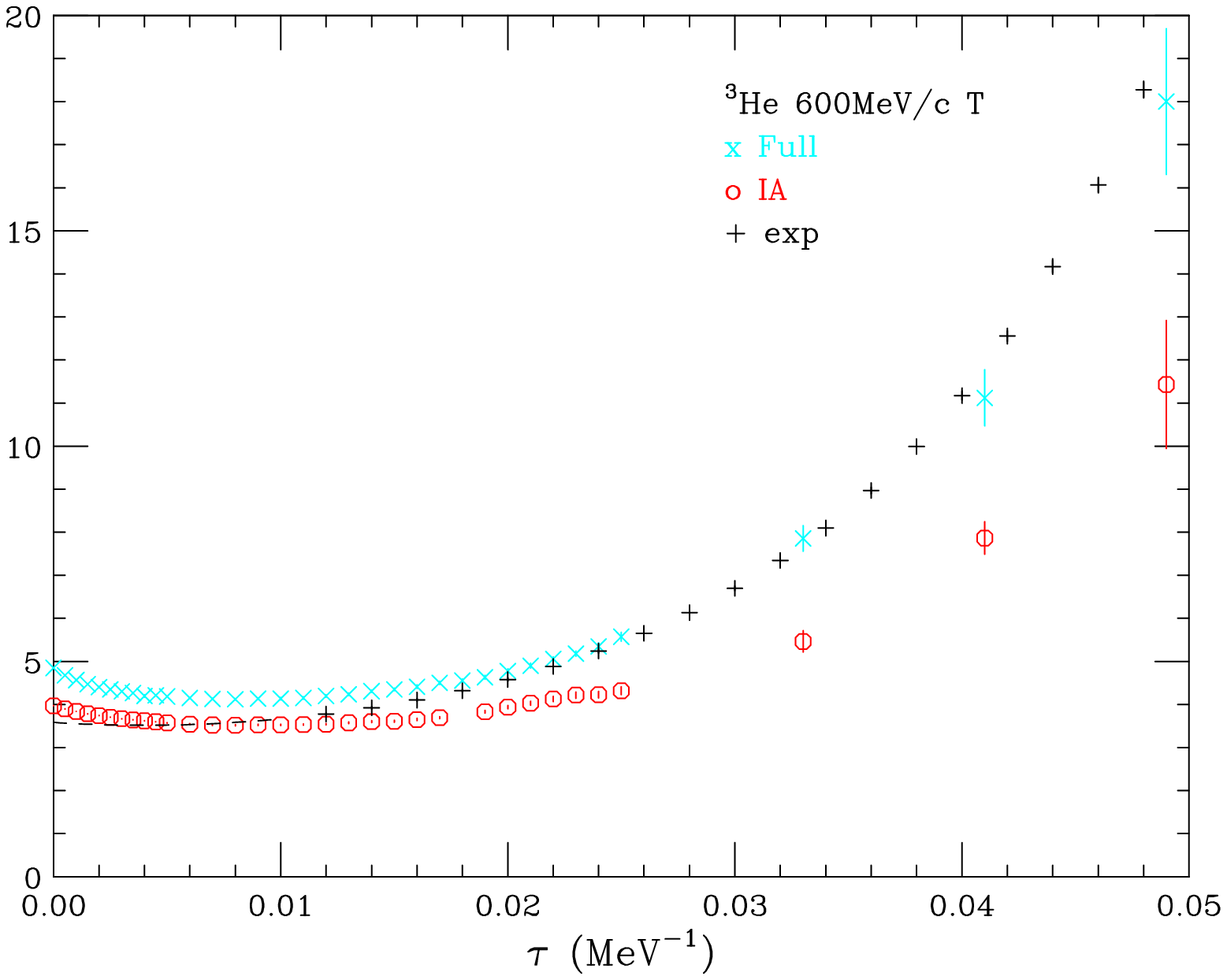}}}
\centerline{ \parbox{12cm}
{\caption[]{Longitudinal (upper half of figure) and transverse Euclidean 
response of $^3 {\rm He}$ for momentum transfers 300--600 MeV/c. 
}\label{euclhe32}}}
\end{figure}
%\clearpage

\begin{figure}[p]
\vspace*{-1cm}
\centerline{\mbox{\epsfysize=53mm\epsffile{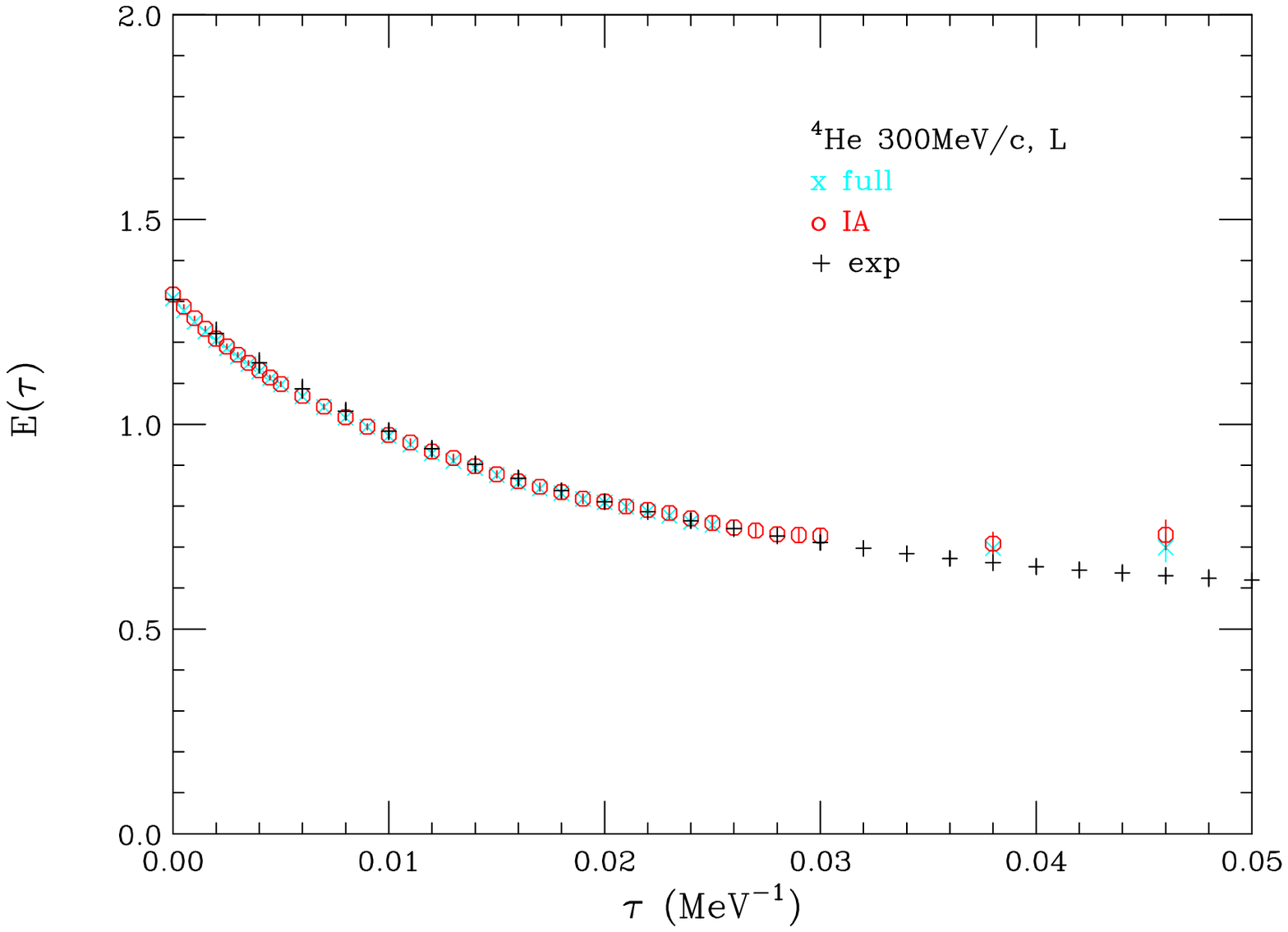}
\hspace{7mm} \epsfysize=53mm\epsffile{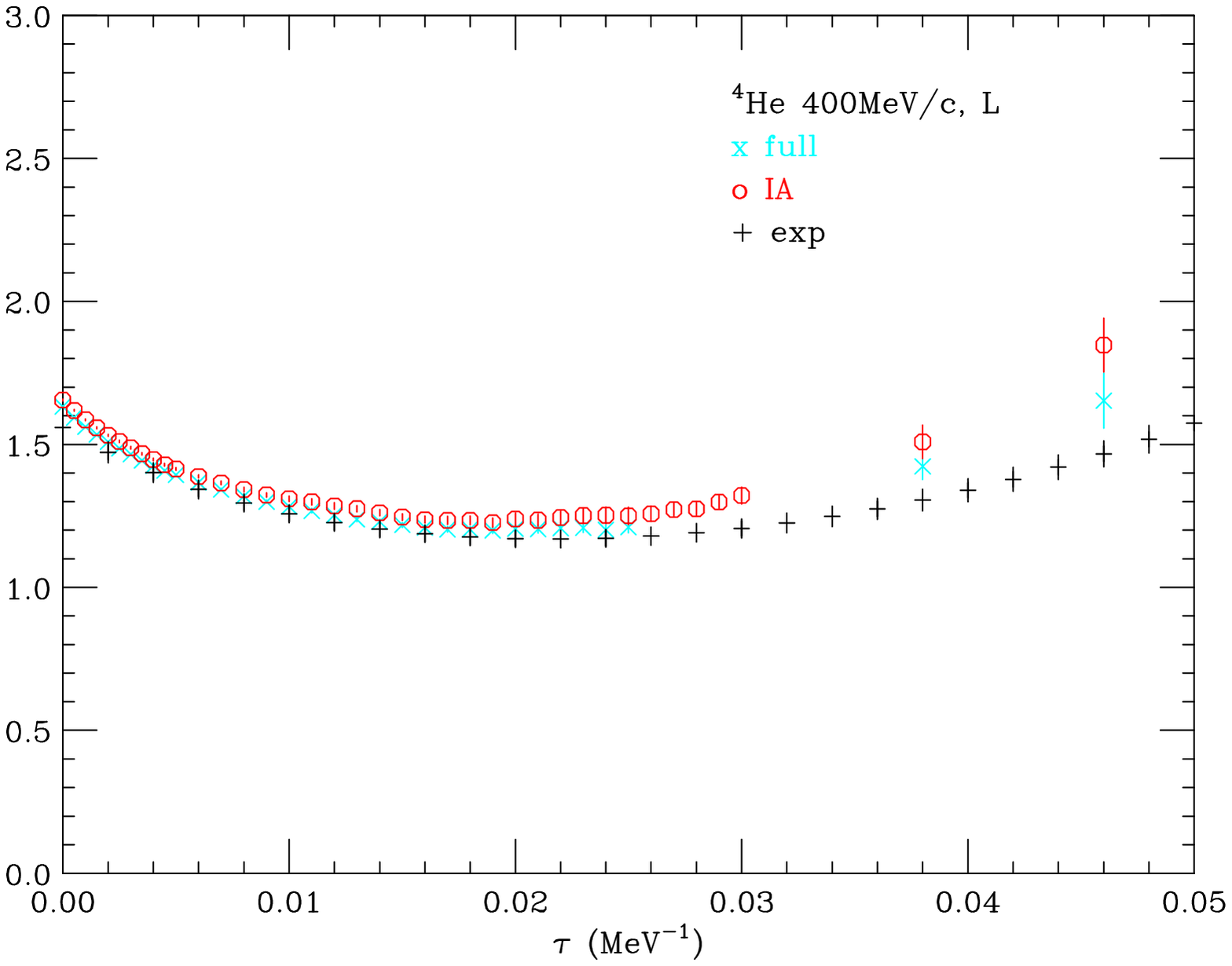}}}
\centerline{\mbox{\epsfysize=53mm\epsffile{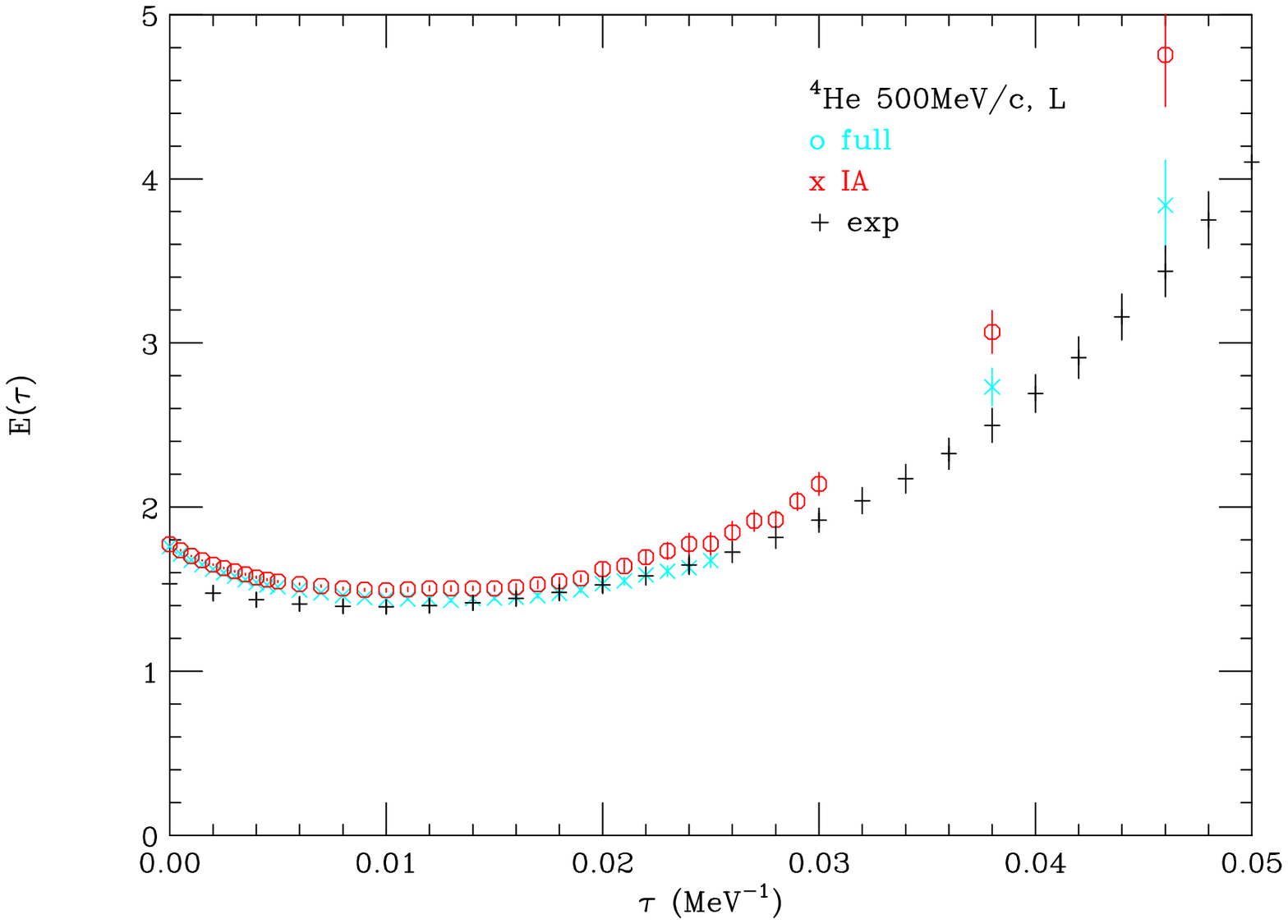}
\hspace{7mm} \epsfysize=53mm\epsffile{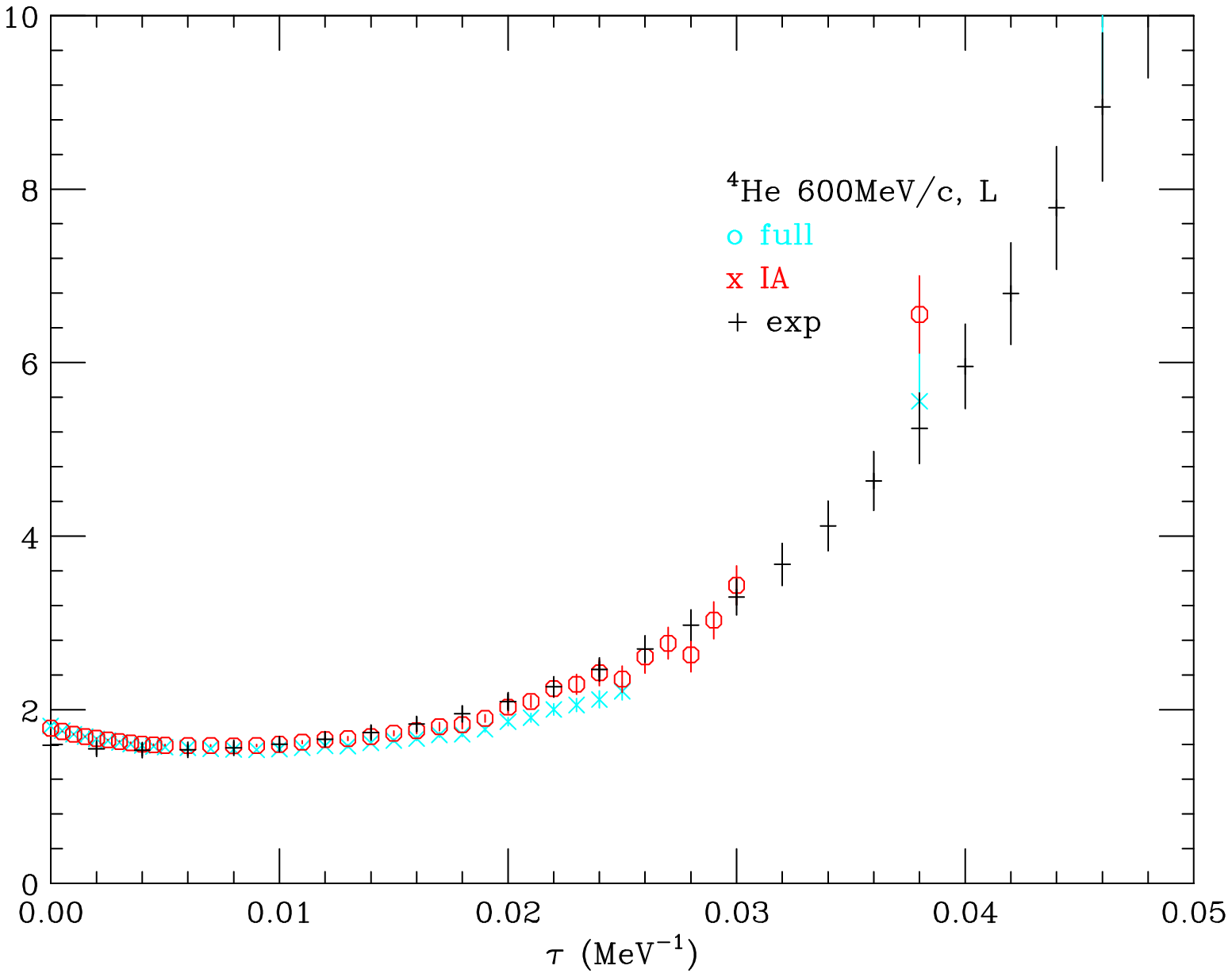}}}
\centerline{\mbox{\epsfysize=53mm\epsffile{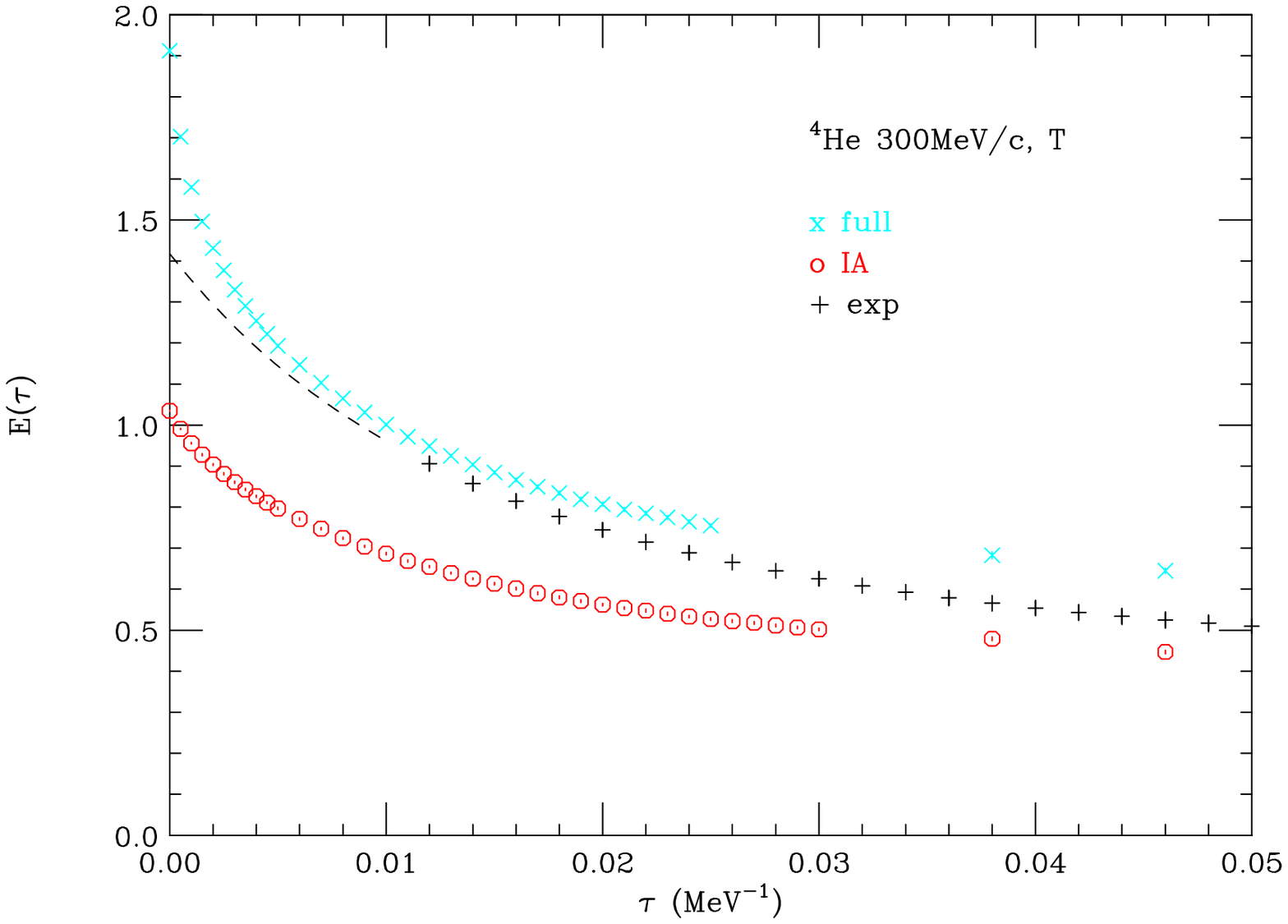}
\hspace{7mm} \epsfysize=53mm\epsffile{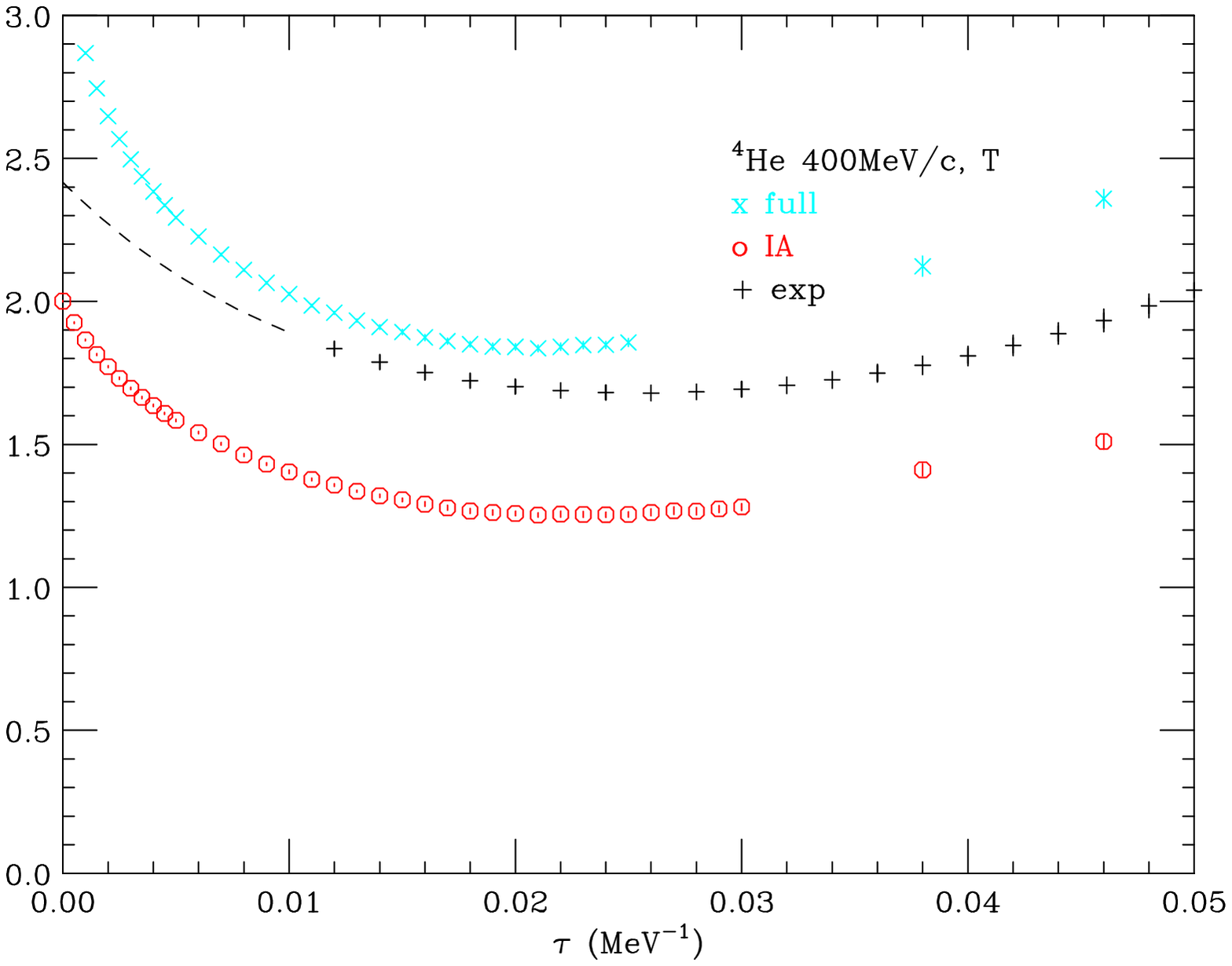}}}
\centerline{\mbox{\epsfysize=53mm\epsffile{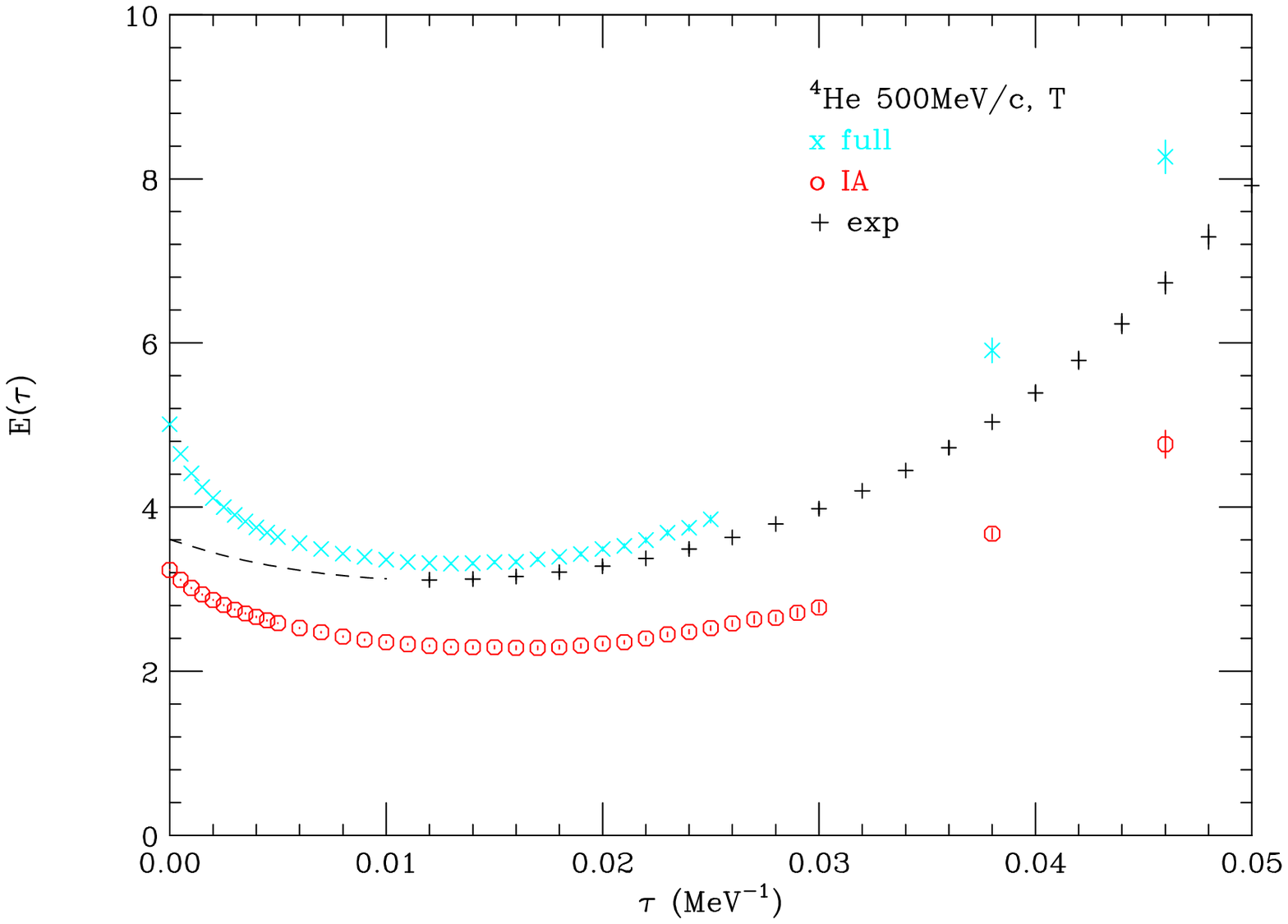}
\hspace{7mm} \epsfysize=53mm\epsffile{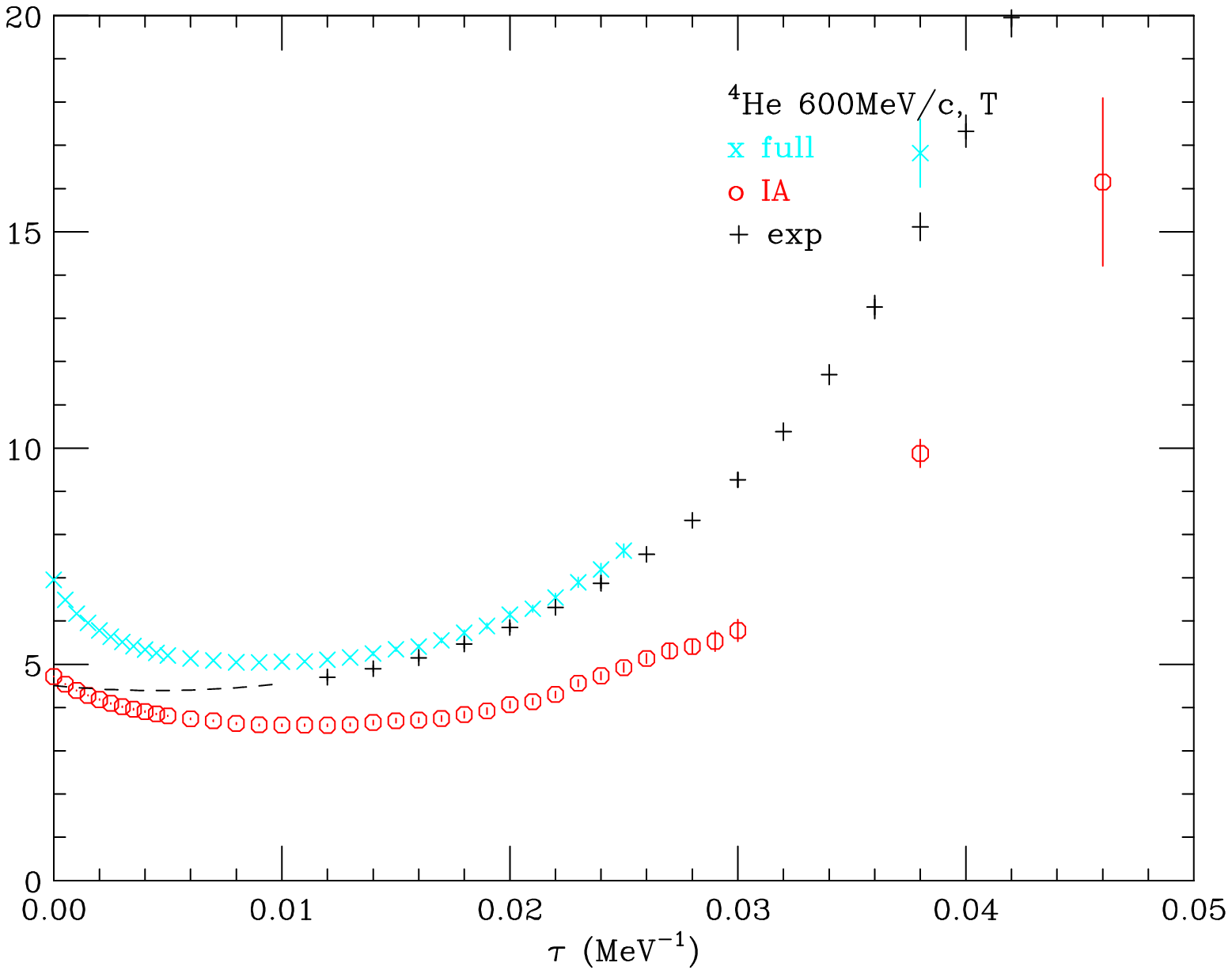}}}
\centerline{ \parbox{12cm}
{\caption[]{Longitudinal (upper half of figure) and transverse Euclidean response of $^4 {\rm He}$ for 
momentum transfers 
300--600 MeV/c. 
}\label{euclhe41}}}
\end{figure}

We have used the longitudinal ($L$) and transverse ($T$)
experimental response functions of Figs.~\ref{rhe3}
and \ref{rhe4} to compute the corresponding experimental Euclidean
responses shown in Figs. \ref{euclhe32} and \ref{euclhe41}.  The nucleon
electromagnetic form factors are divided out using the parameterizations
of H\"ohler {\em et al.} \cite{Hoehler76}.  In order to not include too 
much of the tail of the $\Delta$-resonance, the integration has been 
performed up to the energy loss $\omega$ where the $T$-response
starts to increase significantly with $\omega$ (the corresponding value
of $\omega$ is indicated in Figs. \ref{rhe3} and \ref{rhe4} by a $+$).
Since for the $T$-Euclidean response at very small $\tau$ the tail of the
$\Delta$-peak nevertheless plays a role, the experimental response in this region
is indicated by a dashed line only, and should not be compared to the
theoretical calculations discussed below.

The statistical errors of the experimental Euclidean response are obtained
via the usual error propagation when integrating.  The additional overall
systematic uncertainty, estimated from the $L/T$-separation, amounts to
typically 3\% for $^3$He for both the $L$- and $T$-cross sections.  For $^4$He
a similar uncertainty in the scale factor applies at the lower momentum
transfers and for both $L$ and $T$; at 600 MeV/c the uncertainty in $R_L(\omega)$
increases to 6\%.  These scale errors, which then apply also to the corresponding
Euclidean responses, have not been included in the error bars shown in
Figs.~\ref{euclhe32} and~\ref{euclhe41}.

In Figs.~\ref{euclhe32} and~\ref{euclhe41} we also show the calculated Euclidean
responses, obtained in IA and with inclusion, in addition, of the contributions
associated with the two-body charge and current operators, discussed in
Sec.~\ref{sec:MEC}.  It is immediately apparent that two-body contributions
reduce by a small amount the $L$-responses, while increasing the $T$-responses
very substantially at all momentum transfers.  The enhancement in the $T$-channel
occurs already at low $\omega$, as is seen from the Euclidean response at large
$\tau$.  Two-body effects thus are important over the entire quasi-elastic peak,
and not only --- as was often expected --- in the \lq\lq dip-region\rq\rq on the
large-$\omega$ side of the quasi-elastic peak.  These conclusions are in
agreement with those of an earlier study~\cite{Carlson94}, as well as with
those inferred from the super-scaling analysis of Ref.~\cite{Donnelly99a} for nuclei 
with mass number $A$=12--56. 

When considering the effect of two-body currents as a function of momentum
transfer --- in particular, when studying Fig.~\ref{euclhe41} --- one notes that 
at low $q$ the effect of two-body currents at large $\omega$ (low $\tau$)
is bigger than at low $\omega$ (large $\tau$).  At large $q$, this situation
becomes the reverse.  Figures~\ref{euclhe32} and~\ref{euclhe41} also show 
that theory explains well the rapid increase of two-body contributions
between $^3$He and $^4$He.  In contrast to most published calculations
(for a discussion see Sec.~\ref{intro}), the present calculation does
give the sizeable two-body contribution required by the data. 

Figure~\ref{euclhe41} shows that the $T$-Euclidean response at low
$q$ rises very rapidly towards very small $\tau$, reaching almost twice the 
IA value at $\tau$=0, thus suggesting that part of the two-body strength is located
at very large $\omega$, basically under the $\Delta$-peak (compare to
Fig.~\ref{respw}).  It also implies that this strength is very spread out in 
$\omega$, and presumably best discussed in terms of the sum-rule (see Sec.~\ref{sec:sum}). 

At lower values of momentum transfer, the calculated $^4$He $T$-Euclidean 
response is a bit high at large $\tau$, implying that the corresponding 
calculated cross section would be somewhat too high at low $\omega$.
As emphasized by the sensitivity studies in Sec.~\ref{model}, the low $\omega$
region gets great weight for large $\tau$, so a small increase in the absolute
value of $\sigma(\omega)$ leads to a large increase in $E(\tau)$.

Overall, the agreement between theory and experiment for $^4$He, the nucleus 
which allows us best to study the relative role of one- and two-body 
contributions, is excellent for the $L$-response, thus implying that
an accurate treatment of the nuclear spectrum has been achieved, since
two-body operators give small corrections in the $L$-channel.
For the $^4$He $T$-Euclidean response, the large two-body
effects predicted by theory are confirmed by experiment,
although the associated contributions are a bit too large
in the $q$-range 400--500 MeV/c.   

\section{Longitudinal and transverse sum rules
\label{sec:sum} }
Sum rules provide a powerful tool for studying integral
properties of the response of the nuclear many-body
system to an external electromagnetic probe.  Of particular
interest are those for the longitudinal and transverse
response functions at constant three-momentum transfers,
since they can be expressed as ground-state expectation
values of the charge and current operators and, therefore,
do not require any knowledge of the complicated structure
of the nuclear excitation spectrum.  Direct comparison
between the theoretically calculated and experimentally
extracted sum rules cannot be made unambiguously, however,
for two reasons.  Firstly, the experimental determination
of the longitudinal and transverse sum rules requires measuring
the associated response functions in the whole energy-transfer
range, from threshold up to $\infty$.  Inclusive electron
scattering experiments only allow access to
the space-like region of the four-momentum transfer
($\omega < q$).  While the response in the time-like
region ($\omega > q$) could in principle be measured
via $e^+ e^-$ annihilation, no such experiments have
been carried out to date, to the best of our knowledge.
Therefore, for a meaningful comparison between theory
and experiment, one needs to estimate the strength 
outside the region covered by experiment.  In the past, this
has been accomplished, in the case of the longitudinal
response, either by extrapolating the data~\cite{Jourdan96} or by 
parameterizing the high-energy tail and using
energy-weighted sum rules to constrain it~\cite{Schiavilla87,Schiavilla89b}.
For the $A$=2--4 nuclei, the unobserved strength amounts
to 5--10\% at the most for three-momentum transfers in
the range $q < 1$ GeV/c~\cite{Schiavilla89b}, and both procedures lead
to similar results.  Indeed, the calculated
(non-energy-weighted) longitudinal sum rule --- also
known as the Coulomb sum rule --- appears to be well satisfied
by the data~\cite{Schiavilla89b,Schiavilla93}.

The second reason that makes the direct comparison between theoretical
and \lq\lq experimental\rq\rq \ sum rules difficult lies in the inherent
inadequacy of the present theoretical model for the nuclear
electromagnetic current, in particular its lack of explicit
pion production mechanisms.  The latter mostly affect
the transverse response and make its $\Delta$-peak region
outside the boundary of applicability of the present theory.
The charge and current operators discussed in Sec.~\ref{sec:MEC},
however, should provide a realistic and quantitative description
of both longitudinal and transverse response functions in
the quasi-elastic peak region, where nucleon and (virtual) pion
degrees of freedom are expected to be dominant.  In light nuclei
and at the momentum transfer values of interest here, the
quasi-elastic and $\Delta$-production peaks are well
separated, and it is therefore reasonable to study sum rules
of the transverse response.

While non-energy- and energy-weighted longitudinal sum rules
have been extensively studied in the past (see
Refs.~\cite{Carlson98,Orlandini91} for a review), the number of studies
dealing with sum rules of the transverse response is much
more limited~\cite{Schiavilla89c}.  The present section focuses on the latter,
in particular on the enhancement of transverse strength due
to many-body components of the electromagnetic current, within
the limitations discussed above.  It also addresses, within
the sum-rule context, the issue of the enhancement in the ratio
of  transverse to longitudinal strength, observed in the
quasi-elastic response functions of nuclei.  Finally, it attempts
to provide a semi-quantitative explanation for the observed
systematics in the excess of transverse strength, both as function
of mass number and momentum transfer.  All the calculations are
based on the AV18/UIX Hamiltonian model, and use
correlated-hyperspherical-harmonics (variational
Monte Carlo) wave functions for $A$=3--4 ($A$=6) nuclei.

The (non-energy-weighted) sum rules are defined as
\begin{eqnarray}
S_\alpha(q)&=&C_\alpha \, \int_{\omega_{\rm th}^+}^\infty d\omega \,
S_\alpha (q,\omega) \nonumber \\
&=& C_\alpha \Big[ \langle 0| O_\alpha^\dagger({\bf q}) O_\alpha({\bf q}) |0\rangle 
-|\langle 0|O_\alpha({\bf q})|0\rangle|^2 \Big ] \>\>,
\label{eq:sums}
\end{eqnarray}
where $S_\alpha (q,\omega)$ is the point-nucleon longitudinal ($\alpha$=$L$)
or transverse ($\alpha$=$T$) response function, $O_\alpha({\bf q})$
is either the charge $\rho({\bf q})$ or current 
${\bf j}({\bf q})$ operator divided by the
square of the proton form factor $|G_E^p (\tilde{Q}^2)|^2$ (again, $\tilde{Q}^2$ is
evaluated at the energy transfer corresponding to the
quasi-elastic peak), $|0\rangle$ denotes 
the ground state, and the elastic contribution to the
sum rule has been removed.  An average over the nuclear spin orientations
is tacitly implied in the evaluation of the expectation values.
The constant $C_\alpha$, for $\alpha$=$L$ or $T$, is given by
\begin{eqnarray}
C_L &=& \frac{1}{Z}  \>\>, \\
C_T &=& \frac{ 2\, m^2}{ Z \mu_p^2 + N \mu_n^2 } \frac{1}{q^2}  \>\>,  
\end{eqnarray}
where $Z$ ($N$) and $\mu_p$ ($\mu_n$) are
the proton (neutron) number and magnetic moment, respectively.
It has been introduced in Eq.(\ref{eq:sums})
so that, in the limit $q \rightarrow \infty$ and under the
approximation that the nuclear charge and current
operators originate, respectively, from the charge and spin-magnetization
of the individual nucleons only, $S_\alpha(q \rightarrow \infty) =1$.
Note that the Euclidean response functions calculated in Sec.~\ref{results}
and the sum rules defined here are related via
\begin{equation}
S_\alpha(q)= C_\alpha \, E_\alpha(q,\tau=0) \>\>.
\end{equation}
The expectation values in Eq.~(\ref{eq:sums}) are calculated
with Monte Carlo methods, without any approximations.

\begin{figure}[htb]
\centerline{\mbox{\epsfysize=100mm\epsffile{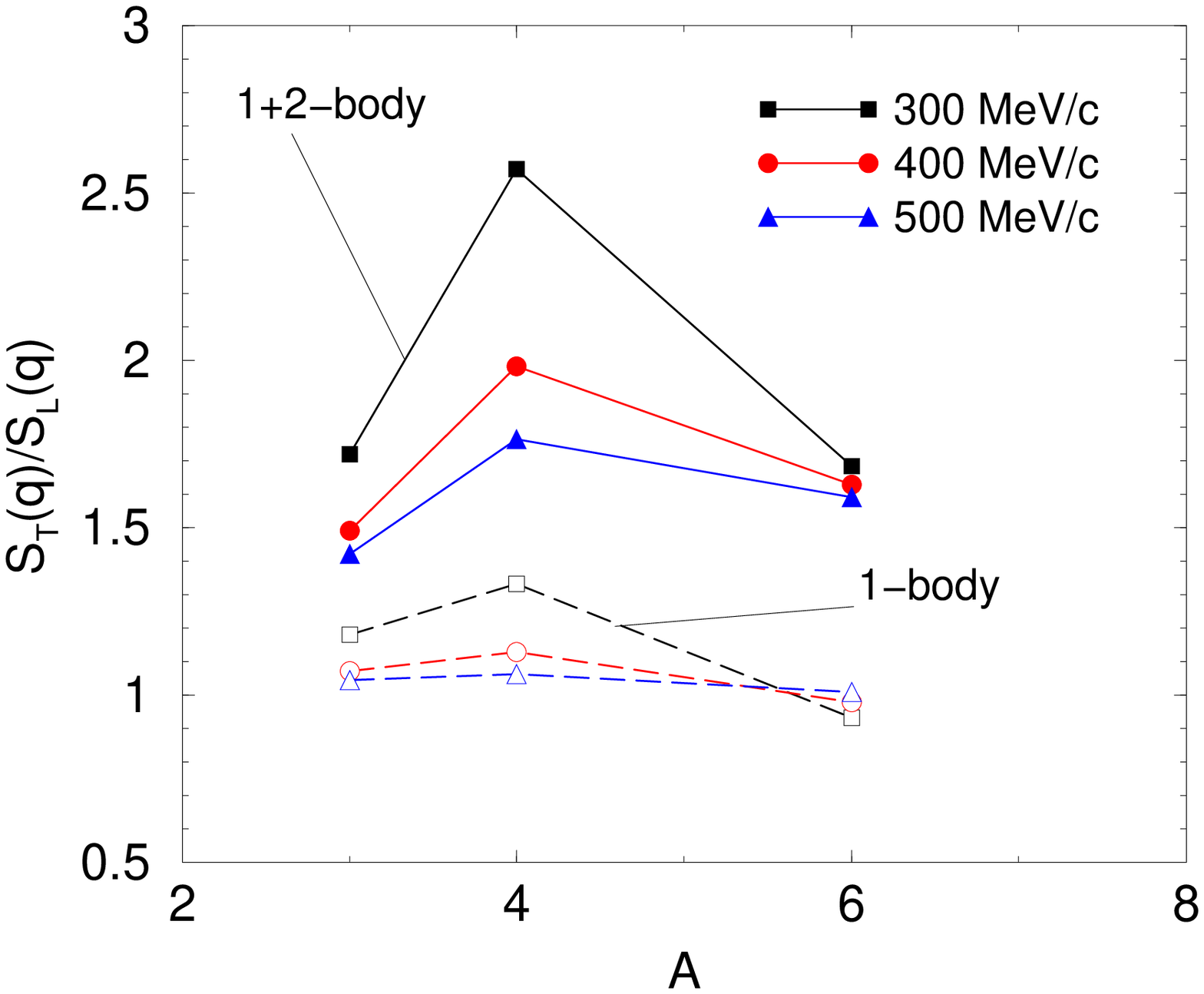}}}
\vspace*{0.5cm}
\caption[]{The ratios $S_T(q)/S_L(q)$, obtained with one-body currents
only and both one- and two-body currents, as function of mass number $A$.}
\label{fig:slta}
\end{figure}

\begin{figure}[htb]
\centerline{\mbox{\epsfysize=100mm\epsffile{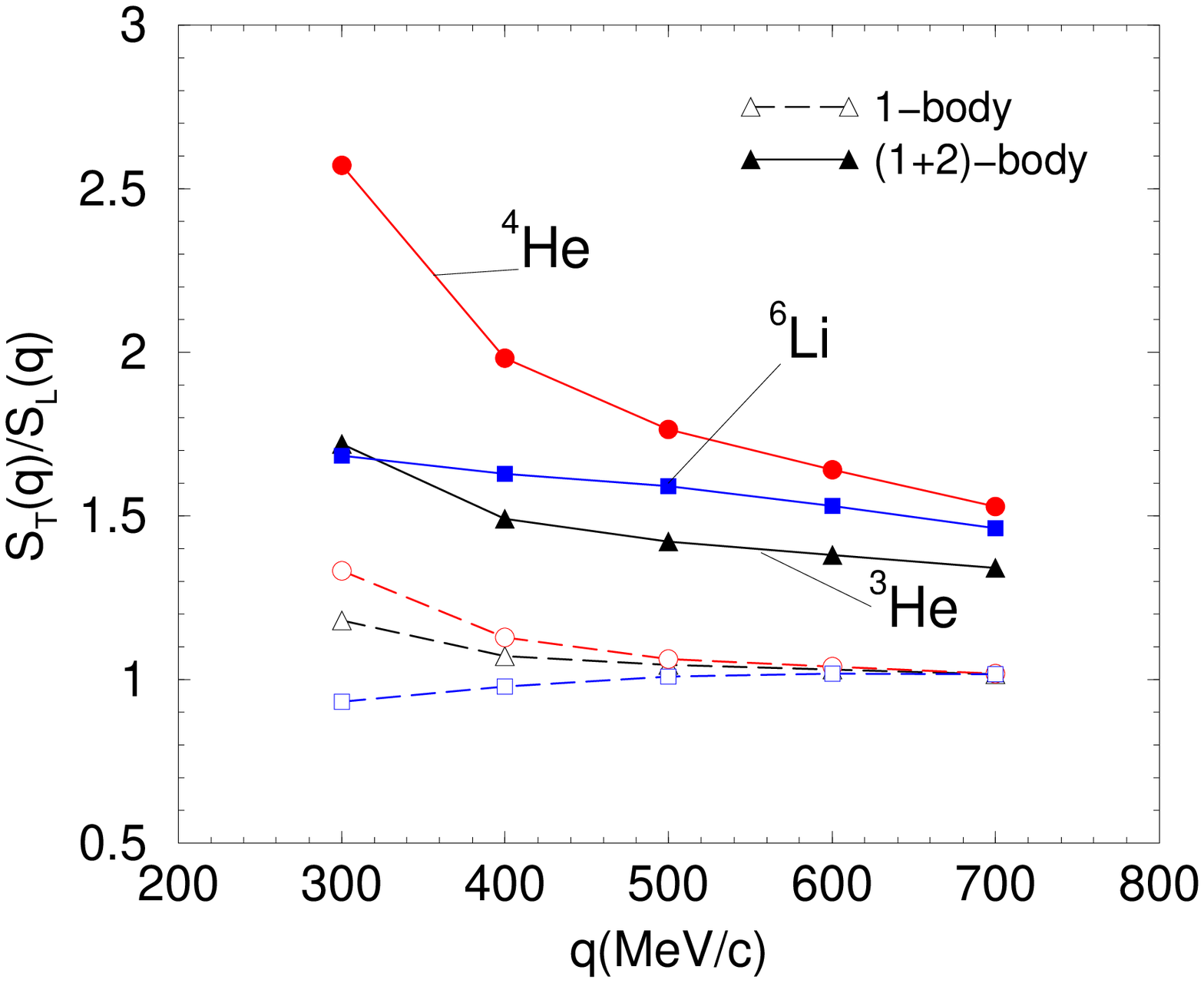}}}
\vspace{0.5cm}
\caption[]{The ratios $S_T(q)/S_L(q)$, obtained with one-body currents
only and both one- and two-body currents, as function
of momentum transfer $q$.}
\label{fig:sltq}
%\vspace*{2cm}
\end{figure}

\begin{table}[htb]
\begin{center}
\begin{tabular}{|ccc|cc|cc|}
\hline
& \multicolumn{2}{c} {$^3${\rm He}} & \multicolumn{2}{c} {$^4${\rm He}} & \multicolumn{2}{c|} {$^6$Li} \\
\hline
q(MeV/c) & 1 & 1+2 & 1 & 1+2 & 1 & 1+2  \\
\hline
300  & 0.787 & 0.763 & 0.670 & 0.649 & 0.977 & 0.933 \\
400  & 0.921 & 0.875 & 0.859 & 0.815 & 0.995 & 0.932 \\
500  & 0.964 & 0.901 & 0.941 & 0.881 & 0.990 & 0.921 \\
600  & 0.982 & 0.908 & 0.973 & 0.910 & 0.990 & 0.924 \\
700  & 0.994 & 0.914 & 0.994 & 0.942 & 0.994 & 0.938 \\
\hline
\end{tabular}
\caption{\label{tbsl}
 The longitudinal sum rule obtained with one-body only
and both one- and two-body charge operators.}
\end{center}
\end{table}
\begin{table}[htb]
\begin{center}
\begin{tabular}{|ccc|cc|cc|}
\hline
& \multicolumn{2}{c} {$^3${\rm He}} & \multicolumn{2}{c} {$^4${\rm He}} & \multicolumn{2}{c|} {$^6$Li} \\
\hline
q(MeV/c) & 1 & 1+2 & 1 & 1+2 & 1 & 1+2  \\
\hline
300  & 0.929 &  1.31 & 0.893 & 1.67 & 0.912 & 1.57 \\
400  & 0.987 &  1.30 & 0.970 & 1.62 & 0.974 & 1.52 \\
500  & 1.01  &  1.28 & 1.00  & 1.55 & 0.999 & 1.46 \\
600  & 1.01  &  1.25 & 1.01  & 1.49 & 1.01  & 1.41 \\
700  & 1.01  &  1.23 & 1.01  & 1.44 & 1.011 & 1.37 \\
\hline
\end{tabular}
\caption{\label{tbst}
 The transverse sum rule obtained with one-body only
and both one- and two-body current operators.}
\end{center}
\end{table}
The calculated sum rules for $^3${\rm He}, $^4${\rm He}, and $^6$Li are
listed in Tables~\ref{tbsl} and~\ref{tbst}.
The longitudinal sum rule $S_L(q)$ is relatively un-influenced
by two-body charge operators, in agreement with the results
of an earlier study~\cite{Schiavilla93}.  The 
transverse sum rule $S_T(q)$ is substantially increased by
two-body current contributions.  The resulting enhancement
has two interesting features: i) it increases, for fixed $q$,
in going from $A$=(2 to) 3 to 4, and decreases from $A$=4 to 6;
ii) it decreases, for fixed $A$, as $q$ increases.  Both
these features are summarized in Figs.~\ref{fig:slta} and~\ref{fig:sltq},
in which the ratios $S_T(q)/S_L(q)$, obtained by including
one-body only and both one- and two-body contributions,
are plotted as function of $A$ for fixed
$q$ and as function of $q$ for fixed $A$.  The former
figure is reminiscent of Fig.~\ref{rat}, in which the
ratio of transverse to longitudinal strength in the
quasi-elastic region is obtained from the measured
response functions.  Obviously, the truncated integrals in Fig.~\ref{rat}
do not include the strength at high $\omega$.

The purpose of the present section is to offer an
explanation of the features mentioned above.  To this
end, three points are worth emphasizing.
Firstly, among the two-body current contributions,
the most important are those associated with
the $PS$ (pion-like) and $\Delta$-excitation currents.
This fact has been explicitly verified by direct calculation,
as shown in Table~\ref{tbpi-dlt} for $^4${\rm He}, as an example.
Note that the results in the 2nd and 3rd columns are
slightly different from those reported above in Table~\ref{tbst},
since they are based on a random walk consisting only of
1,000 configurations, much shorter than that used in 
the calculations of Table~\ref{tbst}.  These calculations, though,
are based upon the same random walk and therefore allow a better
determination of the individual contributions.

\begin{table}[htb]
\begin{center}
\begin{tabular}{|cccc|}
\hline
q(MeV/c) & 1 & 1+2 & 1+2-$\pi$+2-$\Delta$  \\
\hline
300 &  0.915 & 1.65 & 1.58 \\
400 &  0.980 & 1.59 & 1.50 \\
500 &  1.01 & 1.53 & 1.44 \\
600 &  1.01 & 1.47 & 1.38 \\
700 &  1.01 & 1.41 & 1.33  \\
\hline
\end{tabular}
\caption{\label{tbpi-dlt}
The $^4${\rm He} transverse sum rule.}
\end{center}
\end{table}
Secondly, consider expanding the current into one- and two-body 
components $j_l$ and $j_{lm} $,
\begin{equation}
j=\sum_l j_l +\sum_{l<m} j_{lm} \>\>.
\end{equation}
Then, ignoring the very small (and,
with increasing $q$, rapidly vanishing) elastic contribution
to $S_T(q)$, one finds that the first term in Eq.~(\ref{eq:sums}) 
\begin{eqnarray}
j^\dagger j &=& \sum_l j^\dagger_l j_l + \sum_{l \neq m} j^\dagger_l j_m \nonumber \\
            &&+\sum_{l<m} [ ( j^\dagger_l+j^\dagger_m ) j_{lm} + {\rm h.c.} ]
              +\sum_{l<m} j^\dagger_{lm} j_{lm} \nonumber \\
            &&+\>\>{\rm terms\>\>\>involving\>\>\> 3\>\>or
               \>\>4\>\>\> different\>\>\>nucleons} \>\>.
\label{eq:paircont}
\end{eqnarray}
At large momentum transfers, one would expect
terms involving 3 or 4 nucleons to be small, particularly
in light nuclei where Pauli correlations are unimportant.
Dropping the last term corresponds to considering only
incoherent scattering from pairs of nucleons.

This simple expectation is indeed borne
out by a direct calculation, the results of which are listed for
$^4${\rm He} in Table~\ref{tbstmany}.  Thus the transverse sum rule
appears to be saturated by the one- and two-body terms
in the expansion for $j^\dagger j$ above.
\begin{table}[htb]
\begin{center}
\begin{tabular}{|cccc|}
\hline
q(MeV/c) & 1 & 1+2 & 1+2-reduced  \\
\hline
300 &  0.915 & 1.65 & 1.70 \\
400 &  0.980 & 1.59 & 1.59 \\
500 &  1.01 & 1.53 & 1.51 \\
600 &  1.01 & 1.47 & 1.45 \\
700 &  1.01 & 1.41 & 1.39  \\
\hline
\end{tabular}
\caption{ \label{tbstmany}
The $^4${\rm He} transverse sum rule: effect 
of three- or four-nucleon terms.}
\end{center}
\end{table}

Thirdly, the transverse strength associated with two-body currents
is almost entirely due to $pn$ pairs.  To make this observation
more precise, consider the \lq\lq reduced\rq\rq two-body current:
\begin{equation}
j_{lm} \rightarrow j_{lm}( P_l\, P_m+N_l\, N_m) \>\>,
\end{equation}
where $P_l$ and $N_l$ are the proton and neutron projection
operators for particle $l$.  Thus the \lq\lq reduced\rq\rq
two-body current only acts on $p$$p$ or $n$$n$ pairs, and
the transverse sum rule calculated with it should be given
almost entirely by the one-body part of $j$.  This fact
is again confirmed by direct calculation, as it
is evident from Table~\ref{tbstredu}.
\begin{table}[htb]
\begin{center}
\begin{tabular}{|cccc|}
\hline
q(MeV/c) & 1 & 1+2 & 1+2--$p$$p$ or $n$$n$ only \\
\hline
300 &  0.915 & 1.65 & 0.919 \\
400 &  0.980 & 1.59 & 0.987 \\
500 &  1.01 & 1.53 & 1.02 \\
600 &  1.01 & 1.47 & 1.03 \\
700 &  1.01 & 1.41 & 1.03  \\
\hline
\end{tabular}
\caption{ \label{tbstredu}
The $^4${\rm He} transverse sum rule: contribution of $pp$ and $nn$ pairs.}
\end{center}
\end{table}
That $p$$n$ pairs are responsible for the strength due to two-body
currents can also be understood by the following considerations.
The pion-like and $\Delta$-excitation currents have the isospin
structure (see Sec.~\ref{sec:MEC}), again in a schematic notation,
\begin{equation}
j_{lm}(\pi) = ( \tau_l \times \tau_m )_z O_{lm}(\pi) \>\> ,
\end{equation}
\begin{equation}
j_{lm}(\Delta) = \tau_{l,z} O_{lm}(\Delta,{\rm a}) + \tau_{m,z} O_{ml}(\Delta,{\rm a})
                +( \tau_l \times \tau_m )_z O_{lm}(\Delta,{\rm b}) \>\> ,
\end{equation}
while the leading part of the one-body current is given by
\begin{equation}
j_l = \tau_{l,z} O_l(IV) \>\>,
\end{equation} 
where $O_l(IV)$ denotes the isovector part of $j_l$.
Now the term $j^\dagger_{lm} j_{lm}$ (with $j_{lm}$
including pion-like and $\Delta$-excitation
currents) will produce, as far as isospin is concerned, terms like
\begin{equation}
( \tau_l \times \tau_m )_z^2 = 2\, ( 1 - \tau_{l,z} \tau_{m,z} ) \>\>,
\label{eq:iso1}
\end{equation}
\begin{equation}
( \tau_{l,z}\>\>\>{\rm or}\>\>\> \tau_{m,z} ) ( \tau_l \times \tau_m )_z =
        \pm {\rm i} \, ( \tau_l \cdot \tau_m - \tau_{l,z} \tau_{m,z} ) \>\>, 
\label{eq:iso2}
\end{equation}
\begin{equation}
\tau_{l,z} \tau_{m,z} = \frac{T_{lm} + \tau_l \cdot \tau_m}{3} \>\>,
\label{eq:iso3}
\end{equation}
where $T_{lm}$ is the isotensor term
$T_{lm} = 3\, \tau_{l,z} \tau_{m,z} - \tau_l \cdot \tau_m$.

In addition, there will also be isospin-independent terms,
of the type
\begin{equation}
O^\dagger_{lm}(\Delta,{\rm a}) O_{lm}(\Delta,{\rm a}) \>\>.
\end{equation}
However, it is important to note that the operators in Eqs.~(\ref{eq:iso1})
and~(\ref{eq:iso2}) vanish when acting on $p$$p$ or $n$$n$ pairs.
It should also be noted that the isotensor term in Eq.~(\ref{eq:iso3})
vanishes in $T$=0 and $T$=1/2 ground states, namely in $^3${\rm He}, $^4${\rm He} and $^6$Li.
A similar analysis can be carried out for the interference terms
between one- and two-body currents,
\begin{equation}
\sum_{l<m} ( j^\dagger_l + j^\dagger_m ) j_{lm} + {\rm h.c.} \>\>,
\end{equation}
for which one obtains isospin-independent, and type~(\ref{eq:iso2})
or type~(\ref{eq:iso3}) operators.  In any case, the direct calculation
indicates (see Table~\ref{tbstredu}) that $p$$p$ and $n$$n$ pairs
do not contribute appreciably to $S_T(q)$.

On the basis of the above observations and ignoring the
convection term in the one-body $j_l$, one concludes that the excess
transverse strength, defined as
\begin{equation}
\Delta S_T(q) \equiv S_T(q) - S_T(q;1\!-\!{\rm body}) \>\>,
\end{equation}
must be proportional to
\begin{equation}
\Delta S_T(q) = \int_0^\infty dx \, {\rm tr}[ F(x;q)\, \rho(x;pn) ] \>\>,
\label{eq:ra}
\end{equation}
where $F(x;q)$ is a complicated matrix in the spin-space of the
two nucleons depending upon the current operators
alone, and the $A$-dependence
is included in the $p$$n$ elements of the two-nucleon density matrix 
$\rho_2(x;pn,s_l,s_m,s_l^\prime,s_m^\prime)$.
Here $s_l$, $s_m$, etc., are spin projections (up or down) of particles
$l$, $m$, etc.  In fact, one can express these densities in terms
of total spin-isospin $S,T$=0,1 or 1,0 for pair
$lm$.  The crucial point is that, in nuclei, these $p$$n$
densities scale, see~\cite{Forest96}, namely
\begin{equation}
\rho_2(x;T=0;A) = R_A \,\, \rho(x;T=0;{\rm deuteron}) \>\>,
\end{equation}
\begin{equation}
\rho_2(x;T=1;A) = R^\prime_A \,\, \rho(x;T=1;^1{\rm S}_0\>\>{\rm quasi\!-\!bound}) \>\>.
\end{equation}
The scaling factors $R_A$ and $R_A^\prime$ have been calculated in Ref.~\cite{Forest96},
with $R_A^\prime \simeq R_A$ and $R_A$=2.0, 4.7, 6.3, 18.8 for $^3$He, $^4$He, $^6$Li,
and $^{16}$O, respectively, and so one would expect $\Delta S_T(q)$ to scale with
\begin{equation}
\Delta S_T(q) \propto \frac{R_A}{ Z \mu_p^2 + N \mu_n^2 } \>\>,
\label{eq:scale}
\end{equation}
where the factor in the denominator on the r.h.s. is from
the normalization adopted for $S_T(q)$.

\begin{figure}[htb]
\centerline{\mbox{\epsfysize=100mm\epsffile{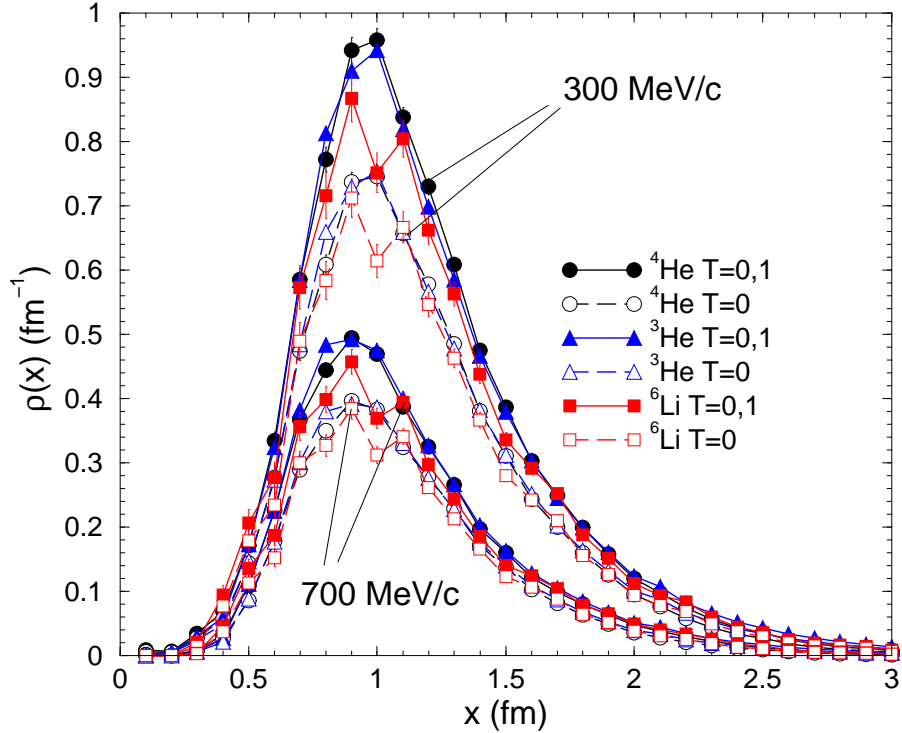}}}
\vspace{5mm}
\caption[]{The integrands in Eq.(\protect\ref{eq:ra}) for
$^3$He, $^4$He, and $^6$Li at momentum transfers of
300 and 700 MeV/c, scaled according to the factor
in Eq.~(\protect\ref{eq:scale}).  Also shown are the
contributions due $p$$n$ pairs in $T$=0 states only.}
\label{fig:rhoa}
%\vspace*{2cm}
\end{figure}

The calculated values for the excess strength $\Delta S_T(q)$
are listed in Table~\ref{tbdst}. 
\begin{table}[htb]
\begin{center}
\begin{tabular}{|cccc|}
\hline
q(MeV/c) & $^3$He & $^4$He & $^6$Li \\
\hline
300 & 0.38  & 0.78  & 0.66 \\
500 & 0.27  & 0.55  & 0.46 \\
700 & 0.21  & 0.43  & 0.36 \\
\hline
\end{tabular}
\caption{ \label{tbdst}
The excess strength $\Delta S_T(q)$ calculated
in $^3$He, $^4$He, and $^6$Li.}
\end{center}
\end{table}
On the basis of the scaling law above one would deduce 
\begin{eqnarray}
\frac{\Delta S_T(q;^4{\rm He})}{\Delta S_T(q;^3{\rm He})}
&\simeq& 0.840 \, \frac{R_4}{R_3} = 1.97 \>\>, \\
\frac{\Delta S_T(q;^6{\rm Li})}{\Delta S_T(q;^4{\rm He})}
&\simeq& 0.667  \, \frac{R_6}{R_4} =0.894  \>\>,
\end{eqnarray}
and these values are reasonably close to those of Table~\ref{tbdst}.
They are also close to those that can be inferred from data,
see Fig.~\ref{rat}.  Finally, in Fig.~\ref{fig:rhoa} 
the integrands in Eq.~(\ref{eq:ra}) are displayed for
$^3$He, $^4$He, and $^6$Li, properly scaled according to the
factor in Eq.~(\ref{eq:scale}).  Note that also shown are the
contributions due to $p$$n$ pairs in $T$=0 states only.
The behavior of the integrands, as illustrated in
Fig.~\ref{fig:rhoa}, is to be expected, since it is
a consequence of the \lq\lq scaling\rq\rq  behavior more
generally observed for the calculated $T,S$=0,1 and 1,0
pair-distribution functions in nuclei~\cite{Forest96}.

Note that the dominant contributions to the excess
strength occur for pair separations around or slightly
less than one fermi.  One would naturally associate
this strength with a significant contribution to
two-nucleon final states of relatively large
relative momenta.  Detailed microscopic calculations
in $A$=4 with full final-state interactions and
two-body currents will be necessary to
make precise predictions.

It should be emphasized that the scaling law
for $\Delta S_T(q)$ can be used to estimate
the excess transverse strength in nuclei, once the
factors $R_A$ are known.  In nuclear
matter, for example, the authors of Ref.~\cite{Akmal98} obtain
$R_\infty=1.59$, the latter being defined
as $R_A/A$ in the limit $A \rightarrow \infty$, and
therefore one would expect a very substantial enhancement
of the transverse sum rule (and, consequently,
transverse response function) due to two-body currents, namely
$\Delta S_T(q;\infty)/\Delta S_T(q;^4{\rm He})\simeq 1.35$.

The nuclear matter transverse
response has been calculated in
Ref.~\cite{Fabrocini97}, by using correlated-basis-function
perturbation theory and including,
in addition to the single-nucleon spin-magnetization
current, the pion-like and $\Delta$-excitation
two-body currents.  Explicit integration of the
response functions~\cite{Fabrocini97,Fabrocini01} indicates
that the transverse strength is
increased by the two-body contributions by roughly 15\%
in the momentum transfer range 300--700 MeV/c.  This
enhancement is significantly smaller than that inferred from
the scaling law above.  The underestimate is presumably
due to the inherent limitations in the calculations carried out
so far, which only retained one-particle--one-hole (1$p$-1$h$)
intermediate states and estimated the contribution of two-particle--two-hole
states by folding the 1$p$-1$h$ response with a width
derived from the imaginary part of the optical potential. 
It should be possible to calculate
the transverse sum rule by direct evaluation
of the ground state expectation value.  Work along
these line is in progress~\cite{Fabrocini01}.

The longitudinal and transverse sum rules in matter can be
estimated in a Fermi gas model in a similar simple manner.
As in calculations of $^4$He (see Sec. \ref{ingredients}), this is useful to help
understand the role of initial-state correlations in
the transverse response of the nucleus.
We again ignore contributions of 3 and 4 nucleon terms
as in Eq. \ref{eq:paircont}.  In matter
this approximation should be valid at high momentum
transfer $q$, but becomes more questionable as $q$ 
is decreased.  A significant enhancement of
the transverse sum rule is expected due to the short-range
part of the two-body currents --- these necessarily involve
large momenta between the pair of nucleons, thus 
broadening the range of validity of this simple approximation.

The Fermi-gas sum rules are decomposed into parts depending only on
the single nucleon currents and the remaining terms which also
involve two-nucleon currents:
\begin{equation}
S_\alpha (q) = S_\alpha(q;1\!-\!{\rm body}) + \Delta S_\alpha(q) \>\>,
\end{equation}
where $\alpha$=$L$, $T$.  The simplest approximation
to the response due to one-body currents
is to assume incoherent scattering from isolated nucleons.  This yields
$S_\alpha(q)=1$, neglecting the neutron charge
and convection current contributions to the longitudinal
and transverse response functions, respectively.

The additional contributions to the sum rules involving
two-nucleon currents can be written as a sum of interference
terms between one- and two-nucleon operators and the square
of the two-nucleon operators, in the same approximation
adopted above.  The short-range nature of the
two-nucleon operators implies that incoherent
scattering from pairs of nucleons should be dominant:
\begin{equation}
\Delta S_\alpha (q) = ( A-1 ) \big[ 2\, \langle\Phi| [ O^\dagger_{\alpha,l}({\bf q})
O_{\alpha,lm}({\bf q}) + {\rm h.c.} ]
+ O^\dagger_{\alpha,lm}({\bf q}) O_{\alpha,lm}({\bf q})|\Phi\rangle\big] \>\>,
\end{equation}
where the factor 2 in the interference term arises because
the pair term can connect to either of the two single-nucleon operators.

To make a simple estimate of the contribution of the two-nucleon
terms of the current, we consider the two-nucleon density matrix
$\rho_2(r_{lm})_{\chi,\chi^\prime}$, which depends only upon the separation
between the pair of nucleons and upon their initial and final
spins and isospins $\chi$ and $\chi^\prime$.
In the Fermi gas approximation, $\rho_2$ is diagonal in the spins
and isospins, and the spatial dependence is given by simple
Slater functions.  We then obtain:
\begin{equation}
\Delta S_\alpha(q) = \rho \sum_{\chi,\chi^\prime}  \int d{\bf r}_{ij} \int \frac{d\Omega_{\hat{\bf q}}}{4 \pi}
\, \rho_2 (r_{ij})_{\chi,\chi^\prime}\, \langle \chi |\big[ 2\, O^\dagger_{\alpha,l}({\bf q})
 + O^\dagger_{\alpha,lm}({\bf q}) \big] O_{lm}({\bf q}) |\chi^\prime \rangle  \>\>,
\end{equation}
where momentum-dependent pieces in the current have again been
dropped.  The excess contributions involving two-nucleon
currents are given in Table~\ref{tbfermi} for the Fermi gas model.
The longitudinal contributions are positive but small, ranging up to $\simeq 0.02$.
The transverse contributions, ranging from $\simeq 0.06$ at 700 MeV/c
to $\simeq 0.11$ at 300 MeV/c, are significant.

Finally, as far as the $q$-dependence is concerned, from
the explicit expressions of the current operators
in Sec.~\ref{sec:MEC} it is evident that the excess
transverse strength should behave as  
\begin{equation}
\Delta S_T(q) \simeq ( \alpha +\beta \, q + \gamma \, q^2 )/q^2 \>\>,
\end{equation}
where the $q^2$-factor in the denominator is due
to the normalization adopted for $S_T(q)$, and so will
approach a constant in the limit of large $q$.

\begin{table}[htb]
\begin{center}
\begin{tabular}{|crr|}
\hline
q (MeV/c) & $\Delta S_L$ & $\Delta S_T$ \\
\hline
300 & 0.004    & 0.114  \\
400 & 0.007    & 0.081  \\
500 & 0.011    & 0.066  \\
600 & 0.017    & 0.060  \\
700 & 0.024    & 0.056   \\
\hline
\end{tabular}
\caption{\label{tbfermi}
Excess-strength contributions $\Delta S_L$ and $\Delta S_T$
to the Fermi gas sum rules from terms involving two-nucleon currents.}
\end{center}
\end{table}
\section{Model studies with simplified interactions, wave functions, and
currents \label{ingredients}}
As a guide to better understanding these results and comparing with
other calculations, it is useful to  contrast
the complete calculations described above with various truncations of the
initial ground state wave function, the current operators, and 
the Hamiltonian.   Of course only the complete calculations
can be meaningfully compared to the data, as they include both
initial state wave functions and current operators which are
consistent with the Hamiltonian used to determine the $\tau$-
(or energy-) dependence of the response.

\begin{figure}[htb]
\centerline{\mbox{\epsfysize=70mm\epsffile{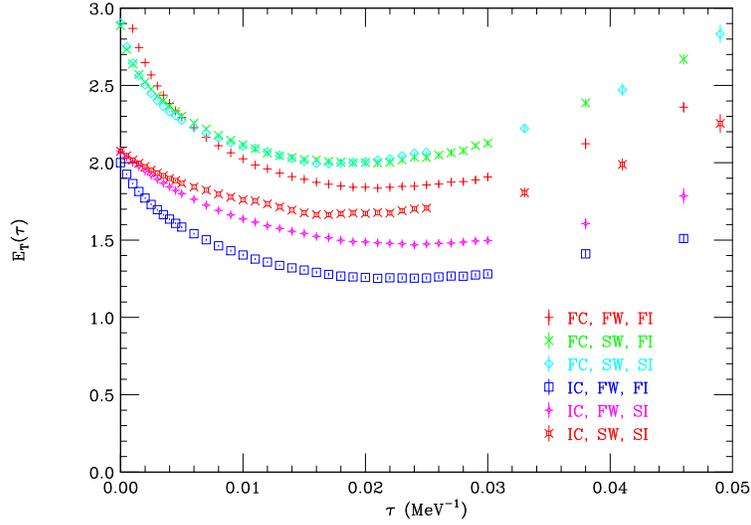}}}
\caption[]{Euclidean transverse response for $^4$He at 400 MeV/c with full (F) 
or IA (I) current (C),  full or simplified (S) wave function (W),
full or simplified interaction (I).
}\label{euclhe4t6}
\end{figure}

The transverse channel is most interesting in this regard,
as it shows a large enhancement from the two-nucleon current
operators. Results for $^4$He at 400 MeV/c with various
truncations are shown in Fig.~\ref{euclhe4t6}. The
truncations include full (FW) and simple (SW) wave functions,
full (FC) and impulse (IC) currents, and full (FI) and
simple (SI) interactions. The simple wave functions and interactions
are described below.  The differences in the
longitudinal channel are much less dramatic.

The full ground-state variational wave function is described above
(Eq. \ref{eq:varwvfn}), it includes strong tensor correlations 
from the pair correlation operators ${F}_{ij}$ and from
the three-nucleon correlation $\tilde{U}_{2\pi}$.  In order to better determine
the origin of the enhancement arising from the two-body currents, we have
also considered a simplified ground-state wave function (SW) where the
tensor correlations $u^t (r)$ and $u^{t\tau} (r)$ (Eq. \ref{eq:fij}))
and the $\tilde{U}_{2\pi}$ correlations arising from the 
two-pion-exchange three-nucleon interaction have been set to zero.

Similarly it is interesting to compare the effect of different
Hamiltonians describing the final-state interactions. In the
Euclidean response this corresponds to using different Hamiltonians
for the imaginary-time propagation of the system.  The Hamiltonian
used in the propagation does not directly affect the sum rules
which depend only upon the initial state and the current operators.
We have constructed a simplified $v_4$ interaction (SI) where the
tensor terms in the full Hamiltonian have been set to zero. This
would, of course, yield a very under-bound alpha-particle ground-state.
To compensate, we add a potential of two-pion exchange range
to both the spin one channels:
\begin{equation}
v'_{S=1;T=0,1} (r)\  =\  v_{S=1;T=0,1} (r)\  -\  1.4 \ T_{\pi}^2 (r) \>\>,
\end{equation}
where
\begin{equation}
T_{\pi}(r) = \left[ 1 - \exp (-c r^2)\right]^2 \ 
\left[ 1 + \frac{3}{\mu r} + \frac{3}{(\mu r)^2}\right]
\frac{\exp (- \mu r) }{ r}
\end{equation}
is taken from the Argonne interaction models and
is a function of two-pion exchange range.  The constant 1.4 MeV-fm has
been set to crudely reproduce the alpha-particle binding.  This
allows us to concentrate on  the spin-dependence of the final-state
interactions as opposed to drastically altering the spectra of the
struck nucleus.  In all cases the full currents (FC) are those obtained from the
Argonne $v_{18}$ interaction (AV18), they have not been reconstructed
to be consistent with the Hamiltonian used for the initial-
or final-states.  The motivation here is to examine the
various contributions to the full calculation.

From the figure it is clear that a dramatic difference remains
between full (FC) and impulse (IC) currents whatever model
is chosen for the wave function and Hamiltonian.  The $\tau=0$ (sum-rule)
difference between full and impulse currents is largest for
the full wave function, but even with a highly simplified
wave function (SW) a large difference remains between the 
results with full currents (FC, SW) and impulse currents alone (IC, SW).

On the basis of Fermi-gas calculations of matter, it had 
been believed that the large enhancement from two-body currents
found in previous calculations of light nuclei\cite{Carlson94} were due to
the presence of strong tensor correlations in the 
ground-state wave function.
While these correlations do make a significant contribution,
even simplified wave functions show a substantial enhancement.
In light nuclei, at least, this is a consequence of the
complete set of final states automatically included in the
sum rule and Euclidean response calculations.
We have also considered more drastic simplifications for the
ground-state wave function, including only central ($f^c$)
correlations.   Even in this case there is a dramatic
enhancement of the response when two-nucleon currents
are included.

Of course the Hamiltonian used for final-state
interactions cannot affect the sum rule at $\tau$=0,
but it can change the energy-dependence of the response.
Calculations with simple wave functions (SW) and the simplified
interactions (SI) are shown as diamonds in the figure.
With simplified wave functions, interactions and impulse
currents (IC, SW, SI), the slope at $\tau$=0 is much more
shallow corresponding to an energy-weighted sum rule much
closer to $k^2/(2m)$ than in the full calculation --- of course, this is
to be expected, since tensor components, missing in the SI model, substantially
enhance the energy-weighted sum rule.  This same
interaction also has a larger low-energy (large-$\tau$)
response than the calculations made with the full current.
This is undoubtedly related to the
choice of modified Hamiltonian, the choice made here
will be more attractive in p-waves than the full Hamiltonian,
and these presumably dominate the low-energy transverse
response in $^4$He.  

With the full currents (FC), there is much less dependence
upon the choice of final-state interactions.  Indeed,
the calculations with the simple wave function and full
currents (FC, SW) are nearly identical over the
range of $\tau$ considered.  The low-energy p-wave
continuum in the more attractive simplified Hamiltonian
yields less overlap with the two-nucleon current operators,
resulting in a very similar full response for the
two different final-state interactions.  The full calculation
(FC, FW, FI) has a much larger contribution at higher
energy, resulting in a steep initial fall-off with $\tau$.
It also a has a somewhat smaller response at low energy
(large $\tau$) than the full calculation.

\begin{figure}[htb]
\centerline{\mbox{\epsfysize=90mm\epsffile{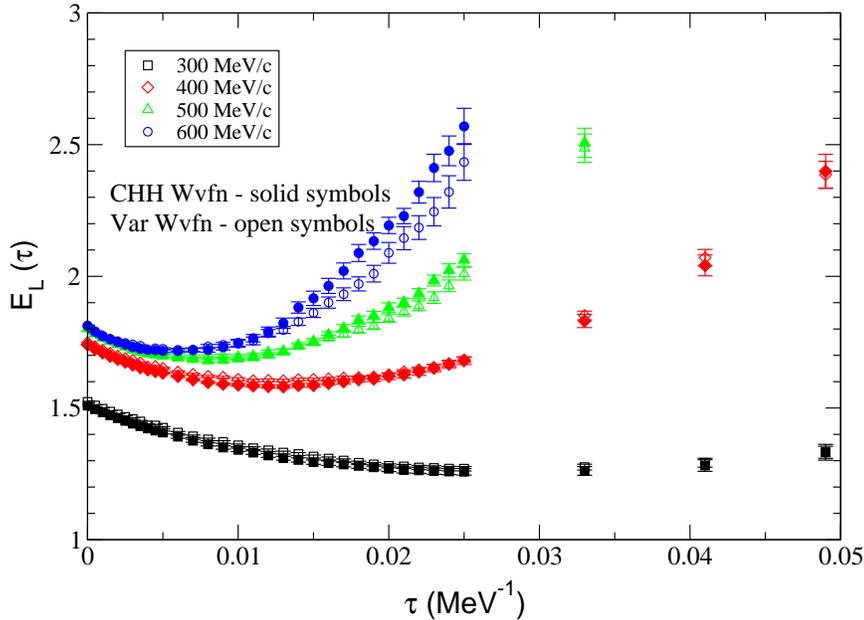}}}
\vspace*{10mm}
\caption[]{Comparison of the $^3$He Euclidean longitudinal response
functions using variational 
(open symbols)   and CHH (solid symbols) wave functions.                                         
}\label{fig:chh}
\end{figure}

Finally, we have calculated the responses in A=3 using the
correlated-hyperspherical-harmonics (CHH)
wave functions obtained by the Pisa group~\cite{Viviani95} for this
same Hamiltonian.  Calculations of the longitudinal
response of $^3$He at various momentum transfers are 
compared in Fig.~\ref{fig:chh}.  The differences between
the variational and CHH wave functions are very small, as
is apparent in the figure.  This is perhaps not surprising,
as the drastic truncations made in the comparisons of simple
(SW) and full (FW) variational wave functions were themselves
somewhat modest.  Differences in the CHH and VMC 
transverse response calculations of $^3$He are also quite small.

These calculations demonstrate that the two-nucleon currents
play a crucial role in the transverse response.  Precise comparisons
with experimental data also require calculations with 
accurate initial-state wave functions and final-state
interactions.  In such realistic calculations, the contributions
of the two-nucleon currents are large both in the integrated
response and in the low-$\omega$ regime.  

\newpage
\section{Conclusions \label{conclusions}}	
We have determined the $^3$He and $^4$He longitudinal and
transverse response functions in the momentum transfer range
300--700 MeV/c from an analysis of the (e,e$^\prime$) world
data.  The corresponding Euclidean response functions have
been derived by direct Laplace transform of the experimental
data, and have been found to be in satisfactory agreement
with those calculated with Green's function Monte Carlo methods
using realistic interactions and currents.  Leading terms of 
the two-body charge and current operators are constructed
consistently with the two-nucleon interaction included in the
Hamiltonian (the Argonne $v_{18}$).  A number of improvements
in the algorithms employed for the Monte Carlo evaluation of the
relevant path integrals have allowed us to reduce, very significantly, the
statistical errors in the Euclidean response calculations.

Two-body charge operators reduce slightly the one-body longitudinal
strength at large $\tau$ (corresponding to the threshold
region of $R_L(q,\omega)$), while two-body currents increase
very substantially, and particularly for $^4$He, the one-body
transverse strength over the whole $\tau$-range considered. 
Thus, in the quasi-elastic region, single-nucleon knock-out processes
are dominant in the longitudinal channel, while both one- and two-body
mechanisms contribute with comparable magnitude 
in the transverse channel.  These qualitative
conclusions are corroborated by the scaling analysis of the data
described in Sec.~\ref{scaling}: the longitudinal and transverse
scaling functions $f_L$ and $f_T$, which would be expected to overlap
if one-body processes alone were to be at play, display in fact drastically
different trends (see Fig.~\ref{flft}).  The enhancement in the ratio
of transverse to longitudinal quasi-elastic strength can be quantified
by considering integrals of $f_L$ and $f_T$ (of course, over the quasi-elastic
peak region alone), as done in Fig.~\ref{rat}.  Experimentally, this
$T$/$L$ ratio is found to increase very significantly from $A$=3 to 4,
to decrease only moderately from $A$=4 to 12, and to remain rather flat
as $A$=12--56.  Of course, the interpretation of the integral of
$f_T$ as reflecting exclusively quasi-elastic strength is not entirely
correct, particularly since, as the momentum transfer becomes larger
and larger, the quasi-elastic and $\Delta$ peaks tend to merge together:
strength from the pion-production region will then necessarily
spill over into the quasi-elastic region, contaminating $f_T$.
Nevertheless, the amount of \lq\lq spurious\rq\rq (non quasi-elastic)
strength contamination should not be too large, at least for
light nuclei, for which the quasi-elastic and $\Delta$ peaks
remain well separated at all momentum transfers considered here. 
The observed enhancement of the $T$/$L$ ratio in $^3$He and
$^4$He is well reproduced by theory, since the Euclidean
response functions derived from data are close to those
obtained in the calculations, over the whole $\tau$-range.

The $T$/$L$ ratio has also been studied in the $A$=3, 4, and 6
nuclei via sum-rule techniques.  Even within the limitations that such
an approach necessarily entails (see Sec.~\ref{sec:sum}), there
are rather clear indications that the present theory is
able to predict its observed dependence upon both mass number
and momentum transfer, see Figs.~\ref{fig:slta} and~\ref{fig:sltq}.
The sum rule study in Sec.~\ref{sec:sum} has also allowed us to
establish quantitatively that the excess transverse strength
associated with two-body currents is almost entirely due to
$p$$n$ pairs.  This fact has then led to the scaling law for
the excess transverse strength $\Delta S_T(q)\simeq R_A/A$, which
derives from the universal scaling behavior obtained for the
calculated $p$$n$-pair distribution functions in nuclei~\cite{Forest96}.

Finally, the role of tensor interactions and correlations
has been investigated via model studies of the $^4$He Euclidean
transverse response function, using simplified interactions,
currents, and wave functions.  In contrast to earlier
speculations~\cite{Carlson94} that the large enhancement
from two-body currents was due to the presence of strong tensor
correlations in the ground state, it is now clear that this
enhancement arises from the concerted interplay of tensor
interactions and correlations in both ground and scattering
states.
A successful prediction of the longitudinal
and transverse response functions in the quasi-elastic region
demands an accurate description of nuclear dynamics, based
on realistic interactions and currents.
\section*{Acknowledgments}
We wish to thank R.B.\ Wiringa for allowing us to use his variational
Monte Carlo wave functions, and A.\ Kievsky, S.\ Rosati, and M.\ Viviani
for providing us with their correlated-hyperspherical-harmonics wave functions.
The work of J.C. is supported by the U.S. Department of Energy under
contract W-7405-ENG-36, while that of
R.S. is supported by the U.S. Department of Energy contract DE-AC05-84ER40150
under which the Southeastern Universities Research Association (SURA)
operates the Thomas Jefferson National Accelerator Facility.
The work of J.J. and I.S. is supported by the
Schweizerische Nationalfonds.
The calculations have been performed at the National Energy Research
Supercomputer Center and at the facilities of the Accelerated Strategic
Computing Initiative at Los Alamos National Lab.
%
%
%

%\bibliography{rocco1,joe,sum2}
%
%\bibliography{[sick.nf3]sick_diff,[sick.scale]susca,ymeyer3:sick}
%\bibliography{/usr/users/jourdan/latex/sick_diff,/usr/users/jourdan/latex/sick}
\bibliographystyle{unsrt}

\end{document}